\documentclass[preprint2]{aastex6}
\usepackage{rotating}
\usepackage{graphicx}
\usepackage{amssymb}
\usepackage[caption=false]{subfig}

\shortauthors{French et al.}
\slugcomment{Submitted to ApJ November 3, 2016}
\shorttitle{Clocking the Evolution of Post-starburst Galaxies}

\newcommand{\chisq}  {$\chi^2$}

\newcommand{\fuv}{{\sl FUV }}
\newcommand{\nuv}{{\sl NUV }}
 \renewcommand{\thefootnote}{\fnsymbol{footnote}}

\begin{document}

\title{Clocking the evolution of post-starburst galaxies: methods and first results}

\author{K. Decker French \altaffilmark{1}  \altaffilmark{2} $^\dagger$, 
Yujin Yang \altaffilmark{3} \altaffilmark{4}, 
Ann I.\ Zabludoff \altaffilmark{1},
Christy A. Tremonti \altaffilmark{5}} 

\altaffiltext{1}{Steward Observatory, University of Arizona, 933 North Cherry Avenue, Tucson AZ 85721}
\altaffiltext{2}{Observatories of the Carnegie Institute for Science, 813 Santa Barbara Street, Pasadena CA 91101}
\altaffiltext{3}{Korea Astronomy and Space Science Institute, 776 Daedeokdae-ro, Yuseong-gu, Daejeon 305-348, Korea}
\altaffiltext{4}{Korea University of Science and Technology (UST), 217 Gajeong-ro Yuseong-gu, Daejeon 34113, Korea}
\altaffiltext{5}{Department of Astronomy, University of Wisconsin-Madison, Madison WI, 53706}

\begin{abstract}
Detailed modeling of the recent star formation histories (SFHs) of post-starburst (or ``E+A") galaxies is impeded by the degeneracy between the time elapsed since the starburst ended (post-burst age), the fraction of stellar mass produced in the burst (burst strength), and the burst duration. To resolve this issue, we combine {\it GALEX} ultraviolet photometry, SDSS photometry and spectra, and new stellar population synthesis models to fit the SFHs of 532 post-starburst galaxies. In addition to an old stellar population and a recent starburst, 48\% of the galaxies are best fit with a second recent burst. Lower stellar mass galaxies (log M$_\star$/M$_\sun<10.5$) are more likely to experience two recent bursts, and the fraction of their young stellar mass is more strongly anti-correlated with their total stellar mass. Applying our methodology to other, younger post-starburst samples, we identify likely progenitors to our sample and examine the evolutionary trends of molecular gas and dust content with post-burst age. We discover a significant (4$\sigma$) decline, with a 117-230 Myr characteristic depletion time, in the molecular gas to stellar mass fraction with the post-burst age. The implied rapid gas depletion rate of 2-150 M$_\sun$yr$^{-1}$ cannot be due to current star formation, given the upper limits on the current SFRs in these post-starbursts. Nor are stellar winds or SNe feedback likely to explain this decline. Instead, the decline points to the expulsion or destruction of molecular gas in outflows, a possible smoking gun for AGN feedback. 
\end{abstract}



\section{Introduction}

\footnotetext[0]{$^\dagger$ Hubble Fellow}
\renewcommand*{\thefootnote}{\arabic{footnote}}
\setcounter{footnote}{0}

Post-starburst (or ``E+A'') galaxies have been caught in the midst of a rapid transition from star-forming to quiescent. They are not currently forming stars, as indicated by their lack of significant nebular emission lines. Yet, their strong Balmer absorption lines reveal a substantial population of A stars, indicating these galaxies have experienced a burst of star formation sometime in the past billion years \citep{Dressler1983,Couch1987}. Post-starburst galaxies show disturbed morphologies and tidal features in at least half of the studied cases, providing evidence that mergers and interactions can drive this transition \citep{Zabludoff1996,Yang2004,Yang2008}. Their range of angular momentum properties is likewise consistent with a variety of possible merger histories \citep{Pracy2009, Swinbank2012, Pracy2013}.

Post-starburst galaxies represent our best candidates for the rapid, non-secular, mode of galaxy evolution \citep{Schawinski2014,Smethurst2015} that half to all red sequence galaxies are expected to experience \citep{Martin2007, Wild2016}.  Post-starburst galaxies are generally found in the ``green valley'' \citep{Wong2012} of the optical color-magnitude diagram, indicating stellar populations that could redden and evolve passively onto the red sequence. Post-starburst galaxy morphologies \citep{Yang2004, Yang2008} and spatially resolved kinematics \citep{Norton2001, Swinbank2012} are also consistent with evolution into early-type galaxies. 

Large spectroscopic surveys have allowed for the study of post-starburst galaxies as a population \citep{Zabludoff1996, Goto2003}. By studying galaxies after their starbursts are complete, and with data sensitive to the newly-formed stellar populations, we can obtain detailed information on the time elapsed since the burst, the mass produced in the burst, the overall burst duration, and whether more than one burst occurred. Previous approaches to age-dating post-starburst galaxies have often been too coarse, suffered from degeneracies between the post-burst age and burst strength, or used uncertain stelar population models.

Simple indices, such as the \ion{Ca}{2} H+H$\epsilon$ vs. \ion{Ca}{2} K lines \citep{Leonardi1996}, or $D_{4000}$ vs. H$\delta_{\rm A}$ \citep{Yagi2006} only crudely constrain the post-burst ages and burst mass fractions, and are only useful for galaxies with very high burst strengths or very young ages.

The post-burst age and burst strength can be determined by fitting the SEDs of post-starburst galaxies with templates from stellar population synthesis (SPS) models \citep{Liu1996, Barger1996, Shioya2002, Falkenberg2009, Du2010, Bergvall2016}.  While this technique benefits directly from the wide wavelength coverage and the spectral resolution of both the models and data, the optical SED cannot suitably break the degeneracy between the post-burst age, burst strength, and burst duration in many cases, and the covariance in adjacent spectral pixels must be considered to produce meaningful errors on the fit parameters. Several studies have used PCA methods on the data surrounding 4000\AA\ \citep{Wild2007, Wild2009, Wild2010, Rowlands2015, Pawlik2015} to both select and age-date post-starburst galaxies, but these methods result in a biased sample of recent SFHs (see \S\ref{sec:thesample}).

Rest-frame UV photometry is essential in breaking the degeneracies. Several studies have employed UV+optical photometry to age-date post-starburst galaxies \citep{Kaviraj2007, Kriek2010, Crockett2011,Melnick2013,Yesuf2014, Ciesla2015}, but are limited by small sample sizes, lack of spectral line information, or poorly resolved time steps or mass fraction/ burst timescale bins.

The properties of the recent starburst are connected to the physical conditions in the galaxy during the starburst, and to the mechanism that ends (``quenches") the burst. The duration of the burst is related to the quenching physics. The amount of mass produced in the starburst is related to how much gas is available and how efficiently it is funneled to the center and forms stars. The number of bursts experienced during the merger is related to the merger progenitors. These burst properties may both drive and be affected by feedback processes, whose impact may depend on the stellar mass of the galaxy \citep{Kauffmann2014, Sparre2015}.

Characterizing the starburst properties of a galaxy is also useful in connecting the starburst progenitors of these systems to their quiescent descendants. There are several proposed methods for selecting such age sequences \citep{Yesuf2014, Rowlands2015, Alatalo2016}, each with their own set of efficiencies and biases. Connecting these galaxies using their detailed SFHs can more finely select sets of galaxies that represent evolutionary sequences, and can be used to track galaxies onto the red sequence of quiescent early-type galaxies. 

The ability to connect time sequences of starbursting and post-starburst galaxies is important to understand the physics of how star formation shuts down, by identifying the likely timescales for various physical mechanisms. While simulations often assume the molecular gas reservoirs are depleted via star formation, stellar feedback, and AGN feedback, ending the starburst \citep[e.g.,][]{Hopkins2006}, recent evidence has emerged that AGN activity may be delayed after the end of the starburst. Large molecular gas reservoirs remain in post-starburst galaxies \citep{French2015}, which are otherwise consistent with evolving to early types in several Gyr. Something else must happen --- another epoch of star-formation or AGN activity --- to deplete the gas reservoirs to match those seen in early type galaxies. QSOs with post-starburst signatures have older stellar populations than some samples of post-starburst galaxies \citep{Cales2015}, which similarly indicates a delay between the end of the starburst and the period of QSO activity. Studies of AGN activity in galaxies with ongoing and recent starbursts \citep{Davies2007,Wild2010} indicate a delay of 50-300 Myr between the onset of star-formation and the onset of AGN activity. The intermediate stellar ages of Seyfert and LINER galaxies \citep{Schawinski2009} also suggest such a delay. In simulations, the delay between the starburst or merger, and the peak of AGN activity or feedback, depends on the details of how AGN feedback is implemented \citep[see e.g.,][]{Pontzen2016, Sparre2016}. To make progress, we need to compare the gas reservoirs of post-starburst evolutionary sequences. Identifying the period over which they lose their gas, and determining whether the loss can be explained by consumption by residual star formation is critical.

Because of the recent sharp end to their star-formation, the details of how star formation progressed and ended can be better constrained in post-starburst galaxies than in galaxies with less eventful star-formation histories, older galaxies, and galaxies still undergoing such a starburst. Here, we present a catalog of 532 galaxies for which we determine details of recent SFHs: post-starburst ages, burst strengths, evidence for multiple bursts, and burst durations. We discuss the data used and post-starburst selection in \S2, the age-dating technique and discussion of biases and sources of systematic error in \S3, results relating to constraints on the recent SFHs in \S4, and results relating to the evolution of gas and dust in the post-starburst phase in \S5.

\section{Sample Selection and Data}
\subsection{Post-Starburst Galaxy Sample}
\label{sec:selectionsdss}

\begin{figure}
\includegraphics[width = 0.5\textwidth]{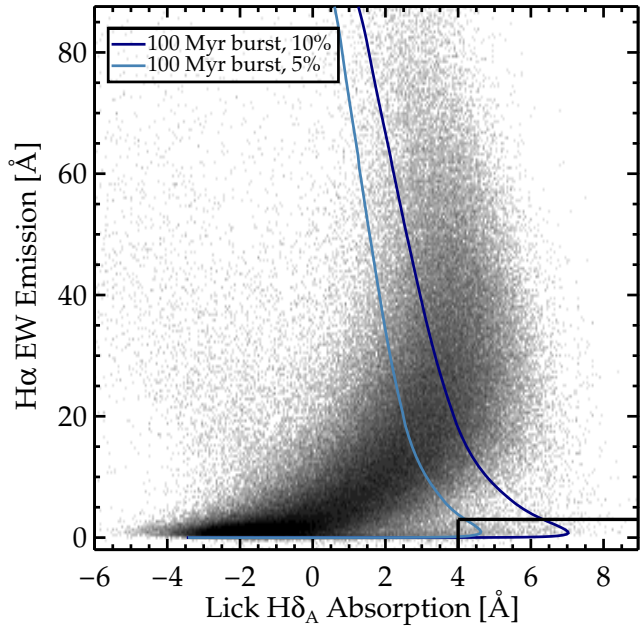}
\caption{H$\delta$ absorption vs. H$\alpha$ emission equivalent width for our parent sample of 595,268 galaxies in the SDSS DR8. The region outlined in black shows our post-starburst selection criteria. To select galaxies with little-to-no current star formation, we require H$\alpha$ EW $<$ 3\,\AA. To select galaxies with a recent ($\lesssim$ Gyr) starburst population indicated by strong Balmer absorption lines, we require  H$\delta_{\rm A}$ $-$ $\sigma$(H$\delta_{\rm A}$) $>$ 4\,\AA, where $\sigma$(H$\delta_{\rm A}$) is the measurement error of the H$\delta_{\rm A}$ index. Post-starburst galaxies are located on this spur of the distribution of blue cloud and red sequence galaxies. We show two example tracks of a 100 Myr burst added to an old stellar population, with a mass fraction of 5 or 10\%. A starburst must form a substantial fraction of the galaxy's stars, and be over a short duration, in order to go through the post-starburst spur. Star-forming galaxies in the parent sample are at higher H$\alpha$ absorption, and moderate H$\delta_{\rm A}$ (the turnover in H$\delta_{\rm A}$ at high H$\alpha$ is due to absorption line filling). Quiescent galaxies have little H$\alpha$ emission and little H$\delta$ absorption. Due to our use of the A-star optimized H$\delta_{\rm A}$ index, these galaxies extend to negative values.}
\label{fig:psbselection}
\end{figure}

Our parent sample is drawn from the DR8 SDSS main galaxy spectroscopic sample \citep{Strauss2002, Aihara2011}, using the galaxy properties from the MPA-JHU catalogs \citep{Brinchmann2004, Tremonti2004}. We exclude galaxies with $z <0.01$ to eliminate those that are much larger than the the 3\arcsec \ diameter of the SDSS fibers (we explore aperture bias \S\ref{sec:apbias}). We also exclude galaxies with unreliable H$\alpha$ equivalent widths (we require \texttt{h\_alpha\_eqw\_err > -1}) or median (over the whole wavelength range) signal-to-noise values of less than 10 per pixel. These cuts ensure that the line index measurements are reliable.  Our final parent sample from DR8 is composed of 595,268 galaxies.

We select post-starburst galaxies from our parent sample by identifying galaxies with strong stellar Balmer absorption lines, characteristic of a recent ($\lesssim$ Gyr) starburst, and little nebular emission, indicating a lack of significant on-going star formation (Figure \ref{fig:psbselection}).  We use the Lick H$\delta_{\rm A}$ index \citep{Worthey1997} to characterize the stellar Balmer absorption. We do not correct the index for filling due to nebular emission. This effect is negligible for our post-starburst galaxies given that they have weak nebular lines by definition and H$\delta$ will be at most $\sim$8\% of the H$\alpha$ emission line. We require H$\delta_{\rm A}$ $-$ $\sigma$(H$\delta_{\rm A}$) $>$ 4\,\AA, where $\sigma$(H$\delta_{\rm A}$) is the measurement error of the H$\delta_{\rm A}$ index. We include $\sigma$(H$\delta_{\rm A}$) in our selection criteria to elliminate spurious objects because H$\delta_{\rm A}$ measurements can be noisy (median $\sigma$(H$\delta_{\rm A}$) $\sim$ 0.48 \AA\ in the parent sample). We select for galaxies that have little on-going star formation by requiring H$\alpha$ EW $<$ 3\,\AA\ in the rest frame. We require that the SDSS spectra have no gaps over H$\alpha$ or H$\delta$ in the rest frame. These selection criteria result in a sub-sample of 1132 galaxies from the parent sample (0.2\%). We discuss the biases that result from this sample selection in \S\ref{sec:thesample}.

Figure \ref{fig:psbselection} shows the distribution of H$\alpha$ EW and H$\delta_{\rm A}$ in our parent sample. The region delineated by the black lines represents our selection criteria for post-starburst galaxies.  Star-forming galaxies in the parent sample are at higher H$\alpha$ emission and moderate H$\delta_{\rm A}$ absorption. For star-forming galaxies with more H$\alpha$ emission, the Lick H$\delta_{\rm A}$ index will be partially filled, resulting in the ``turnover" in H$\delta_{\rm A}$ at high H$\alpha$ seen in Figure \ref{fig:psbselection}. Quiescent galaxies have little H$\alpha$ emission and little H$\delta$ absorption. Due to our use of the A-star optimized H$\delta_{\rm A}$ index, these galaxies extend to negative values. This correlation is expected for galaxies forming stars roughly continuously, because H$\alpha$ emission and H$\delta$ absorption both trace star formation, but on different timescales: H$\alpha$ emission is powered by O stars which have lifetimes shorter than 10 Myr, whereas stellar Balmer absorption is produced by A stars with lifetimes of 0.3 - 1.3 Gyr. Post-starburst galaxies are visible as a distinct spur of points with small H$\alpha$ EW and large H$\delta_{\rm A}$, well-separated from the populations of passively-evolving red galaxies and actively star forming galaxies. We also show two example tracks of a 100 Myr burst, with 5\% (or 10\%) of the current stellar mass produced in the recent burst and 95\% (or 90\%) from an old stellar population. A starburst must form a substantial fraction of the galaxy's stars over a short duration to go through the post-starburst spur.

\subsection{Optical Photometry and Spectroscopy}

We use the $ugriz$ magnitudes and Lick indices from the SDSS in our age-dating procedure. The emission line and absorption line index measurements are described in \citet{Tremonti2004} and \citet{Brinchmann2004}. We use the following Lick indices:  {\tt D4000\_Narrow, Lick\_CN2,  Lick\_Ca4227, Lick\_G4300, Lick\_Fe4383, Lick\_Ca4455, Lick\_Fe4531, Lick\_C4668, Lick\_Hb, Lick\_Fe5015, Lick\_Mg1, Lick\_Mg2, Lick\_Mgb, Lick\_Fe5270, Lick\_Fe5335, Lick\_Fe5406, Lick\_Fe5709, Lick\_Fe5782, Lick\_TiO1, Lick\_TiO2, Lick\_Hd\_A, Lick\_Hg\_A}. We eliminate NaD due to concerns about possible contamination from interstellar absorption. We only use the A-star defined Lick H$\delta$ and H$\gamma$ indices, and exclude the F-star defined quantities to avoid duplicating lines. Similarly, we exclude Lick CN1, using CN2 instead.

For the photometric data, we adopt the {\tt modelmag} magnitudes, which provide stable colors while containing most of the galaxy light \citep{Al.2004}. We make small corrections to the $u$ and $z$ bands ($-$0.04 and 0.02 mag) to put them on the correct AB magnitude system\footnote{{\tt http://www.sdss3.org/dr8/algorithms/fluxcal.php}}.  In addition to photometry errors given in the SDSS catalog, we add the magnitude zero-point errors (5\%, 2\%, 2\%, 2\%, and 3\% in $ugriz$, respectively) in quadrature to ensure that we obtain a realistic \chisq\ values in the SED fitting  \citep{Blanton2007}\footnote{{\tt http://kcorrect.org}}.

\subsection{UV Photometry}

For each of the post-starburst galaxies selected from the SDSS, we search for matching {\it GALEX} \citep[Galaxy Evolution Explorer;][]{Martin2004} \nuv and \fuv detections using the {\it GALEX} GCAT All-Sky Survey Source Catalog (GASC) and GALEX Medium Imaging Survey Catalog (GMSC) catalogs\footnote{{\tt https://archive.stsci.edu/prepds/gcat/}}. We search for galaxies within 4\arcsec\ of the SDSS positions. This radius is similar to the FWHM of the \nuv PSFs and much larger than the {\it GALEX} astrometric uncertainties (0.59\arcsec\ in {\sl FUV}). In the GCAT catalog, we find 729 galaxies (64\%) with \nuv detections of at least $3\sigma$, and 532 galaxies (47\%) with additional \fuv detections. In this paper, we use only the galaxies with both \nuv and \fuv detections, as the UV data is essential for breaking the age-burst strength-burst duration degeneracy. 

We use the {\tt MAG\_FUV} and {\tt MAG\_NUV} magnitudes from the GCAT catalogs, which were determined from the SExtractor {\tt AUTO} magnitudes. These magnitudes should represent the total galaxy light, and thus are comparable to the SDSS\, {\tt modelmag} magnitudes \citep{Al.2004}.  The zero-point calibration errors of 0.052 and 0.026 mag \citep{Morrissey2007} are added to the \fuv and \nuv photometry errors respectively to reflect the uncertainties in photometric uniformity. The addition of these zero-point errors to the formal photometric errors is critical in our study, because we combine photometric measurements from different databases to construct SEDs.  Under-estimated errors could lead us to inflated \chisq\ values and therefore to unrealistic uncertainties in the post-burst ages and the burst mass fractions derived from our SED fitting procedure.

We test whether the {\it GALEX} photometry is affected by a lack of deblending, by comparing the SDSS and {\it GALEX} centroids. The mean difference is 1.1\arcsec, and only 3\% have centroids more than 3\arcsec apart. The lack of cases where the centroids are more offset than expected from the PSFs indicates that blending from other sources is not a significant effect.

\section{Age Dating Technique}

\subsection{Stellar Population Synthesis Models}

The details of the SPS models used in our analysis are especially important given the presence of evolving low to intermediate mass stars after the end of the starburst. The thermally pulsing asymptotic giant branch (TP-AGB) phase is especially uncertain, so different SPS model treatments of the TP-AGB phase can have large impacts on the predicted post-starburst SEDs. Specifically, the \citet{Bruzual2003} models feature a less prominent contribution from this phase than the \citet{Maraston2005} models. The \citet{Maraston2005} models predict younger ages than the BC03 models \citep{Maraston2006}. 

Analyses of post-starburst galaxies \citep{Kriek2010} have found that the \citet{Maraston2005} models generally overpredict the TP-AGB luminosity. Similar results are found by \citet{Zibetti2012} and \citet{Melnick2013}. Thus, previous studies of the ages of recently quenched galaxies \citep{Schawinski2007} that use the \citet{Maraston2005} models may be subject to error. The FSPS (Flexible Stellar Population Synthesis) set of SPS models \citep{Conroy2009, Conroy2010} includes two parameters to tune the shift in $T_{\rm eff}$ and $L_{\rm bol}$ of the TP-AGB track on the HR diagram, from the tracks predicted by \citet{Marigo2008}. These models include a best fit of $\Delta T_{\rm eff}$ and $\Delta L_{\rm bol}$ to data from the SDSS and 2MASS \citep[2 Micron All Sky Survey][]{Skrutskie2006}.

We age-date our sample of post-starburst galaxies using both the BC03 and FSPS models and do not find a difference in the resulting ages above the statistical error. This is likely due to the greater influence of age on the {\it GALEX} UV data points, as the TP-AGB contribution is primarily in the NIR, which is less sensitive to the details of the recent star formation histories. The remainder of this work uses only the FSPS models. 

Here, we use the FSPS models to generate stellar population synthesis models for several families of physically-motivated SFHs for post-starburst galaxies. We use the Miles stellar libraries and Padova isochrones. We consider several possibilities for a younger stellar population from the recent starburst (\S\ref{sec:sfh}) on top of and older, pre-starburst, stellar population (\S\ref{sec:old}). 

\subsection{Recent Star Formation Histories}
\label{sec:sfh}

\begin{figure*}[!htb]
\includegraphics[width = 1.0\textwidth]{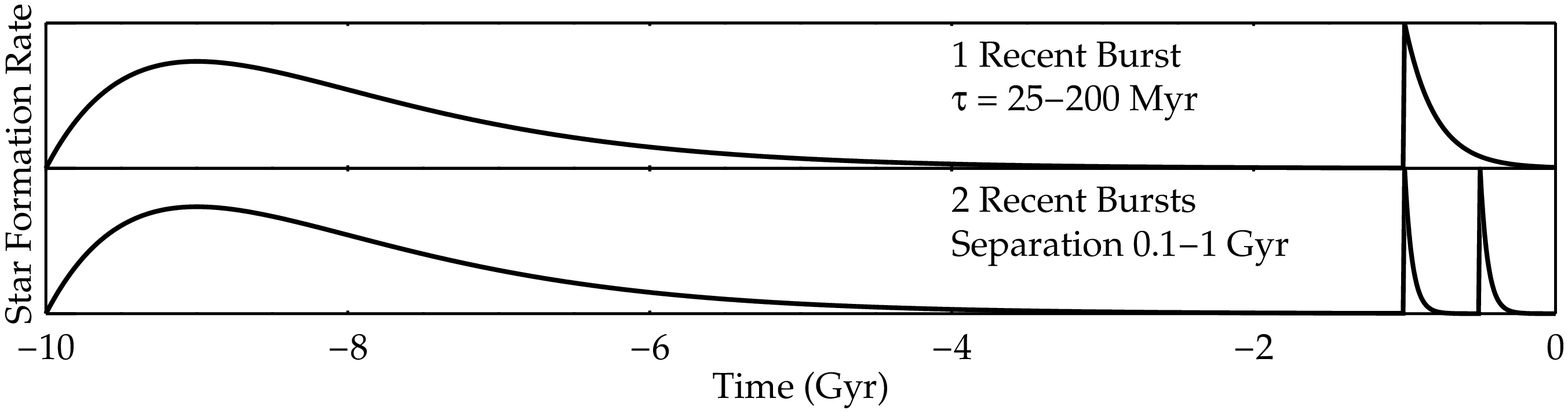}
\caption{Star Formation History (SFH) models used here in fitting the post-starburst SFHs. The old stellar population is modeled by a linear exponential. The young stellar population is fit to two different classes of SFHs, those with 1 recent burst, varying the burst duration, and those with 2 recent bursts, varying the separation between bursts. Post-starburst galaxies are likely post-merger, so we use simulations of gas-rich major mergers to motivate the range of recent SFH used (see \S\ref{sec:sfh}). The burst durations range from 25 Myr to 150 Myr for the single burst models. We do not have the sensitivity to distinguish the burst mass fractions or durations of the individual bursts in the double recent burst model, so we assume the bursts are each 25 Myr exponentially declining models, and form equal stellar masses. We instead vary the separation between each burst, from 100 Myr to 1 Gyr.}
\label{fig:sfh1}
\end{figure*}

Post-starburst galaxies are likely the result of recent major mergers \citep[e.g.][]{Zabludoff1996, Yang2004, Yang2008, Pawlik2015}, showing evidence of disturbed morphologies and the young stellar populations we are interested in here. We motivate the details of our assumed SFHs using simulations of gas-rich major mergers. Simulations of merging galaxies predict the triggering of a starburst, with one \citep[e.g.][]{Snyder2011,Hayward2014} or two \citep[e.g.][]{Mihos1994,Cox2008,Renaud2014} exponentially declining starburst events. In a merger, gas is funneled into the center of the galaxy, causing an initial burst during the first galaxy-galaxy tidal passage, and another upon coalescence. If the galaxies already have a bulge in place, it acts to stabilize the gas from collapse, and most of the new stars will be formed in a single burst. If the bulge is not in place, some of the gas forms stars during the first passage, resulting in two bursts of star formation. The timescale of the separation between bursts depends on the initial conditions of the galaxies' relative positions and velocities before the merger. 

Multiple starburst episodes can also occur when stellar feedback is important \citep[e.g.,][]{Muratov2015}. Galactic outflows observed in post-starburst galaxies \citep{Tremonti2007} have been proposed to arise from stellar feedback \citep{Diamond-Stanic2012}. Multiple episodes of star-formation are expected to occur in low mass galaxies, with observations constraining the duty cycles of these ``bursty'' star formation histories \citep{Lee2009,Geha2012}. However, some observations of long duration starbursts in low mass galaxies have shed doubt on whether these bursty SFHs can be explained with stellar feedback alone \citep{McQuinn2010}.

Because post-starburst galaxies are primarily found in poor groups \citep{Zabludoff1996, Hogg2006}, we do not include ``truncated" SFHs in our stellar population fitting method, as in \citet{Ciesla2015}, as such models are motivated by processes unique to dense cluster environments, such as ram pressure stripping.

We test several different models for the recent SFH (Figure \ref{fig:sfh1}). Motivated by the simulations described above, we choose two different classes of SFHs: one or two recent bursts, with an early period of star formation at high redshift. We then vary the duration of the exponentially declining burst in the former and the separation between the bursts in the latter. Exponentially declining (``$\tau$") bursts are seen in simulations, but we also test a gaussian-shaped recent burst, at different durations. The post-burst ages fit using the two models are consistent, using the definition below. The $\tau$ model bursts produce better fits to the data than the gaussian shaped bursts, so we use these in all cases.

For the two recent burst model, we do not have the sensitivity to fit separate ages, burst strengths, durations, and the time between bursts, due to degeneracies in the spectra. We choose five burst separations between 100 Myr and 1 Gyr, typical of those seen in the simulation results discussed above. We set the burst durations to $\tau=$25 Myr and require the mass fractions to be equal between the two recent bursts. 

To properly compare the post-burst ages from different SFH fits, we define the ``post-burst" age to be the time elapsed since the burst was complete, instead of the time elapsed since the burst began. Because the SFHs we consider have different burst durations and numbers of recent bursts, this convention allows the post-burst ages to be compared in a physically meaningful way. While we are interested in the post-burst age for studying how post-starburst galaxies evolve through this phase, a different convention may be desirable for other purposes, such as timing the onset of various evolved stellar populations. We include both the age since the starburst began and the post-burst age in Table \ref{table:ages}. 

We define the ``post-burst age'' to be the time elapsed since the majority (90\%) of the stars formed in the recent burst(s). For the case of two recent bursts, the post-burst age is thus the time since 90\% of the stars in the two recent bursts were formed. We choose this convention by fitting single population recent bursts $+$ a 10 Gyr old single stellar population to synthetic UV-optical SEDs for the single and double burst SFHs described above, with typical uncertainties added, then comparing the age of the single stellar population model to that of the single or double exponentially declining bursts. The age since 90\% of the stars formed, or ``post-burst age", is most consistent with the age obtained by using a single stellar population to model the recent burst. This definition is thus comparable to a light-weighted young stellar population age, as derived using a ``K+A" model \citep[e.g.,][]{Quintero2004}; it also allows comparisons between post-starbursts with single vs. double burst recent SFHs, to explore how the galaxies evolve after the burst has ended (see \S5). 

\begin{figure*}
\includegraphics[width = 1\textwidth]{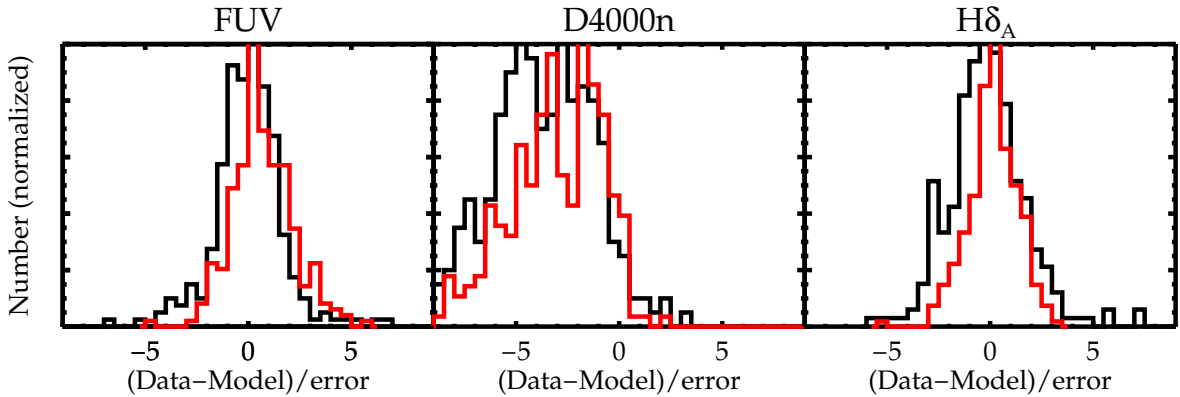}
\caption{Histograms of fitting residuals for the {\it FUV} flux, D$_n$(4000) index, and Lick H$\delta_{\rm A}$ index, for galaxies best fit by a double recent burst. The double burst model fit residuals are shown in red, and the (not preferred) single burst model fit residuals are shown in black. For galaxies that strongly prefer the double recent burst model, the fitting residuals are narrower, and less biased than the single burst model fitting residuals for the same galaxies. }
\label{fig:chiplot}
\end{figure*}

We demonstrate how the double burst model is a better fit to many of the galaxies, by plotting the fitting residuals for the {\it FUV} flux, D$_n$(4000) index, and Lick H$\delta_{\rm A}$ index in Figure \ref{fig:chiplot}. For galaxies that strongly prefer the double recent burst model, the fitting residuals are narrower and less biased than the single burst model fitting residuals for the same galaxies. The addition of an intermediate F star population allows for these indices, as well as many of the iron-influenced Lick indices, to be better fit, while still providing a good fit to the UV colors and Balmer absorption indices. We ultimately find that 50\% of the galaxies prefer the single burst model, and 48\% prefer the double burst model. Only 11 galaxies (2\%) do not show a statistical preference given the error in $\chi^2$. For these galaxies, we nonetheless assign the model with the lower $\chi^2$, although this does not affect our conclusions.

\subsection{SED fitting}

To determine the time elapsed since the starburst ended, what fraction of the stellar mass was produced during the burst, the duration of the recent burst, and whether there was more than one burst, we model the SEDs of post-starburst galaxies as a combination of old and new stellar population. The old stellar population is modeled as a linear-exponential star formation rate over time $t_{old}$:
\begin{equation}
\Psi \propto t_{old} e^{-t_{old}/\tau_{old}}
\label{eqn:old}
\end{equation}
beginning 10 Gyr ago and characterized by the timescale $\tau_{old} = 1$ Gyr. 
The young stellar population's star formation history is modeled as either one or two exponential declining components in the star formation rate over time $t_{young}$. For one recent burst:
\begin{equation}
\Psi \propto e^{-t_{young}/\tau}.
\label{eqn:young}
\end{equation}
We vary the time $t_{SB}$ since this recent period of star formation began\footnote{Equations \ref{eqn:old} and \ref{eqn:young} are related by $t_{old}-t_{young} = 10$ Gyr $- t_{SB}$.} as well as the characteristic timescale $\tau$. 

 For two recent bursts, 
\begin{equation}
\Psi \propto e^{-t_{young}/\tau} + e^{-(t_{young} - \Delta t)/\tau}, 
\label{eqn:young2}
\end{equation}
where $\tau=25$ Myr for each, and the separation $\Delta t$ between the two recent bursts is 0.1-1 Gyr.

We use the flexible stellar population synthesis (FSPS) models of \citet{Conroy2009, Conroy2010} to construct model template spectra. We assume a metallicity $Z$ using a stellar mass prior (\S\ref{sec:metallicity}), a Calzetti reddening curve, and a Chabrier IMF. The effects of these assumptions, as well as the assumed SFHs, are discussed below (\S\ref{sec:metallicity}, \S\ref{dust}, \S\ref{imf}).

The observed spectrum is modeled as a linear combination of the young and old stellar templates:
\begin{equation}
f_{model}  = [y f_{young} + (1-y) f_{old}] \times 10^{-0.4 k(\lambda) A_V},
\end{equation}
where $k(\lambda)$ is the reddening curve as a function of the wavelength $\lambda$, $A_V$ is the amount of extinction expressed in magnitudes of $V$-band absorption, $f_{young} (\lambda; t_{SB}, \tau/\Delta t, Z)$ is the young stellar population spectrum (arising from Eqs. \ref{eqn:young} and \ref{eqn:young2}) with an SFR decay rate of $\tau$ (or the separation $\Delta t$ between 2 recent bursts), and $f_{old}(\lambda; Z)$ is the old stellar population spectrum (arising from Eq. \ref{eqn:old}). $Z$ is the stellar metallicity assumed, using the priors described below (\S\ref{sec:metallicity}). Each spectrum is normalized within the rest-frame 5200--5800 \AA\ wavelength window, and $y$ represents the fraction of the total galaxy light in the young stellar template. The mass fraction of new stars $m_f$ is derived from $y$ and $t_{SB}$. Thus, we parameterize the spectrum using four free parameters, $t_{SB}$, $A_V$, $y$, and $\tau$. 

The priors on these four parameters are as follows. $A_V$: [0,2] magnitudes, spaced linearly, $t_{SB}$ (age since starburst began): [30, 2000] Myr, space logarithmically, $y$: [0.01, 1], spaced logarithmically, $\tau$ = [25,50,100,150,200] Myr \footnote{We find that we cannot distinguish statistically between $\tau = 150$ Myr and $\tau = 200$ Myr, so this last option is not used in the following analysis.} or $\Delta t$ = [100,200,300,500,1000] Myr. The priors on $A_V$ were set by the typical dust attenuation seen in SDSS galaxies \citep{Brinchmann2004}. The minimum age prior is set to be smaller than the minimum time any galaxy would take to enter the post-starburst selection criteria. The maximum age prior is set to be well after all galaxies would have exited the post-starburst selection critera. Similarly, starbursts with $y<0.01$, or $\tau>200$ Myr will never be selected as post-starbursts. We tested smaller values of $\tau$, but found this method cannot distinguish statistically amongst burst durations shorter than 25 Myr. 

We compare the SDSS $ugriz$ and {\it GALEX} $FUV, NUV$ photometry, and 22 Lick indices (29 total data points) to synthetic photometry and Lick indices calculated from the model spectra. While we would ideally make use of every spectral data point available, the use of the full SDSS spectra is complicated by the covariance in the spectral data points. Thus, we extract information from the SDSS spectra using the Lick indices alone \citep{Worthey1994, Worthey1997}. While these data points are not truly independent due to the astrophysical processes underlying their relative strengths, we avoid instrumental and calibration uncertainties in interpreting the spectra. We determine the best fit model using $\chi^2$ minimization and marginalize over all other parameters to determine the 68\% likelihoods for each parameter. The errors on the data and Lick indices are taken from the SDSS catalogs described above, and we assume the error distributions are Gaussian.

\subsection{Early Star Formation History}
\label{sec:old}

The overall star formation history of galaxies can be approximated as a linear-exponential, $\Psi_{old} \propto t e^{-t/\tau}$ \citep{Simha2014}. We use this model for the old stellar population in our stellar population fitting, with 1 Gyr as the characteristic duration $\tau$ (see Figure \ref{fig:sfh1}). To test how this choice of old stellar population model affects our results, we replace the linear-exponential model with a single-age stellar population, placed at different times 10 Gyr to 1 Gyr before the starburst. We find that the post-burst ages derived do not change by more than the formal fit error, so long as the old stellar population is $>4$ Gyr before the starburst. Thus, the linear-exponential model with $\tau$=1 Gyr is indistinguishable from a single stellar population at any time $>$4 Gyr before the starburst, in terms of its effects on our results. Old SFHs with substantial star-formation extending to the present produce poor fits to the data, as they are not flexible in varying the amount of light from old vs. intermediate vs. young stars.

However, the SFR before the burst is expected to be non-zero, and typical of gas-rich disk galaxies before a major or minor wet merger. We test the effect of adding an additional SFH component, with at a constant SFR over the 10 Gyr prior to the starburst. For consistency with the prior analysis, we define the post-burst age as the age since 90\% of the A stars were formed. As long as the constant SFR component does not dominate the light ($y_{constant} <0.5$), the difference in derived post-burst age is less than the statistical + systematic error.

\subsection{Metallicity}
\label{sec:metallicity}

To fit the post-starburst data to SPS models, we must assume or fit stellar metallicities. The choice of metallicity is important in determining the post-burst ages, as changes in Z will result in systematic differences in the post-burst ages. Traditionally, the age-metallicity degeneracy can be broken using the Lick index system \citep{Worthey1994,Worthey1997}, which we make use of in our fitting procedure. However, we cannot simultaneously break the age-burst strength-burst duration degeneracy and the age-metallicity degeneracy to the age precision required to track galaxies across the post-starburst stage. The photometry of these galaxies is unhelpful in fitting the stellar metallicities, as the extremely blue colors bias the metallicities low. Ignoring the photometry and fitting the models only to the Lick indices is not sufficient, as the difference in ages in our fitting grid is small compared to the range of metallicities allowed. 

We test two priors on metallicity in our method: a constant solar value or a stellar-mass dependent value using the \citet{Gallazzi2005} mass-metallicity relation. We use the second case throughout, as it is more physically motivated, and produces smaller average $\chi^2$ values. To estimate the systematic error in post-burst age, we propagate the range in likely metallicities for each stellar mass from \citet{Gallazzi2005} through the SED fits. We draw 10 randomly-selected metallicities from the mass-metallicity relationship, using the error bounds in \citet{Gallazzi2005}, assuming a Gaussian distribution, and using the stellar masses from the SDSS MPA-JHU catalogs \citep{Brinchmann2004, Tremonti2004}. We determine the median scatter in the age and burst mass fraction in these 10 trials for each post-starburst galaxy in the sample. The median error resulting from this analysis is 14\% of the post-burst ages and 23\% of the burst mass fractions. We discuss the treatment of these systematic errors in \S\ref{sec:systematic}. We note that the stellar masses predicted at the end of this process are within the fit errors of those by the MPA-JHU group ($\sim 0.15$ dex for this sample).

Should post-starburst galaxies lie on the mass-metallicity relation generated from all SDSS absorption line galaxies? If post-starburst galaxies are to evolve to the early-type galaxies that lie on this relation, without further episodes of star formation, they should experience no evolution on it. Short bursts of star formation can increase the stellar mass before the ISM is enriched for further star formation, resulting in a bias low in the M-Z relation. However, galaxies with longer duration bursts, multiple bursts, or those that start out at a higher Z for their stellar mass, will compensate for this effect.

\subsection{Dust Extinction}
\label{dust}
\begin{figure*}
\includegraphics[width = 0.5\textwidth]{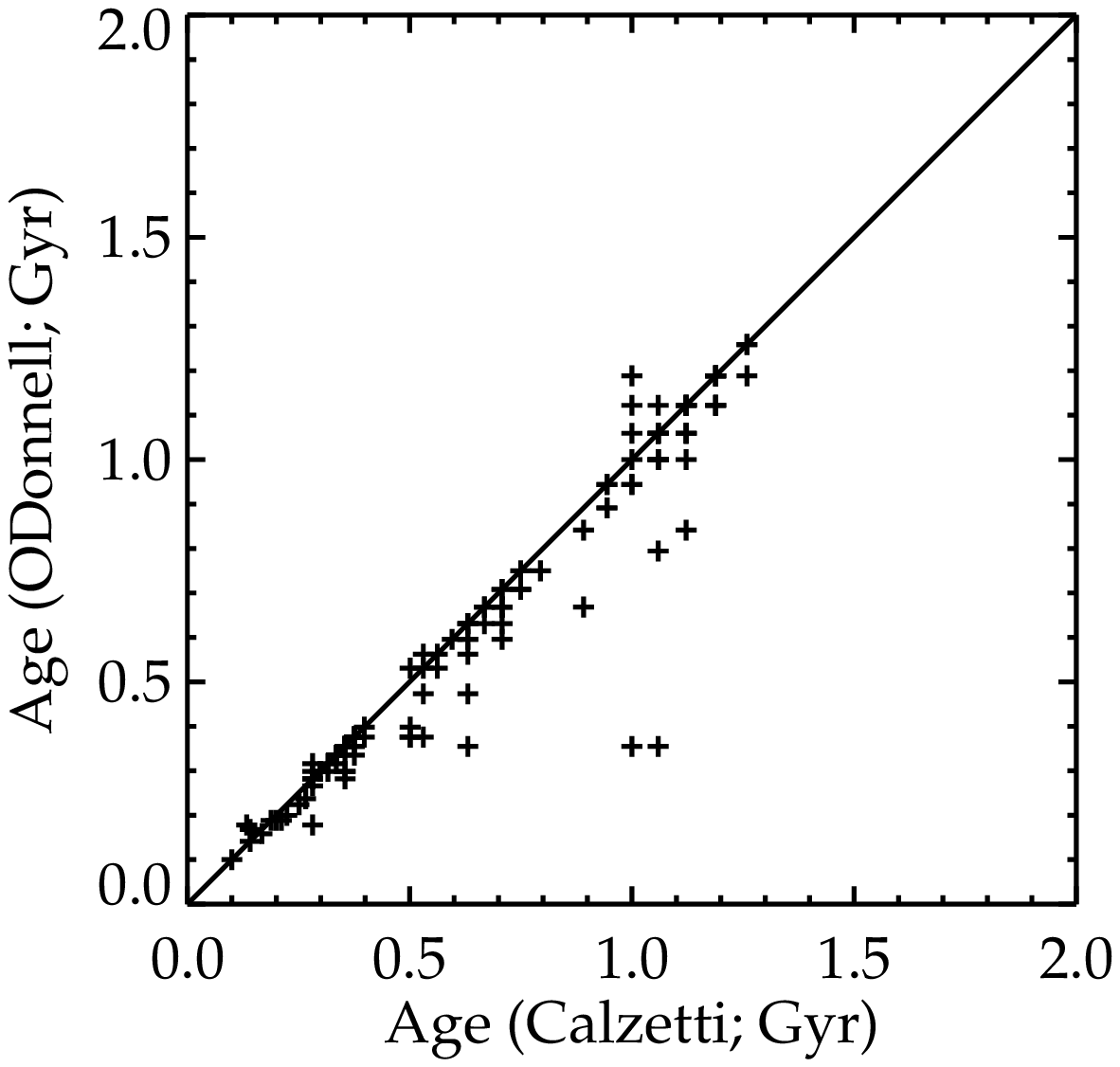}
\includegraphics[width = 0.5\textwidth]{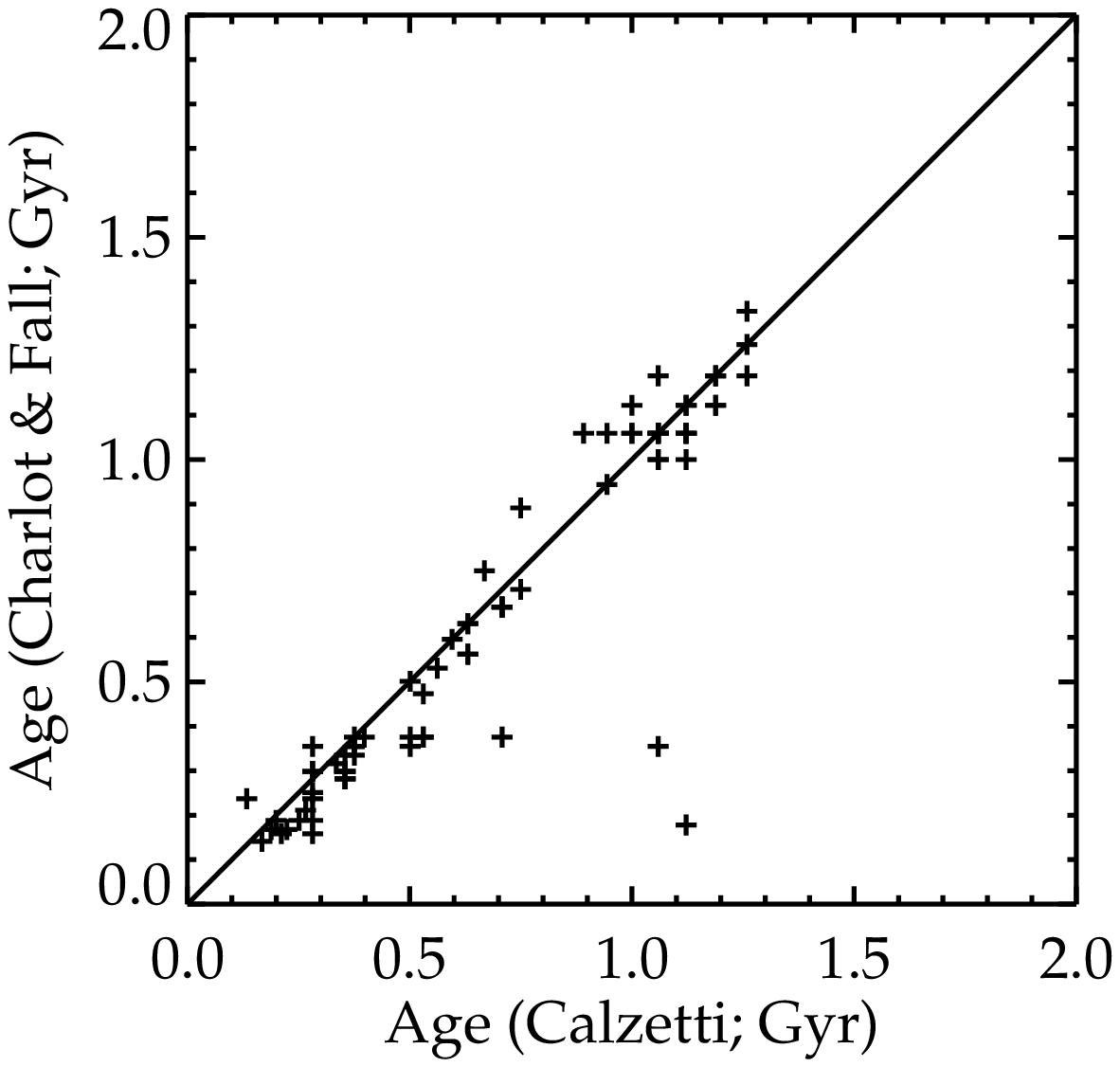}

\includegraphics[width = 0.5\textwidth]{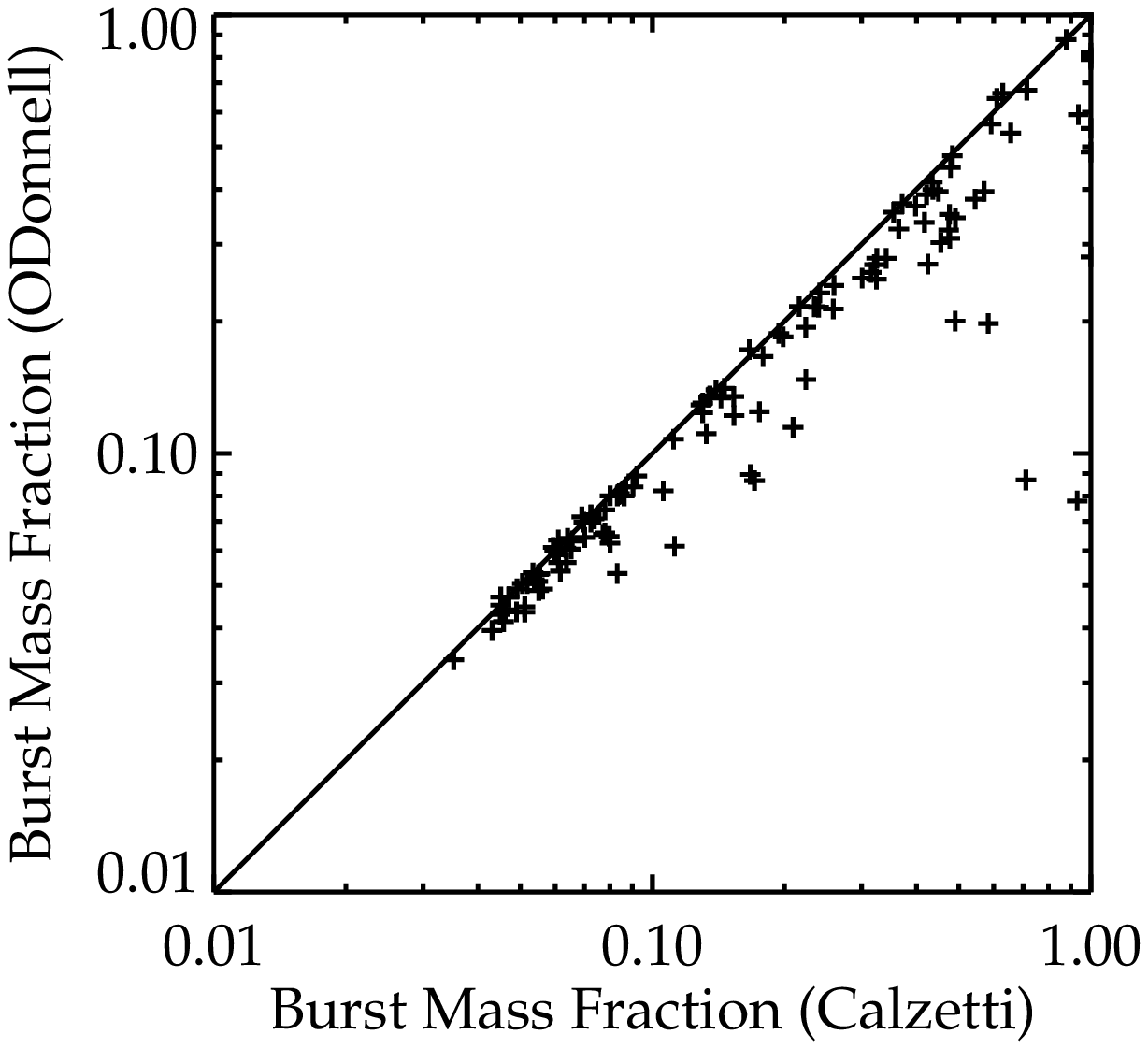}
\includegraphics[width = 0.5\textwidth]{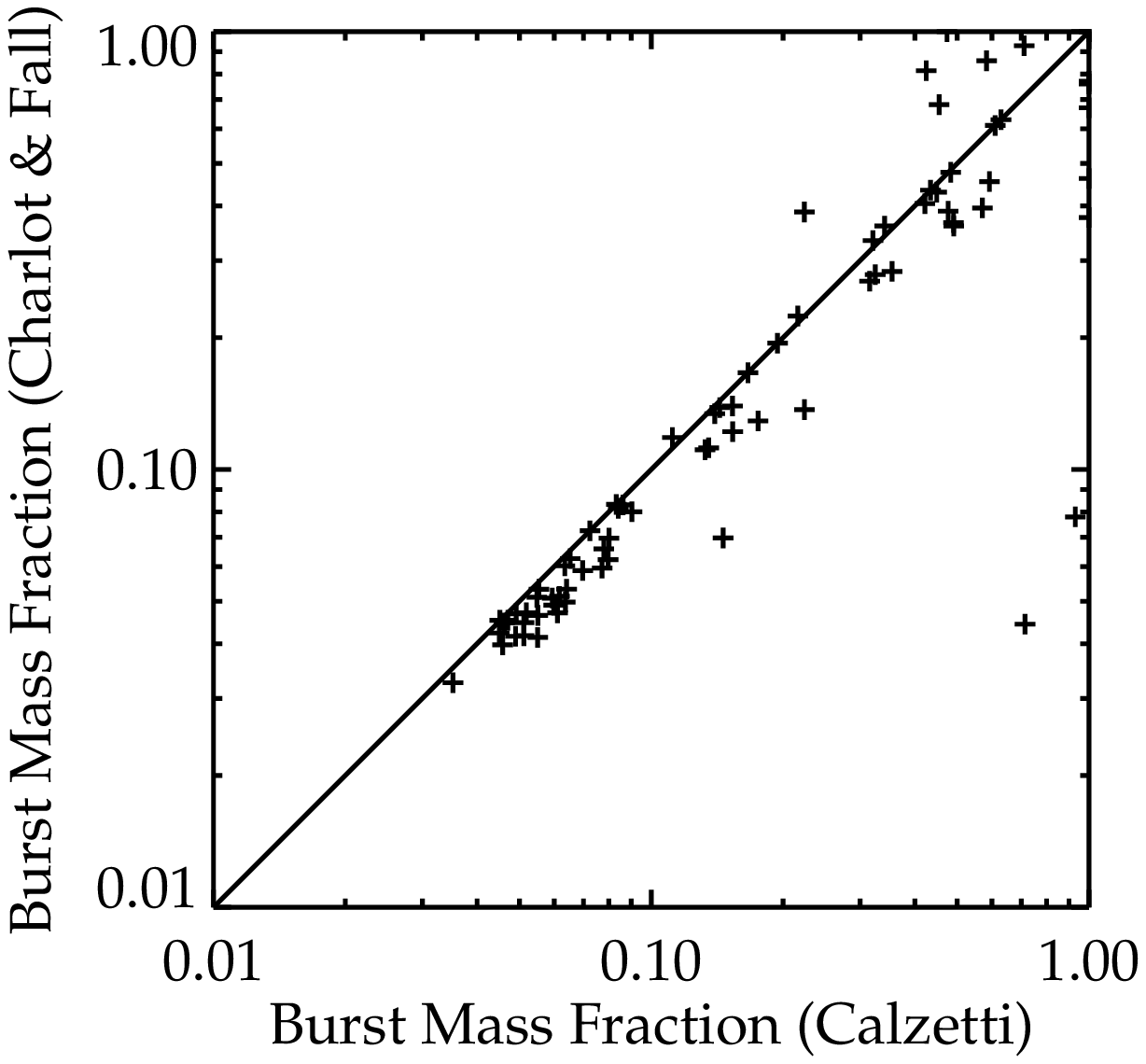}
\caption{Ages and burst mass fractions fit using either Calzetti \citep{Calzetti2000}, Charlot \& Fall \citep{Charlot2000}, or O'Donnell \citep{O'Donnell1994} extinction laws, for a test of 100 post-starburst galaxies. Our derived parameters are robust to the choice of extinction law, and the only significant outliers are those with large fit errors on the derived parameters.}
\label{fig:dust}
\end{figure*}

We have assumed a \citet{Calzetti2000} extinction law in our model fits so far. The characteristic extinction in this model has uncertainty R$_V = 4.05 \pm 0.80$, but because we fit A$_V$ directly, only A$_V$ is sensitive to a change in R$_V$. Assuming the reddening curves of Charlot \& Fall \citep{Charlot2000} or O'Donnell \citep{O'Donnell1994}, we do not find significant changes in the extinction (E(B-V)) or post-burst ages or burst mass fractions measured. The results of these tests are plotted in Figure \ref{fig:dust}.

We have an independent constraint on A$_V$ through the Balmer decrement. We examine the H$\alpha$ and H$\beta$ fluxes from the MPA-JHU dataset \citep{Aihara2011}.  We assume the standard case B recombination at T=$10^4$ K and an intrinsic value of 2.86. In general, the extinction fit from the SPS models is less than that derived from the Balmer decrement, although consistent with the errors propagated from uncertainties in the emission line fluxes. This is not unexpected \citep{Hemmati2015}, as the dust surrounding the A star population (measured in our SED fits) may cause less extinction that the dust surrounding the nebular line emission regions. We allow for the foreground Galactic extinction to be fit alongside the extragalactic extinction. The median measured extinction of A$_V$= 0.5 mag is much larger than the typical foreground Galactic extinction of A$_V$=0.08 mag from the MPA-JHU catalogs.

\subsection{IMF}
\label{imf}
\begin{figure}
\includegraphics[width = 0.5\textwidth]{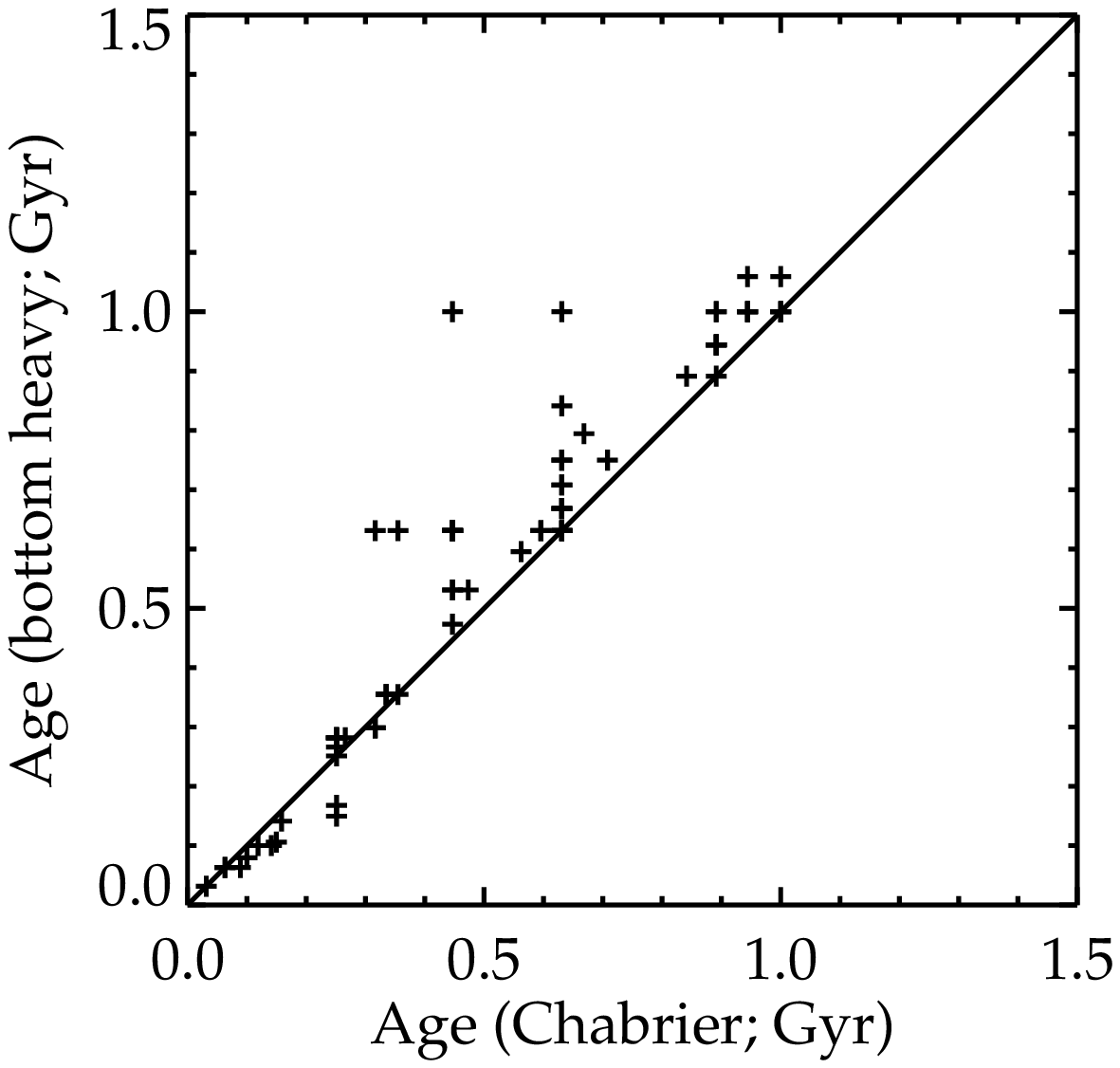}
\includegraphics[width = 0.5\textwidth]{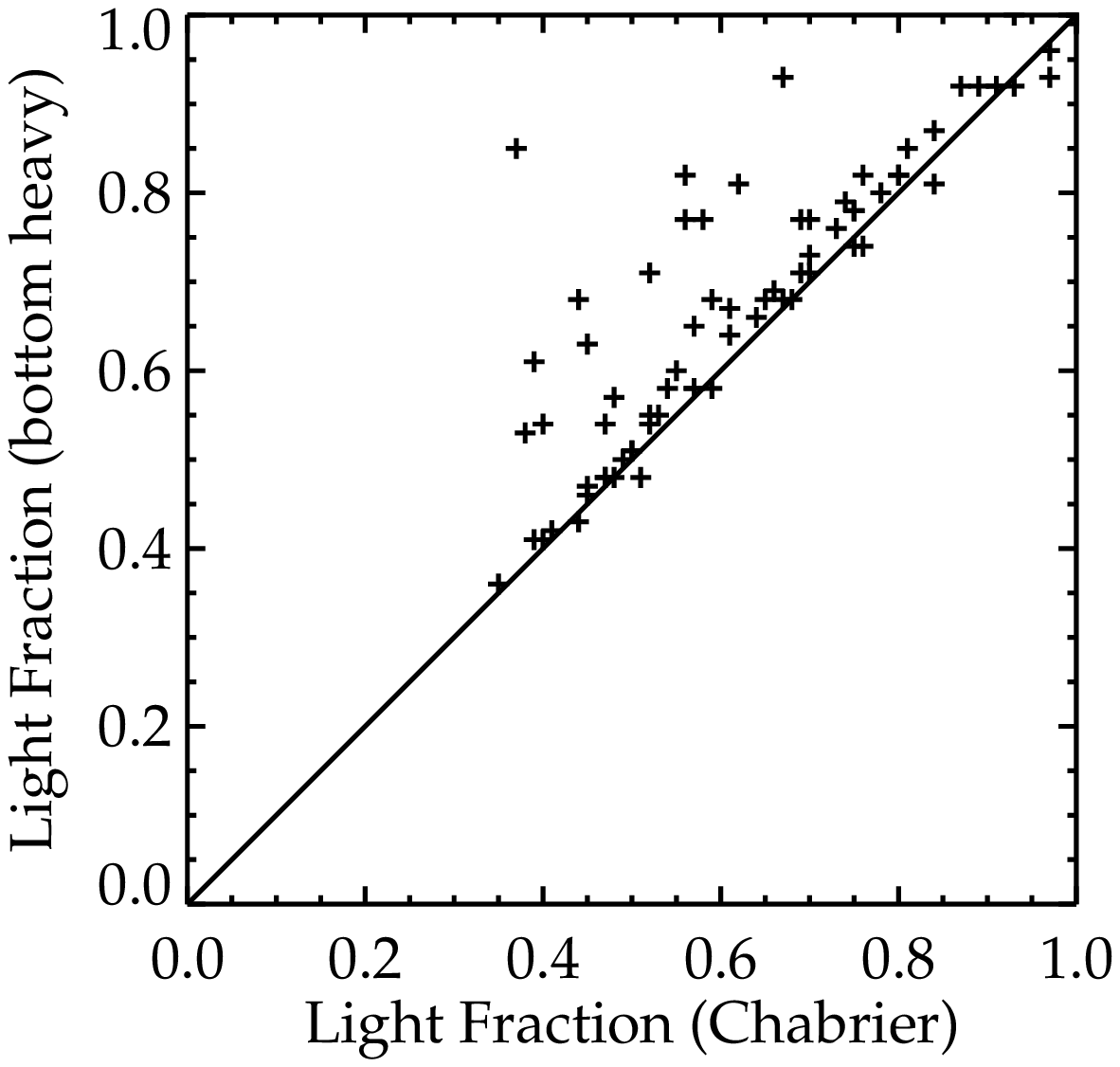}
\caption{Ages and burst light fractions when assuming a Chabrier IMF vs. a bottom-heavy IMF with slope $x=-3$ from stellar masses $0.1-100 M_\sun$, for a test of 100 post-starburst galaxies. The change in IMF primarily affects the light or mass fraction inferred, as the IMF effectively re-weights the old and young stellar contributions in each burst. A more bottom-heavy IMF (fewer bluer high-mass stars) looks similar to a lower burst mass fraction (which would also have fewer bluer high-mass stars). The difference in derived ages is greater than the fit errors in only 29\% of cases, and the difference in derived light fractions greater than the fit errors in 18\% of cases, both within the expected number for the 68\% error ranges. 
}
\label{fig:imf}
\end{figure}

In our analysis, we have assumed a Chabrier IMF \citep{Chabrier2003}. However, there is some evidence that the IMF varies between galaxies, especially in early-types \citep{Cappellari2012, vanDokkum2012}. We test the effect on the age-dating fits if we assume instead a bottom-heavy IMF, with slope $x=-3$ from stellar masses $0.1-100 M_\sun$. The change in IMF primarily affects the light or mass fraction measured, as the IMF effectively re-weights the old and young stellar contributions in each burst. The results of this test are plotted in Figure \ref{fig:imf}. We compare the difference in fit results to the fit errors estimated using each IMF. While a bottom heavy IMF would lead to measuring systematically younger ages and lower light fractions, the difference in derived ages is greater than the fit errors in only 29\% of cases, and the difference in derived light fractions greater than the fit errors in 18\% of cases, both within the expected number for the 68\% error ranges. 

\subsection{Aperture Bias}
\label{sec:apbias}

Because the post-starburst galaxies have larger angular sizes than the 3\arcsec\ fibers with which they are selected and characterized, these methods may suffer from aperture bias. A key concern in post-starburst selection is to select only truly post-starburst galaxies, excluding post-starburst nuclei surrounded by a star forming disk. We select only galaxies with $z>0.01$ to avoid this case. Indeed, IFU observations of more local post-starbursts \citep{Pracy2014a} have found pockets of star formation outside of the post-starburst nucleus of galaxies. However, these star forming regions are only $\sim500$ pc from the center, and would have been included inside a 3\arcsec\ diameter fiber for galaxies with $z>0.017$ (or 99\% of our sample). 

\begin{figure}
\includegraphics[width = 0.5\textwidth]{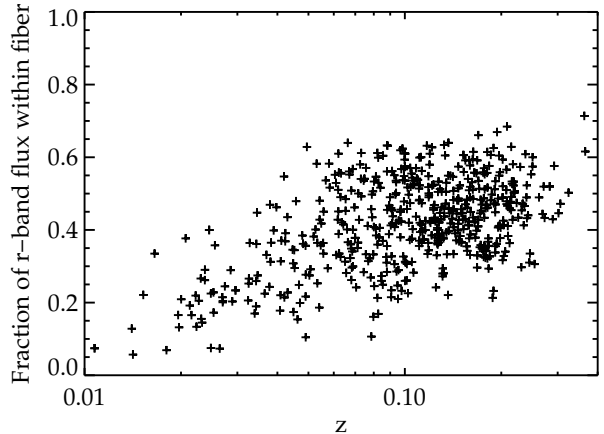}
\caption{Fraction of $r$-band flux that falls within the 3\arcsec\ SDSS fiber vs. redshift. Only 5\% of the sample has $<20$\% of its total flux within the SDSS fiber, where errors in determining global galaxy properties become large \citep{Kewley2005}. We do not remove these galaxies from our sample, as they do not have different distributions in any of the derived properties. }
\label{fig:apbias}
\end{figure}

Fiber-based estimates of global galaxy properties are known to have large errors due to aperture bias when the total flux gathered by the fiber is less than 20\% of the total flux \citep{Kewley2005}. In Figure \ref{fig:apbias} we plot the fraction of the $r$-band flux captured by the SDSS fiber for our post-starburst sample. Only 5\% have $<20$\% of the total flux within the SDSS fiber (for any band), and we do not observe trends in any of the derived galaxy properties with the fraction of flux inside the fiber. The centrally concentrated light in post-starbursts results in less aperture bias than in our parent sample of SDSS galaxies: the median flux within the fiber is 43\% for our post-starburst sample. For the parent sample of SDSS galaxies described above (selected to have $z>0.01$), the median flux within the fiber is 30\%, and 25\% have $<$20\% of the total flux captured by the SDSS fiber.

One concern is the dilution of the burst signatures, when the fiber samples an area greater than the extent of the starburst \citep{Snyder2011}. This effect acts to decrease the H$\delta$ index, such that higher redshift galaxies will have weaker Balmer absorption lines. We thus caution that the parameters fit here represent the area of the fiber sampled. The area over which the starburst took place is both observed \citep{Swinbank2012} and predicted \citep{Snyder2011} to vary, depending on the progenitors and configuration of the triggering merger. To fully resolve the issue of how aperture affects measurements of the post-starburst properties of galaxies, spatially resolved spectra are needed. Future IFU surveys such as MANGA \citep{Drory2014} will contribute to the growing number of post-starburst galaxies with resolved spectroscopy. 

Another concern is that a mismatch between the vital {\it GALEX} photometry and SDSS fiber may bias the results. The FWHM of the {\it GALEX} PSFs are 4.9\arcsec\ and 4.2\arcsec\ for the {\it NUV} and {\it FUV} bands respectively, making it difficult to determine whether the flux is as centrally concentrated as the optical light. We estimate the effect this may have by comparing the $u-r$ colors from the {\tt modelmag} and {\tt fibermag} magnitudes. The difference between these colors is on average only 34\% of the uncertainty in these colors, so the mismatch in apertures is unlikely to have a severe impact on the quantities derived from the SED fitting.

\subsection{Error Estimation}
\label{sec:systematic}

\begin{figure}
\includegraphics[width = 0.5\textwidth]{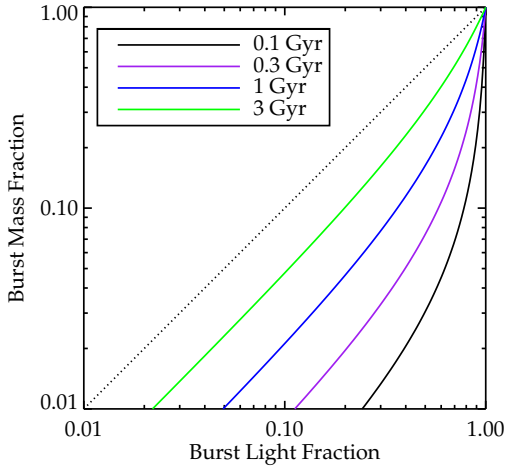}
\caption{Burst light fractions vs. burst mass fractions for four post-burst ages, for a single recent burst with $\tau=25$ Myr. At very young post-burst ages, the light from the young stellar population will dominate the total light for any mass fraction $>10$\%. As a result, large uncertainties exists in many of the burst mass fractions, especially for short post-burst ages. }
\label{fig:yfracmfrac}
\end{figure}

While we parameterize our models using the light fraction of the recent burst(s) compared to the old population, we are interested in the burst mass fraction, which we derive from the light fraction. The conversion from light fraction to mass fraction depends on the post-burst age and burst duration ($\tau$ or $\Delta t$), as younger populations will have lower mass to light ratios. The error in the burst mass fraction is calculated by determining the likelihood function from the light fraction and age likelihood grid.  We plot the light fraction vs. mass fraction for several post-burst ages for the single burst models (Figure \ref{fig:yfracmfrac}). At very young post-burst ages, the light from the young stellar population dominates the total light for any mass fraction $>10$\%. As a result, large uncertainties exist in many of the burst mass fractions, especially for short post-burst ages. 

\begin{figure*}
\includegraphics[width = 0.33\textwidth]{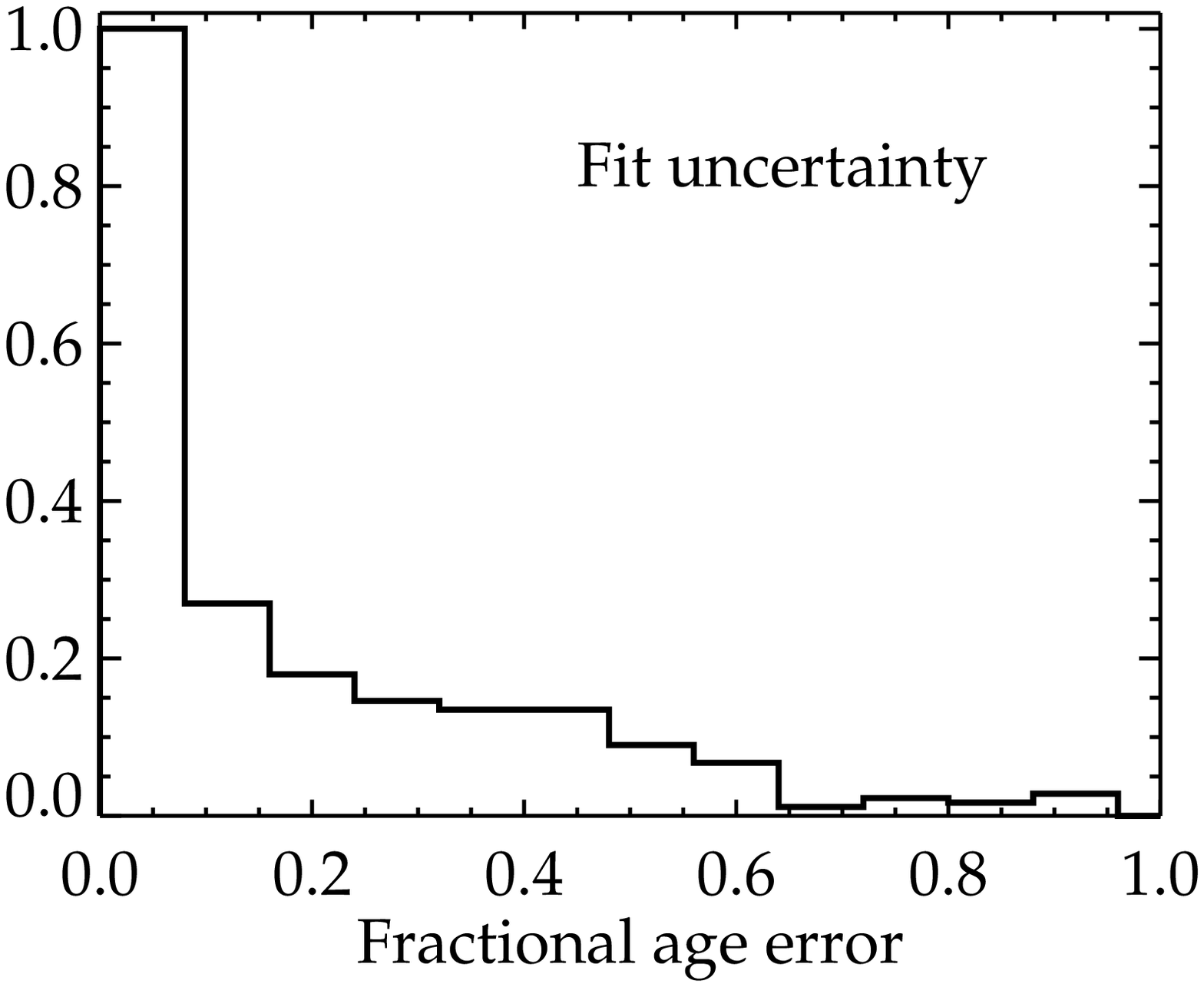}
\includegraphics[width = 0.33\textwidth]{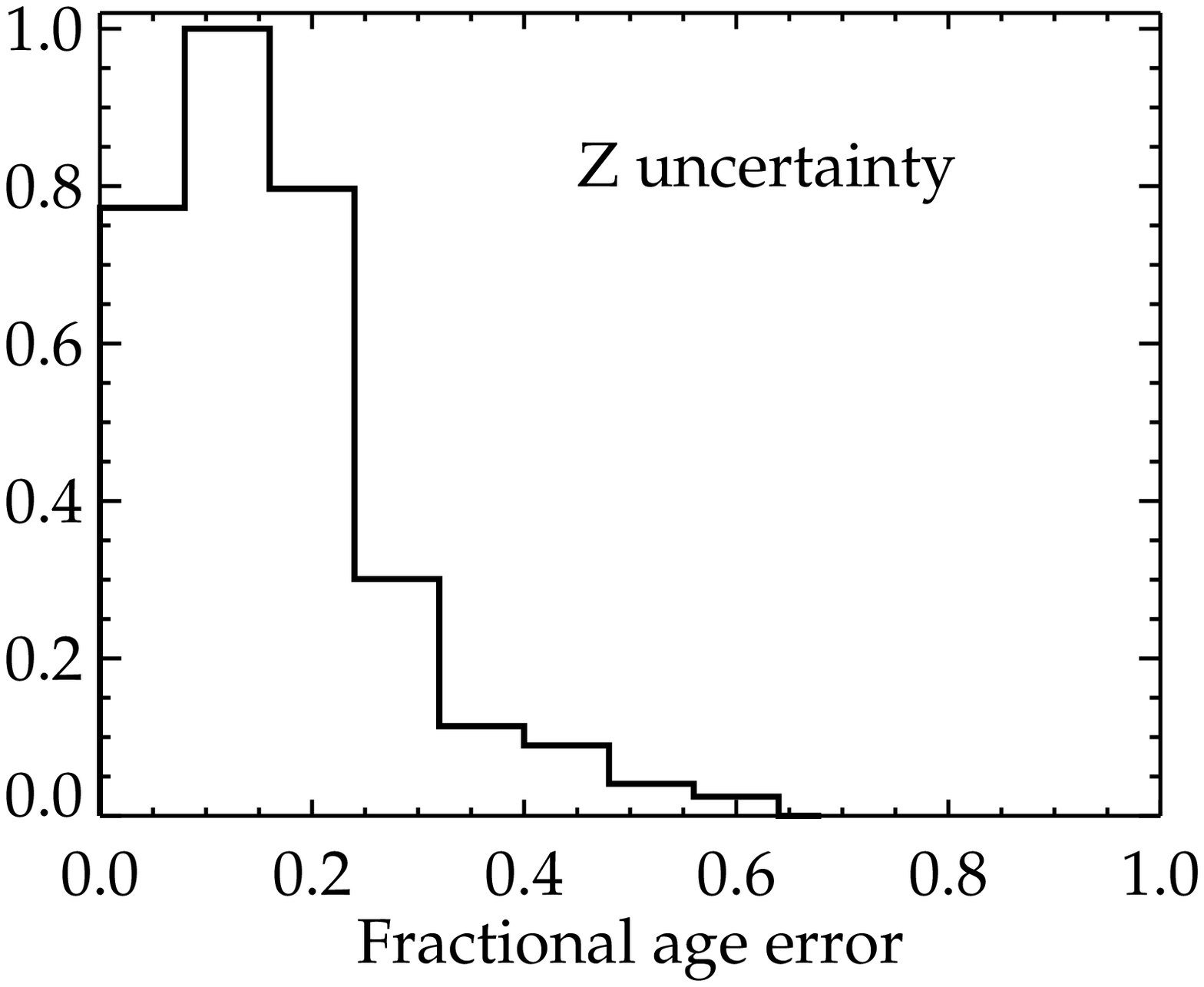}
\includegraphics[width = 0.33\textwidth]{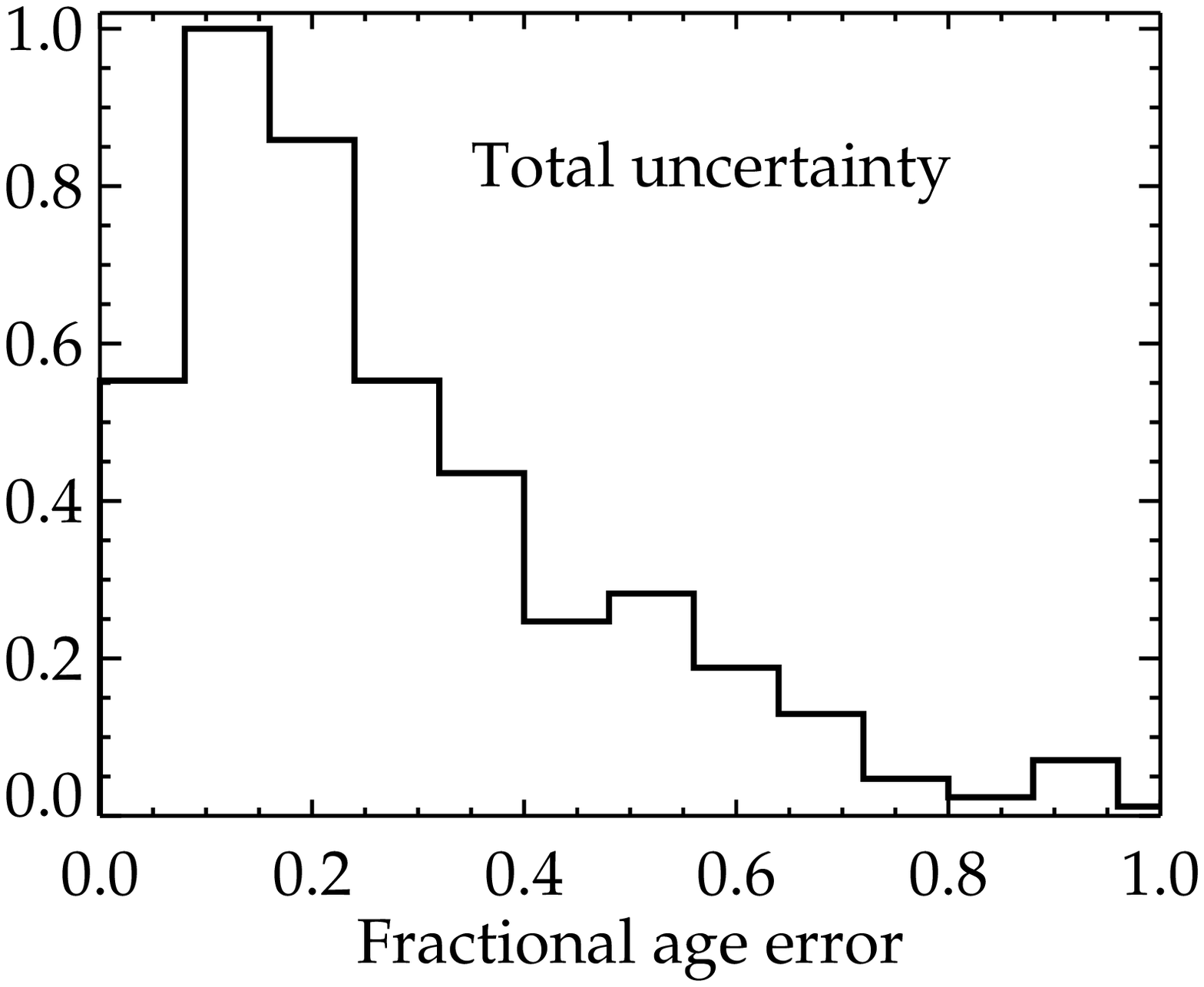}

\includegraphics[width = 0.33\textwidth]{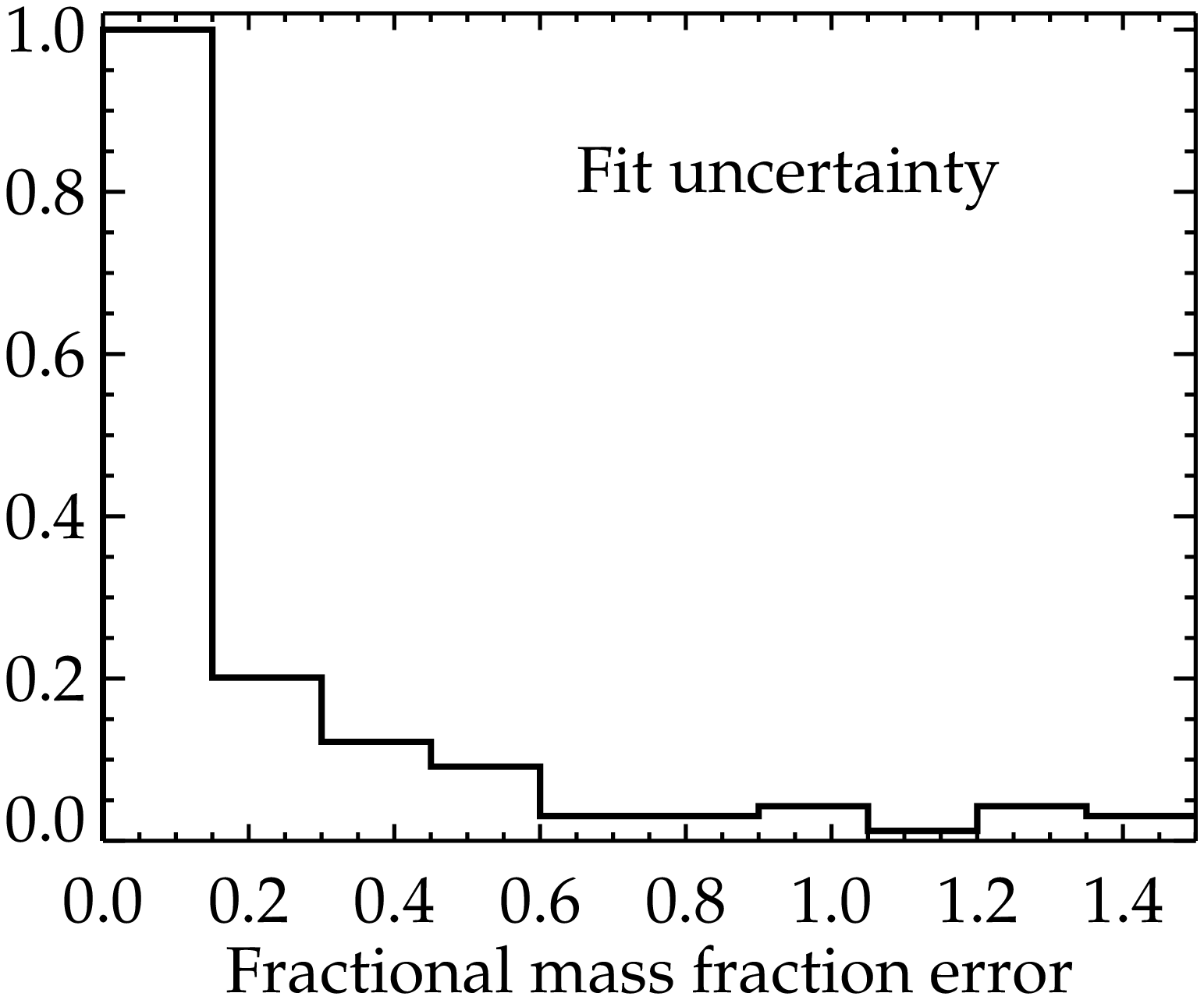}
\includegraphics[width = 0.33\textwidth]{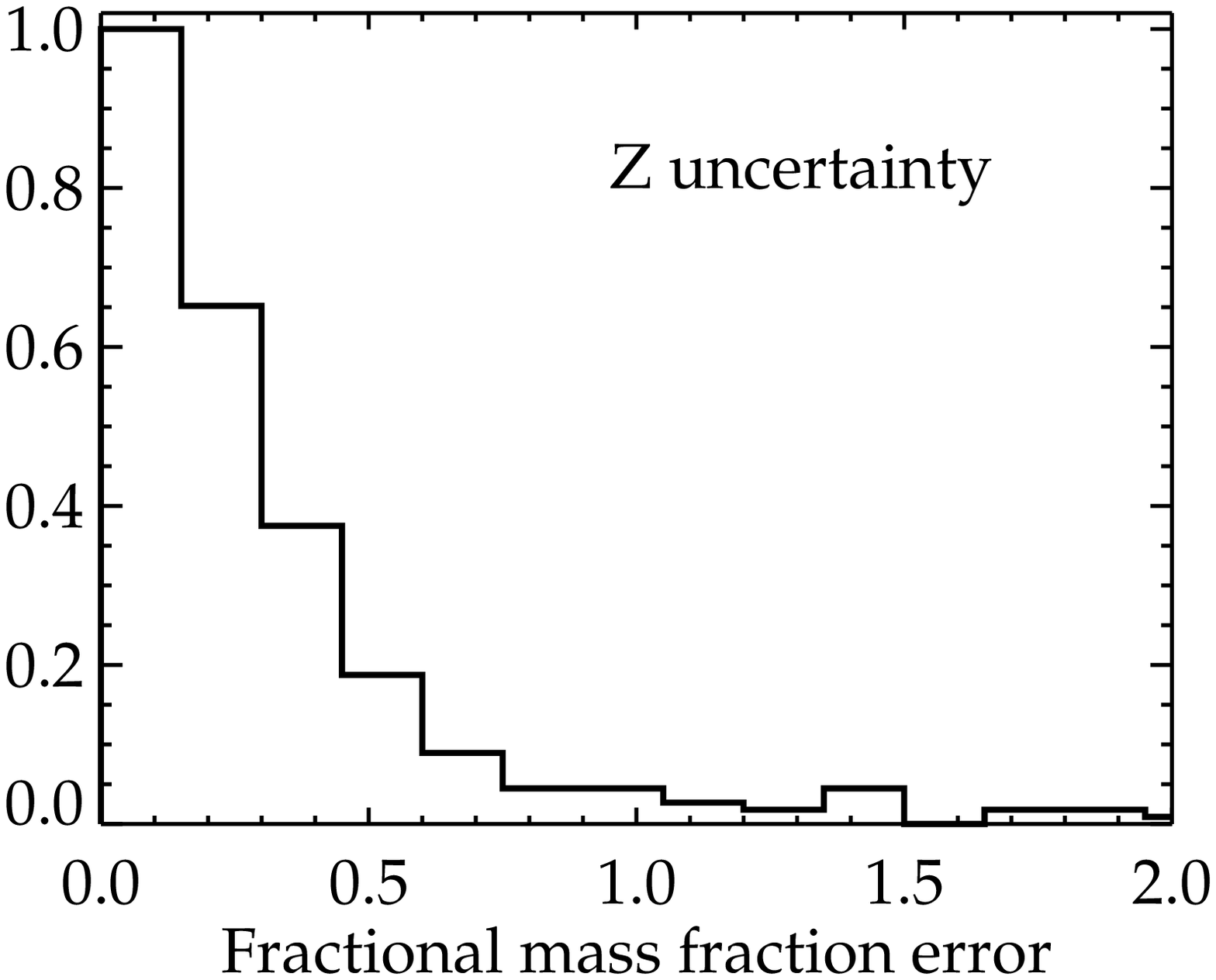}
\includegraphics[width = 0.33\textwidth]{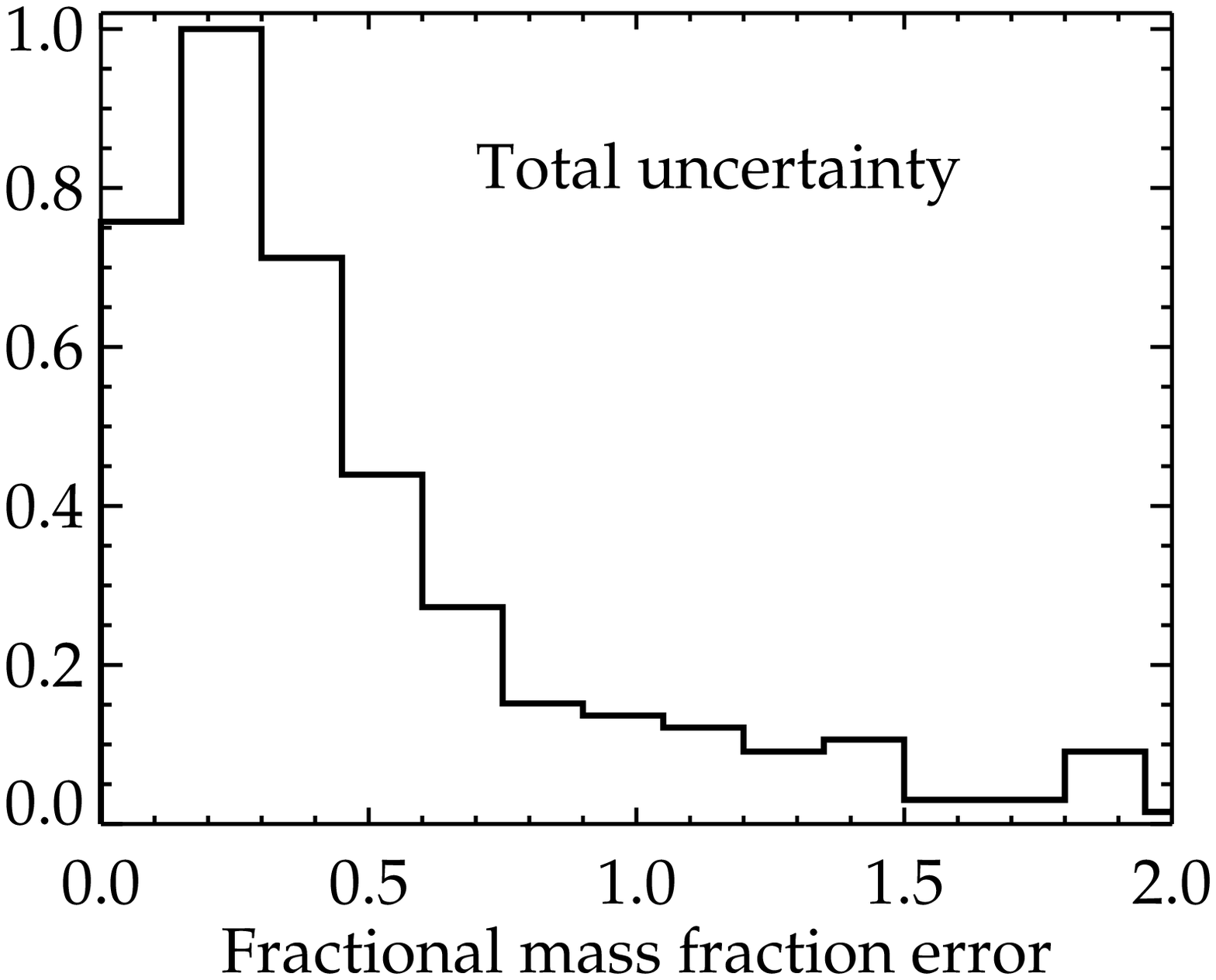}
\caption{Normalized histograms of fractional errors on post-starburst age and burst mass fraction (see \S\ref{sec:systematic}). {\bf Left}: fit error, with data uncertainties propagated through. The median errors on the post-starburst ages are 10\%, and the median errors on the burst mass fractions are 13\%. {\bf Middle:} systematic errors due to metallicity uncertainties. The median errors on the post-starburst ages are 14\%, and the median errors on the burst mass fractions are 23\%. {\bf Right:} combined errors. The median errors on the post-starburst ages are 22\%, and the median errors on the burst mass fractions are 38\%. These trends are not significant functions of either the post-burst age or burst mass fraction.}
\label{fig:errors}
\end{figure*}

We consider three main sources of error in our derived properties: the fit uncertainty (including errors on the data propagated through), our metallicity assumption, the SFH uncertainty. The errors from the fit uncertainty are shown in Figure \ref{fig:errors} (left). The median errors on the age are 10\%, and the median errors on the burst mass fraction are 12\%. The systematic errors due to our metallicity assumption, as described in \S\ref{sec:metallicity}, are shown in Figure \ref{fig:errors} (middle). The median errors on the ages due to the metallicity uncertainties are 14\%, and the median error on the mass fractions is 23\%. Histograms of combined errors are shown in Figure \ref{fig:errors} (right). The median combined errors on the ages are 22\%, and the median combined errors on the burst mass fractions are 38\%. These trends are not significant functions of either the post-burst age or burst mass fraction. We consider additional tests of the parameter errors in Appendix \ref{sec:appdxa}.

We have two physically motivated classes of SFHs: one or two recent bursts. The data available are not sensitive enough to discern among more complex models of the recent SFH, such as varying the relative contributions or durations of the two recent bursts, thus we do not consider external estimates of the systematic error caused by the assumed SFH history. We consider an external check on the derived ages using star cluster measurements in \S\ref{sec:starclusters}.

\subsection{External Check Using Star Cluster Ages}
\label{sec:starclusters}

As an external check on our method, we compare our derived ages to the star cluster ages measured by \citet{Yang2008}. The star cluster ages directly measure the age since their formation in the starburst, because they are well-modeled by single stellar populations, and ages can be measured without the uncertainties of modeling the older stellar population and without the need for disentangling the age from the burst mass fraction. Using \textit{HST} imaging, \citet{Yang2008} measure star cluster ages for four post-starburst galaxies selected from the \citet{Zabludoff1996} sample of post-starbursts in the Las Campanas Redshift Survey (LCRS). Two of these galaxies (EA01A and EA18) have \textit{GALEX} photometry in both bands. For these galaxies, we fit the Lick indices and photometry as for the SDSS sample. In both cases, our derived ages since the starburst began are consistent to within the 95\% likelihoods of the star cluster ages found by \citet{Yang2008}. \citet{Yang2008} find an age range of 10--450 Myr for EA01A, and 400 Myr -- 1 Gyr for EA18. For EA01A, we determine the recent starburst began 32 Myr ago, with a duration of 200 Myr; for EA18, we determine the recent starburst began 1.06 Gyr ago, with a duration of 200 Myr. Thus, the star cluster ages are consistent with being formed during the recent starbursts we measure and there is no evidence of a systematic shift between the two methods.

\subsection{Breaking the Age-Burst Fraction-Burst Duration Degeneracy}
\label{sec:synth}

A key feature in our method is breaking the degeneracies of the post-burst age with the burst mass and burst duration. Doing so relies on the UV photometry from {\it GALEX}. In Figure \ref{fig:nouv}, we demonstrate the effect of the UV photometry and optical lines on decreasing the fit parameter uncertainties and their degeneracies by age-dating a galaxy with and without these data. The post-burst age, burst fraction, and burst duration have highly correlated errors, and higher uncertainties result. Including the UV photometry and optical lines successfully breaks these degeneracies, reducing the uncertainties in the fit parameters.

To test that our method successfully breaks these degeneracies, we add noise to simulated data and apply the age-dating method to recover the input parameters. We draw randomly from the fit grid in age, burst light fraction, and burst duration, keeping a constant A$_V$ and metallicity. Random errors are drawn from the distribution of uncertainties of the real data, and applied to these simulated data. We apply the age-dating method to these synthetic data, testing whether the method recovers the original parameters or adds any systematic biases. The fit ages are within the 68\%ile uncertainty range of the input ages 94\% of the time, the fit burst light fractions are within the 68\%ile uncertainty range of the input burst light fractions 91\% of the time, and the fit burst durations are within the 68\%ile uncertainty range of the input burst durations 92\% of the time. The SFH (single or double burst) is the same as the input 77\% of the time. From the uncorrelated initial set of ages and burst light fractions, no correlation is introduced during the fitting process, and there is no significant ($>3\sigma$) correlation between the recovered ages and burst light fractions.

\begin{figure*}
\includegraphics[width = 0.5\textwidth]{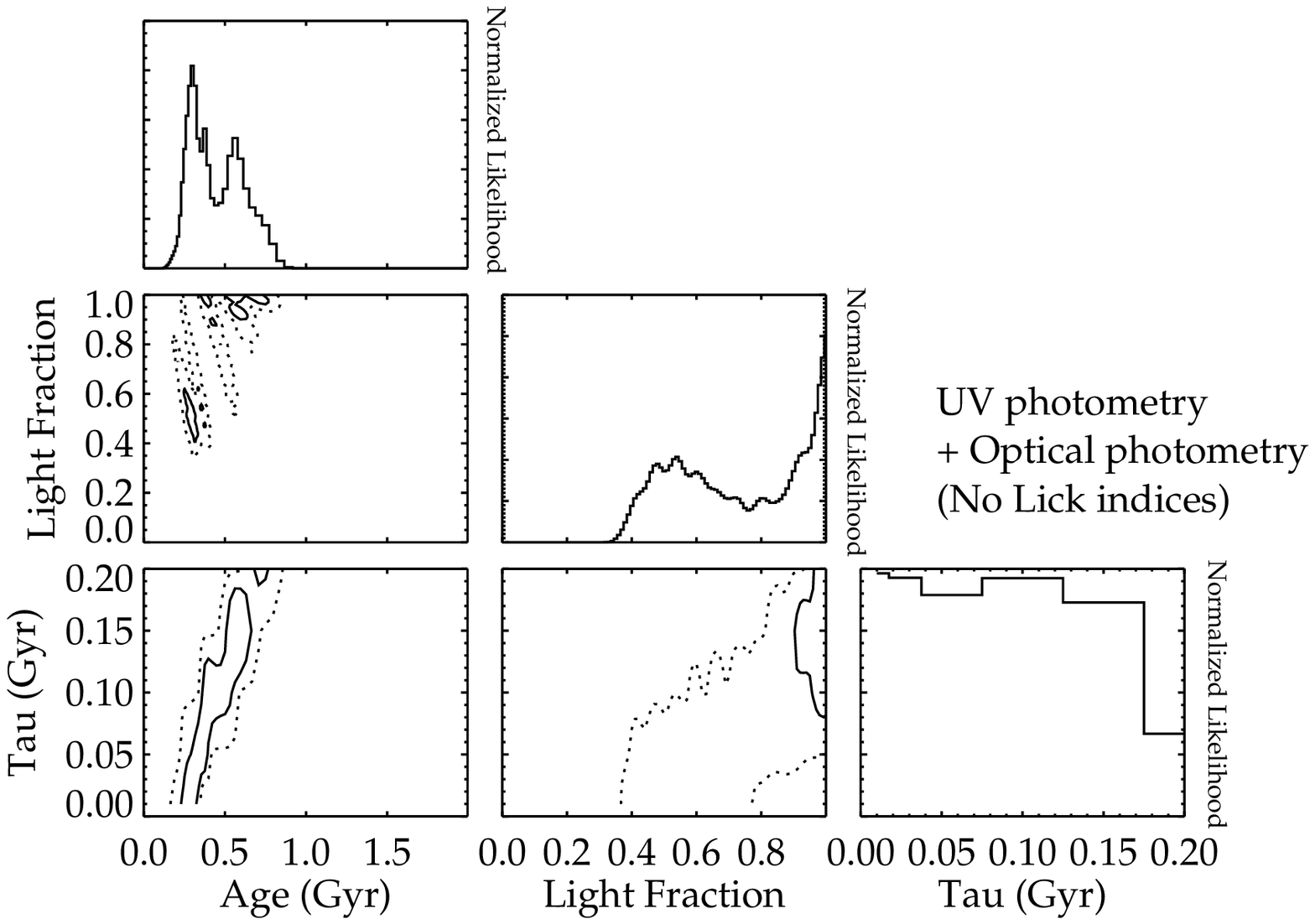}
\includegraphics[width = 0.5\textwidth]{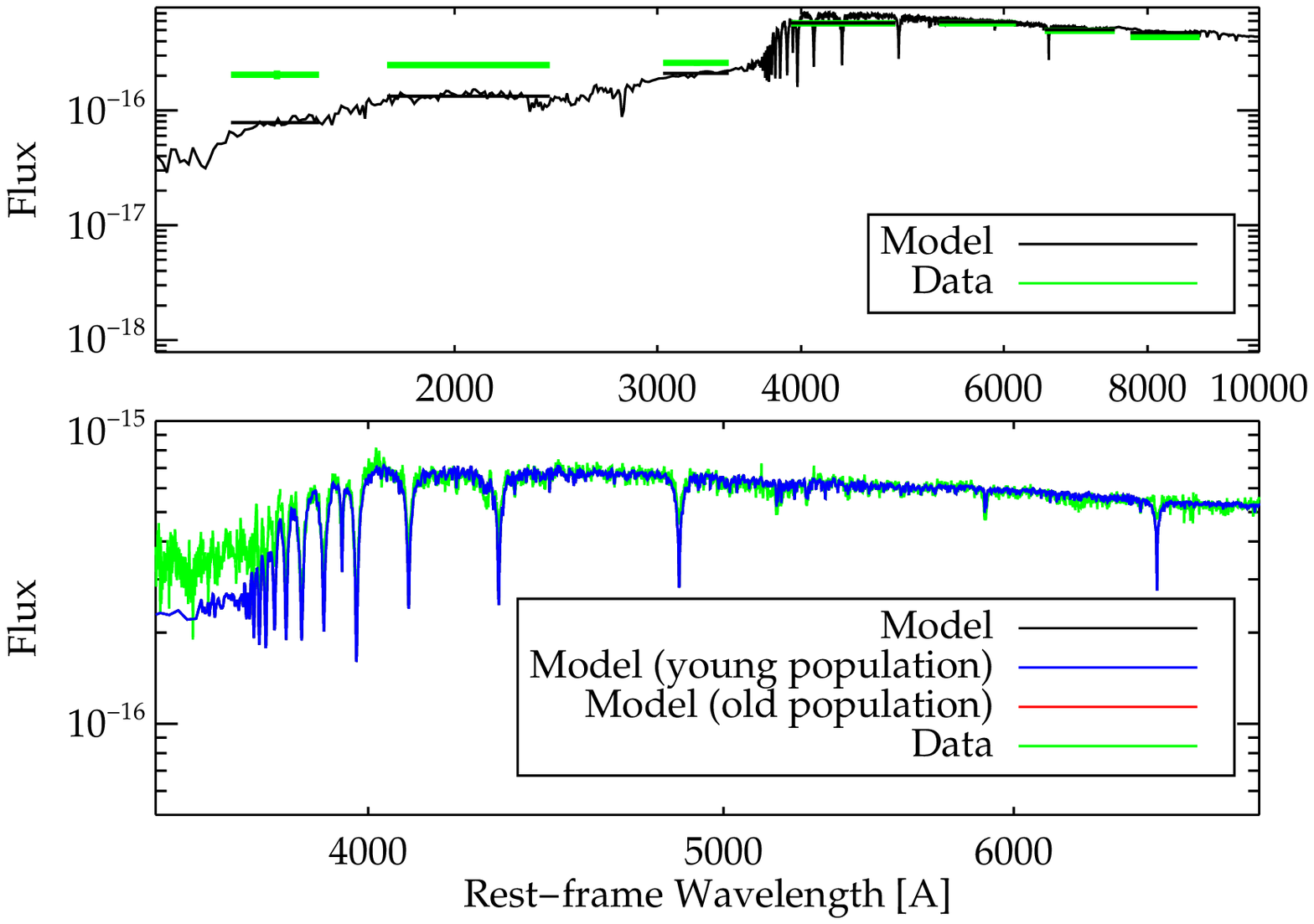}
\includegraphics[width = 0.5\textwidth]{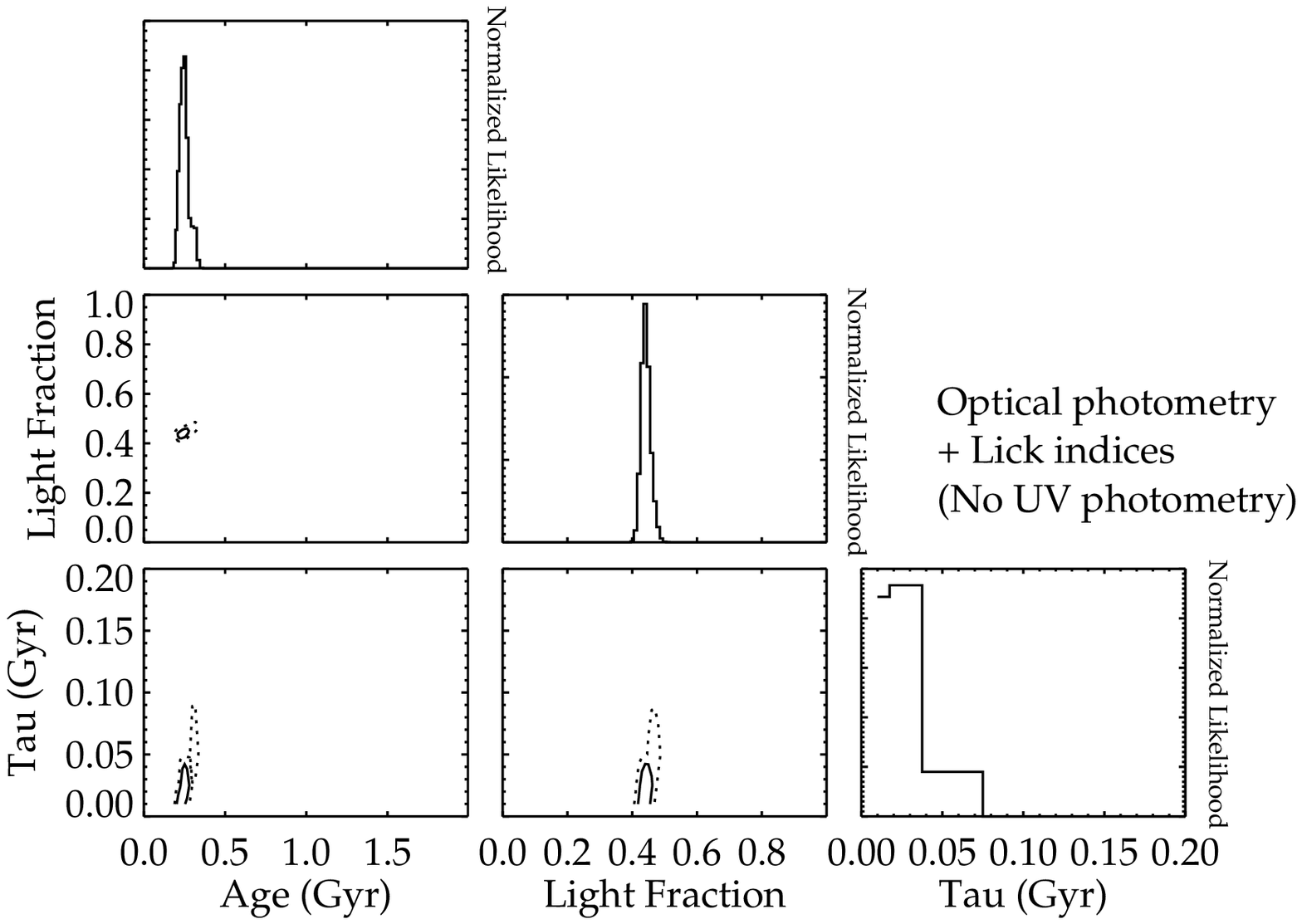}
\includegraphics[width = 0.5\textwidth]{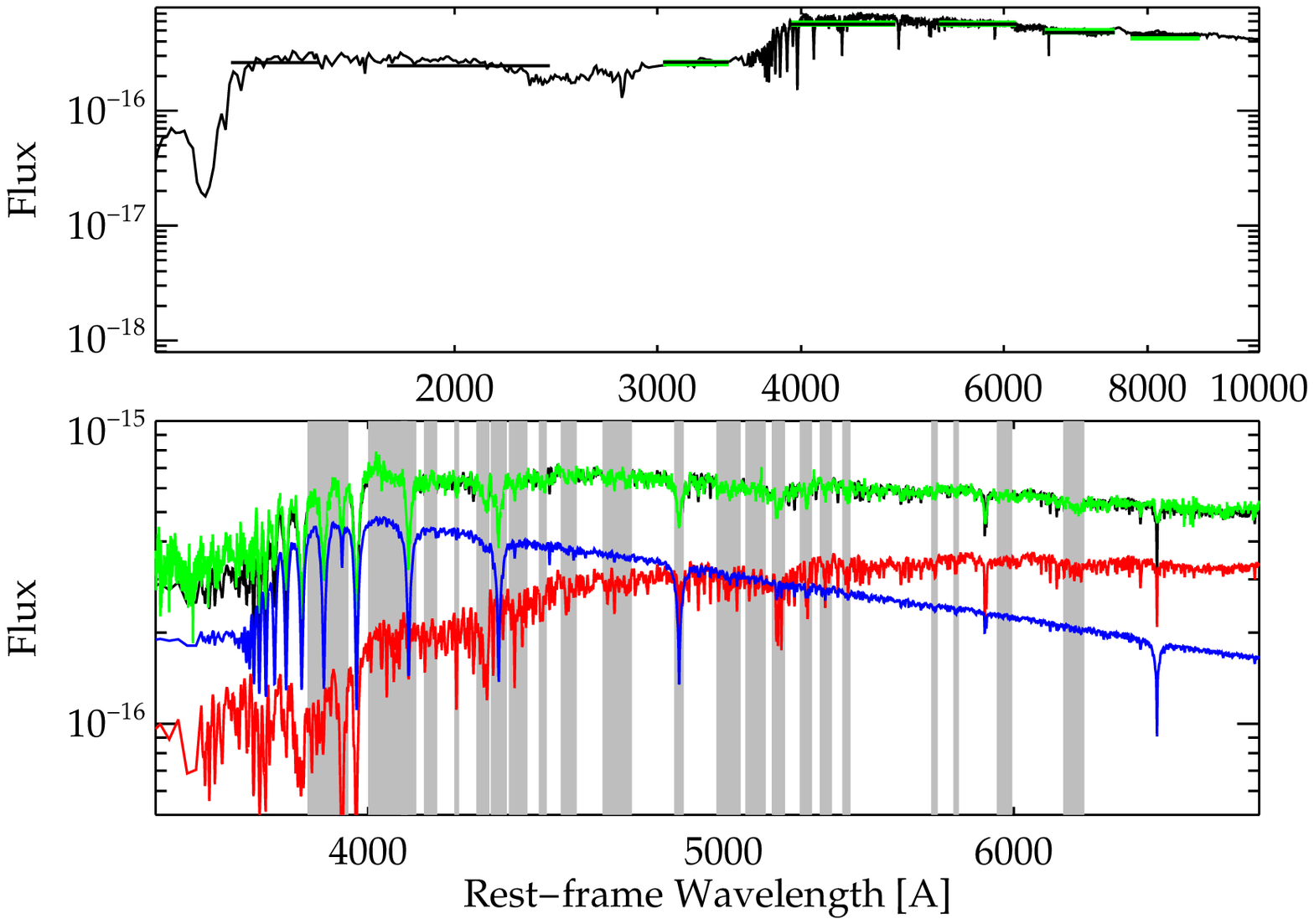}
\includegraphics[width = 0.5\textwidth]{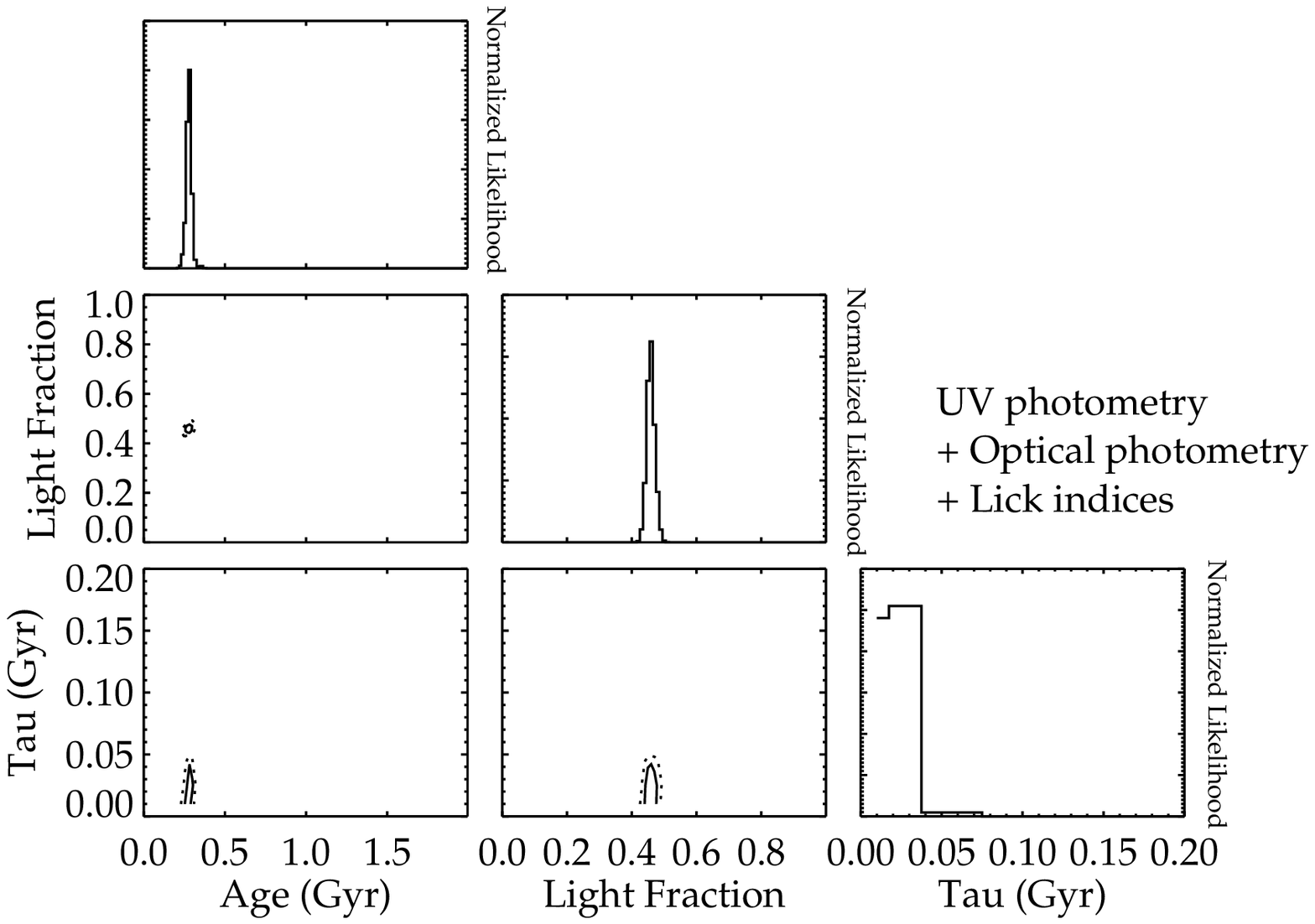}
\includegraphics[width = 0.5\textwidth]{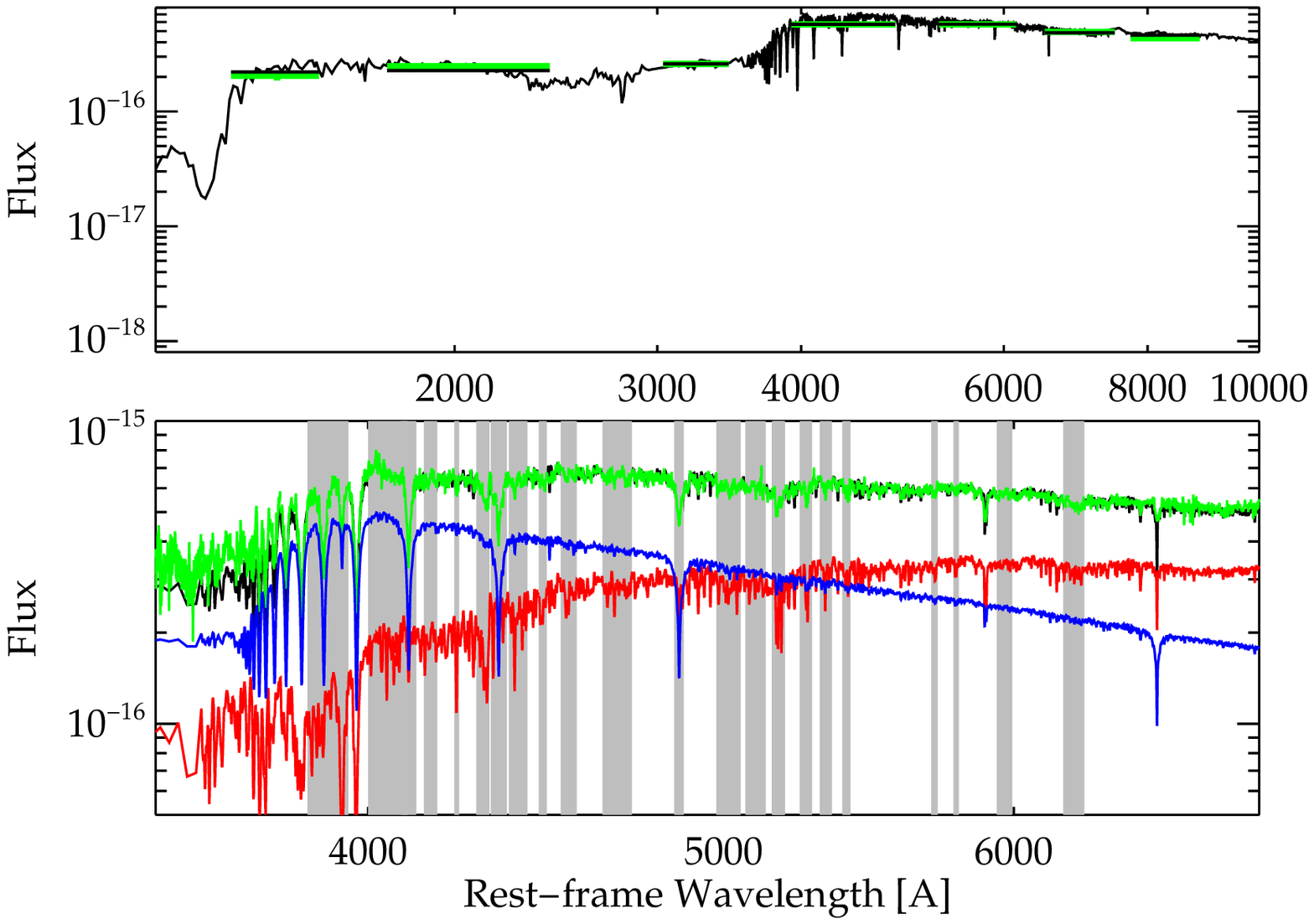}
\caption{Example likelihood contours of model parameters from this age-dating method (left column), and comparisons of the data and model spectra (right column). This case shows the derived properties for the single recent burst SFH model. We marginalize over A$_V$, then plot the 68\% (solid lines) and 95\% (dotted lines) likelihood contours for the remaining parameter pairs, marginalizing over the third parameter. Normalized likelihoods for each parameter are shown at the top of each column.  In the right hand column, we plot the associated model and data spectra and photometry for the best fit given each set of data. The grey bars indicate the location of the Lick indices used to parameterize the spectra. The bottom row shows the results for a galaxy, using the full set of UV-optical photometry and optical line indices. The middle row shows the consequently worse parameter degeneracies and uncertainties, if the UV photometry is not included in the fit, and the top row, if the optical lines are not included in the fit. The redshift of this example galaxy is $z=0.090$.}
\label{fig:nouv}
\end{figure*}

\begin{figure*}
\ContinuedFloat
\includegraphics[width = 0.5\textwidth]{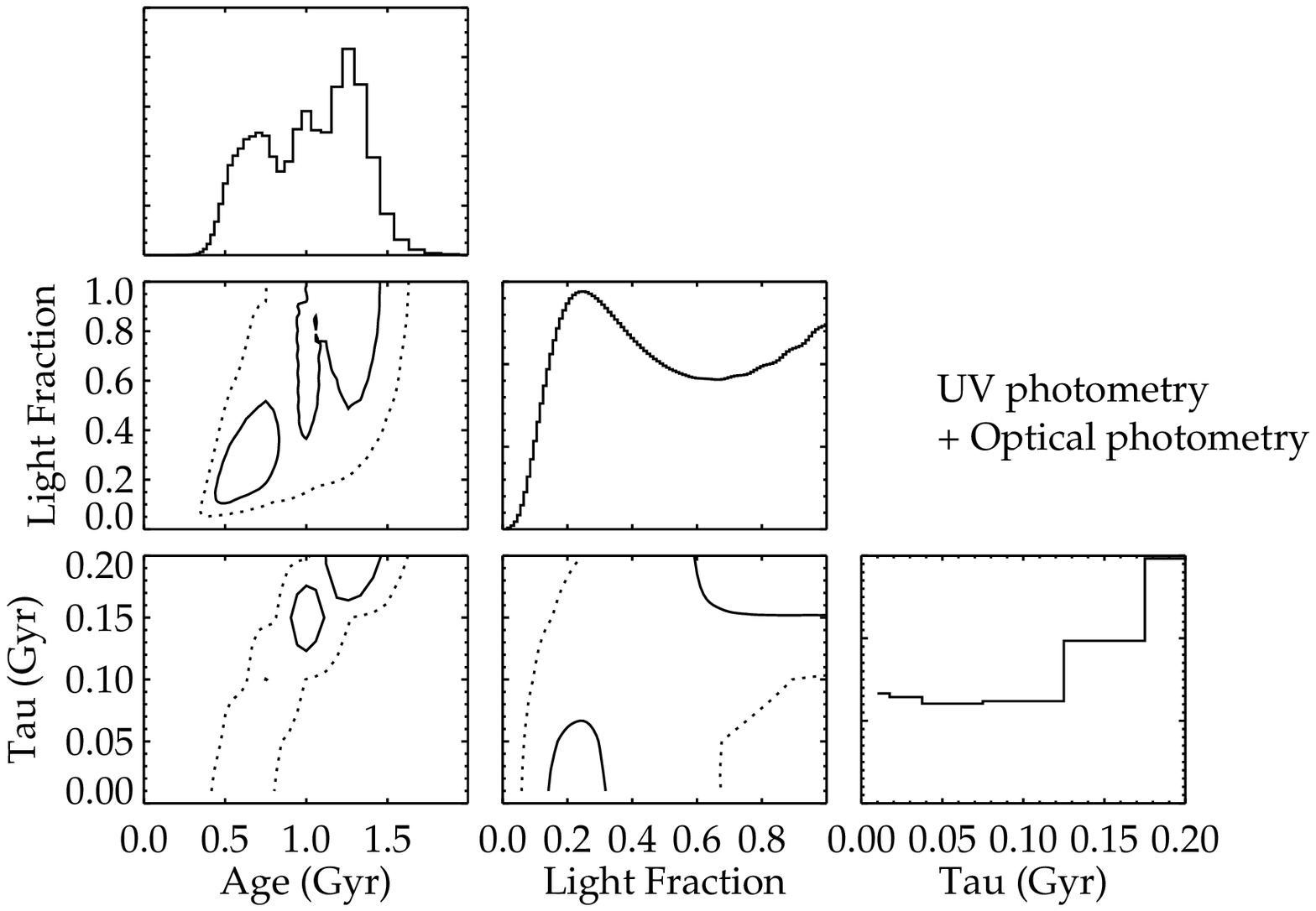}
\includegraphics[width = 0.5\textwidth]{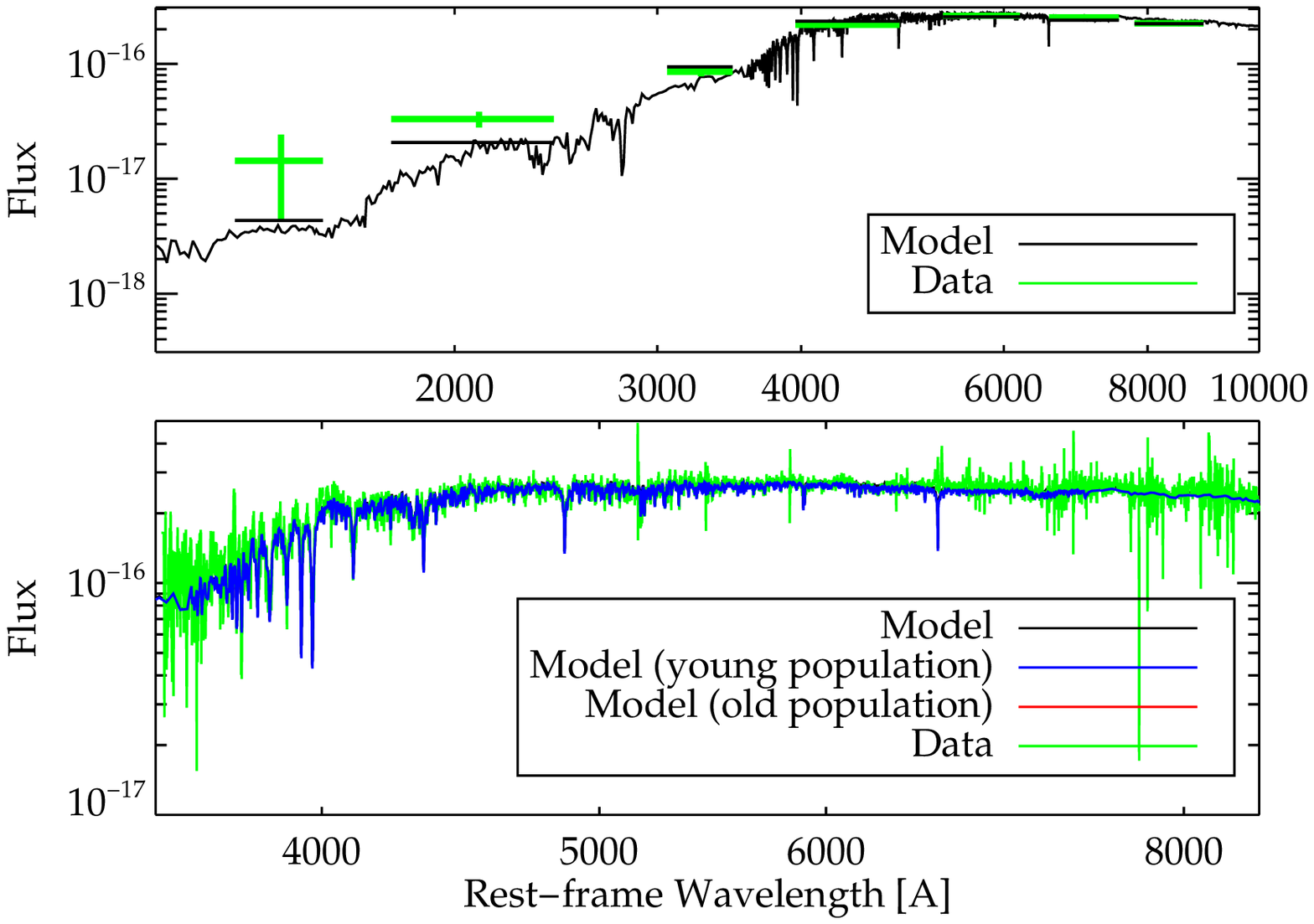}
\includegraphics[width = 0.5\textwidth]{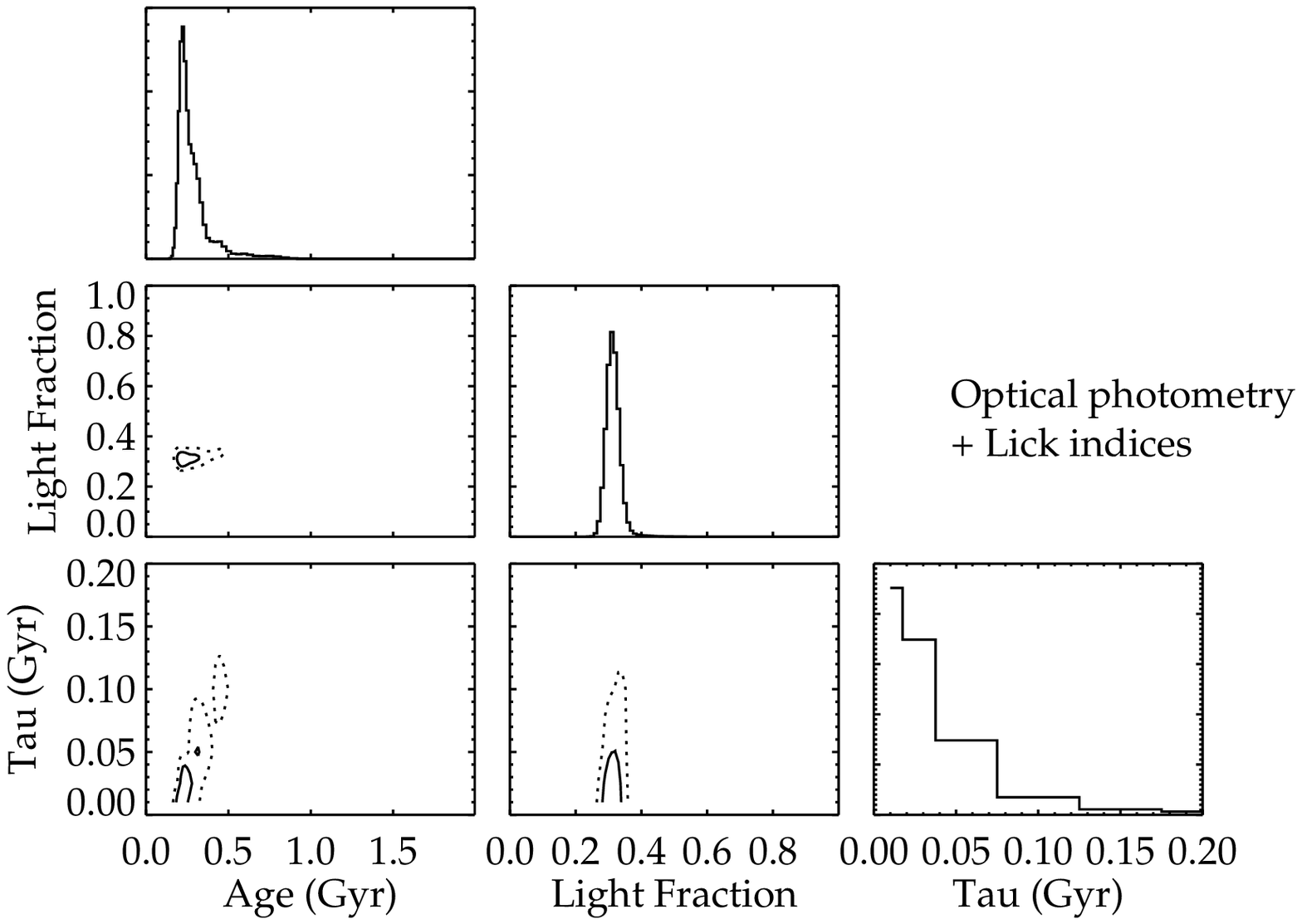}
\includegraphics[width = 0.5\textwidth]{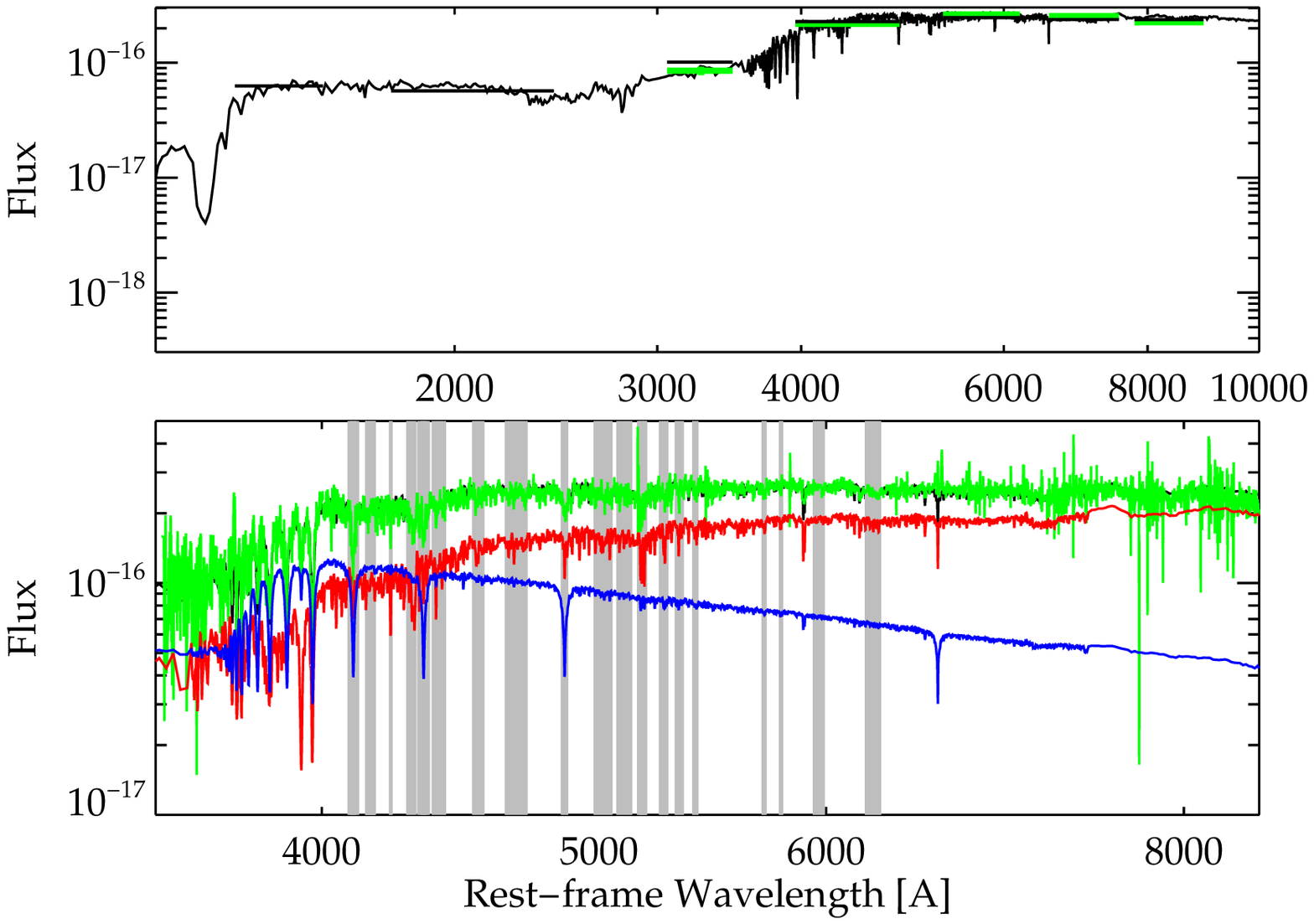}
\includegraphics[width = 0.5\textwidth]{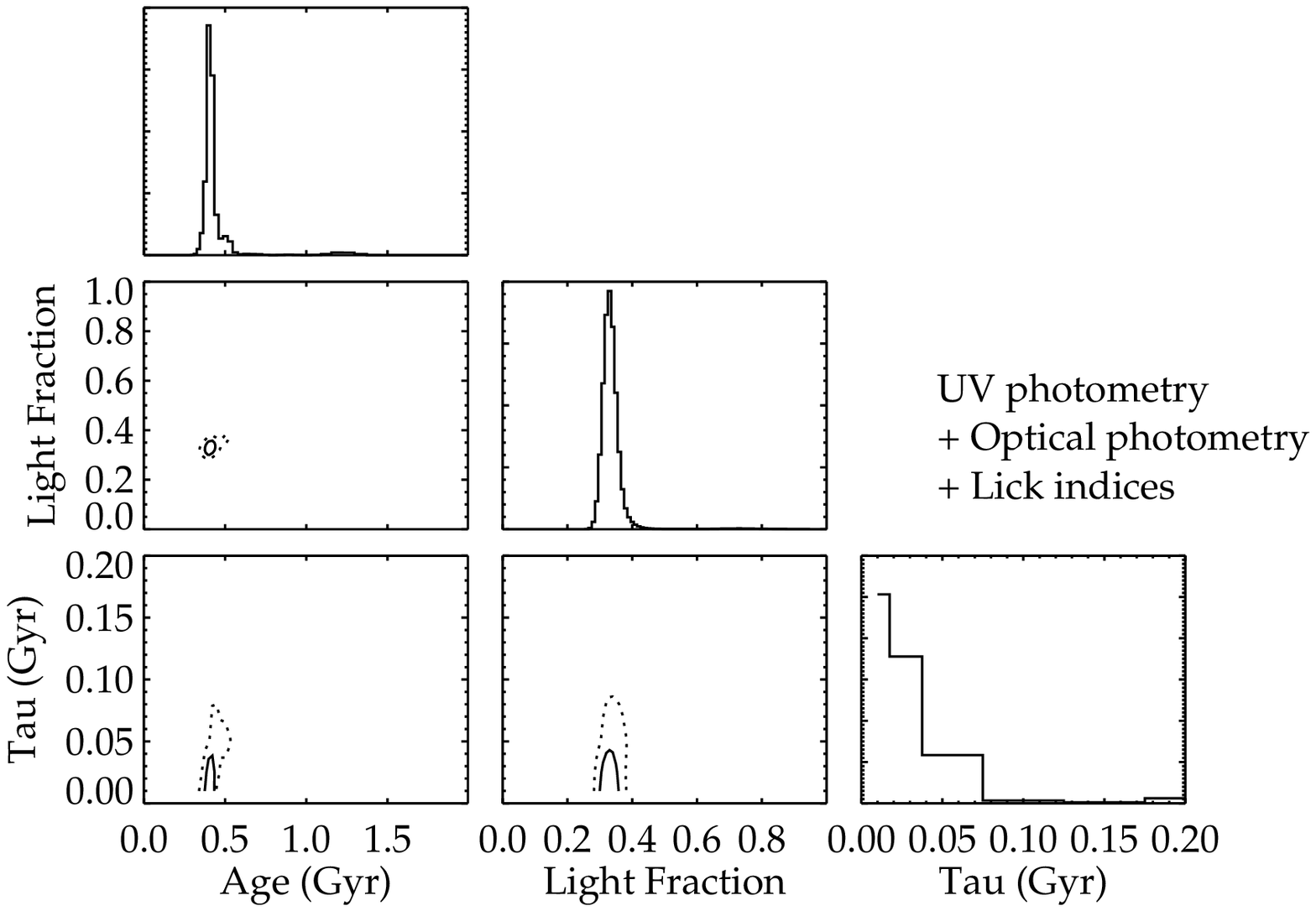}
\includegraphics[width = 0.5\textwidth]{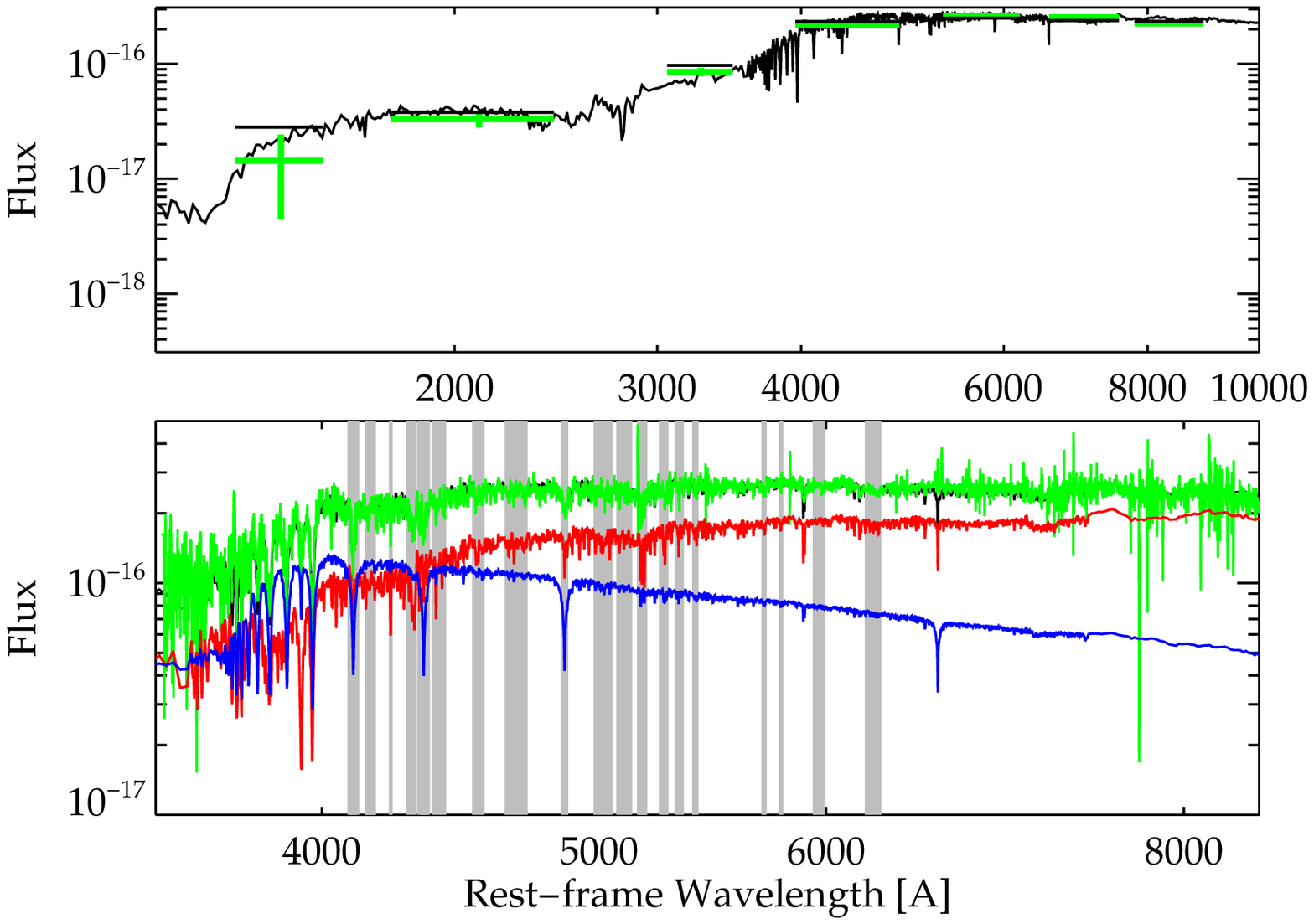}
\caption{(continued) Example likelihood contours of model parameters from this age-dating method (left column), and comparisons of the data and model spectra (right column). This case shows the derived properties for the single recent burst SFH model. We marginalize over A$_V$, then plot the 68\% (solid lines) and 95\% (dotted lines) likelihood contours for the remaining parameter pairs, marginalizing over the third parameter. Normalized likelihoods for each parameter are shown at the top of each column.  In the right hand column, we plot the associated model and data spectra and photometry for the best fit given each set of data. The grey bars indicate the location of the Lick indices used to parameterize the spectra. The bottom row shows the results for a galaxy, using the full set of UV-optical photometry and optical line indices. The middle row shows the consequently worse parameter degeneracies and uncertainties, if the UV photometry is not included in the fit, and the top row, if the optical lines are not included in the fit. The redshift of this example galaxy is $z=0.081$.}
\label{fig:nouv}
\end{figure*}

\subsection{Star Formation History Selection Effects}
\label{sec:thesample}

\begin{figure}
\includegraphics[width = 0.5\textwidth]{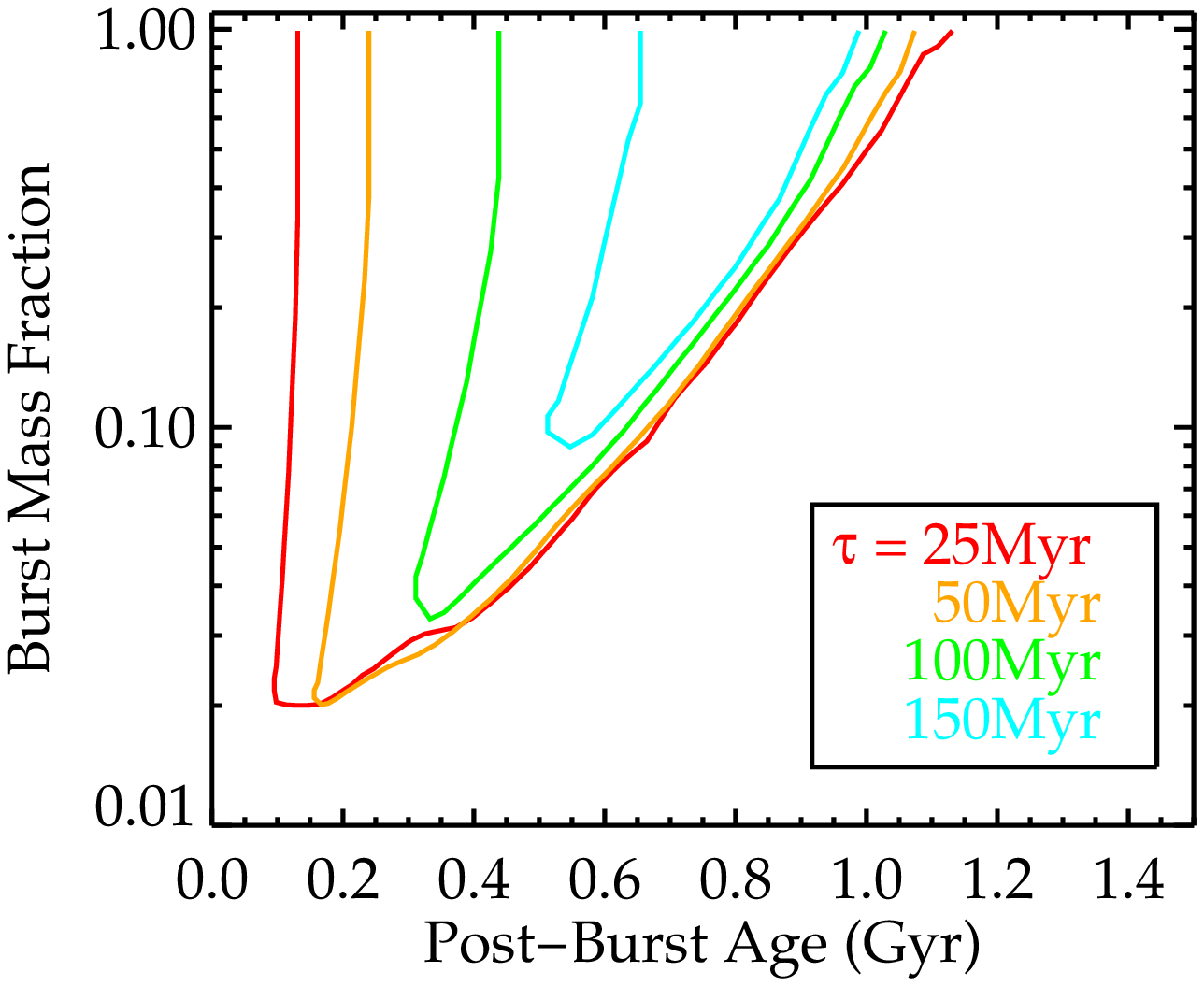}
\caption{Burst mass fraction and post-burst age space for post-starburst galaxies selected with the H$\alpha$-H$\delta$ method described in \S\ref{sec:selectionsdss}, divided up by the burst duration, $\tau$. Colored contours mark where galaxies enter and leave the post-starburst H$\alpha$ emission and H$\delta$ absorption criteria. After the starburst ends, the post-burst age at which the H$\alpha$ emission EW is low enough to enter the post-starburst phase will depend on how much stellar mass is produced in the recent starburst, and the burst duration. As the stellar populations age, the H$\delta_{\rm A}$ absorption will decrease, and galaxies will leave the post-starburst phase at different post-burst ages depending on burst fraction of the starburst. As a result, the descendants of starbursts with longer durations will only be seen at older post-burst ages, and only if the burst fraction is high. This selection must be understood in order to assess physical features in the distribution of starburst properties.}
\label{fig:selection_buckets}
\end{figure}

By selecting against galaxies with significant H$\alpha$ emission, we select only galaxies that are truly \textit{post-}starburst. However, galaxies with different burst durations and burst mass fractions will go through our post-starburst selection criteria at different post-burst ages. In Figure \ref{fig:selection_buckets}, we plot the regions of the burst parameter space that we expect our sample of post-starbursts to occupy, as they evolve in and out of the post-starburst selection criteria. The Lick H$\delta_{\rm A}$ limits are obtained from the FSPS models. The H$\alpha$ EW limits are obtained by converting the model SFR to an H$\alpha$ flux using the \citet{Kennicutt1994} relation and by comparing to the continuum level from the FSPS models. For each burst duration and burst mass fraction, it will take some time before the H$\alpha$ EW has subsided enough for these galaxies to enter our selection criterion, even after waiting for all but 10\% of the stars to be made in the burst. Similarly, the post-burst age at which the strong H$\delta$ absorption fades out of our selection criterion will differ, depending on the burst mass fraction. It is clear from this plot that we are more sensitive to weaker bursts (lower burst mass fraction) at younger ages, for shorter bursts. The effects of our selection must be taken into consideration when examining the statistical properties of post-starbursts.

From the way the post-burst age, burst mass fraction, and burst duration map to the observed post-starburst signatures, it is clear that different selection methods can produce different populations of post-starbursts. The method by \citet{Wild2007, Wild2009}, used for example by \citet{Yesuf2014} and \citet{Rowlands2015}, aims to select post-starburst galaxies with the strongest (i.e., highest burst mass fraction) bursts, using a PCA analysis of spectra around the 4000\AA\ region, especially the D$_n$(4000) index, excess Balmer absorption, and excess H$\epsilon$ vs. Ca II H+K absorption. By selecting only the strongest bursts, this method selects \textit{only} post-starburst galaxies with burst durations of $\gtrsim 150$ Myr, not the population with shorter durations. Figure \ref{fig:selection_buckets} demonstrates that for shorter starbursts, lower burst mass fractions produce similarly strong post-starburst signatures, because the SFRs and sSFRs during the burst are similarly high. Although this method has other strengths, such as identifying galaxies transitioning from starbursting to post-starburst \citep{Rowlands2015}, it neglects a significant number of post-starburst galaxies.

\subsection{Effects of Magnitude-Limited Parent Sample}
\label{maglimit}

Our sample is subject to potential biases from magnitude limits in two ways. The first is the magnitude-limited nature of the SDSS parent sample. The second is the requirement that the galaxies be detected in both the {\it GALEX} \fuv and \nuv bands. 

As detailed in Section \ref{sec:thesample}, the selection on H$\alpha$ emission and H$\delta$ absorption has a strong influence on the distribution of the properties mentioned (especially age since starburst, and the strength and duration of the starburst). This selection biases the distributions of these quantities much more significantly than the magnitude cuts (this can be seen in the predicted vs observed distributions given the selection cuts alone), and as such we caution against making claims about the distributions of these properties independent of their selection.

We test whether the distribution of derived ages is subject to bias from the magnitude limit. The distribution of ages is the same (passes a KS test) for different bins of stellar mass. If a large number of galaxies were brightened in \fuv into our sample, we would have expected (given that the \fuv flux should fade with time) that the lower mass galaxies should have younger age distributions.

We test whether the main difference in the galaxies with \fuv detections and those without is stellar mass or redshift. The distribution in redshift between these samples is different (fails a KS test), but the distribution of stellar masses is the same (passes a KS test). Thus, we do not expect the \fuv-limited nature of the sample to bias the sample, except to exclude farther away galaxies in the SDSS parent sample. We assume the Lyman $\alpha$ flux will contribute negligibly to the \fuv flux, as the H$\alpha$ fluxes are typically $<50\times$ the FUV fluxes.

Nonetheless, we test whether our stellar-mass dependent conclusions might be affected by the magnitude-limited nature of the sample. We define a volume-limited subset, and find that much of the power in our conclusions is driven by the intermediate mass galaxies, rather than a small number of low mass galaxies. More detail is presented in the relevant section: \S\ref{singleordouble}.

\subsection{Comparison to $D_n(4000)$-H$\delta$ Method}

\begin{figure*}
\includegraphics[width = 1\textwidth]{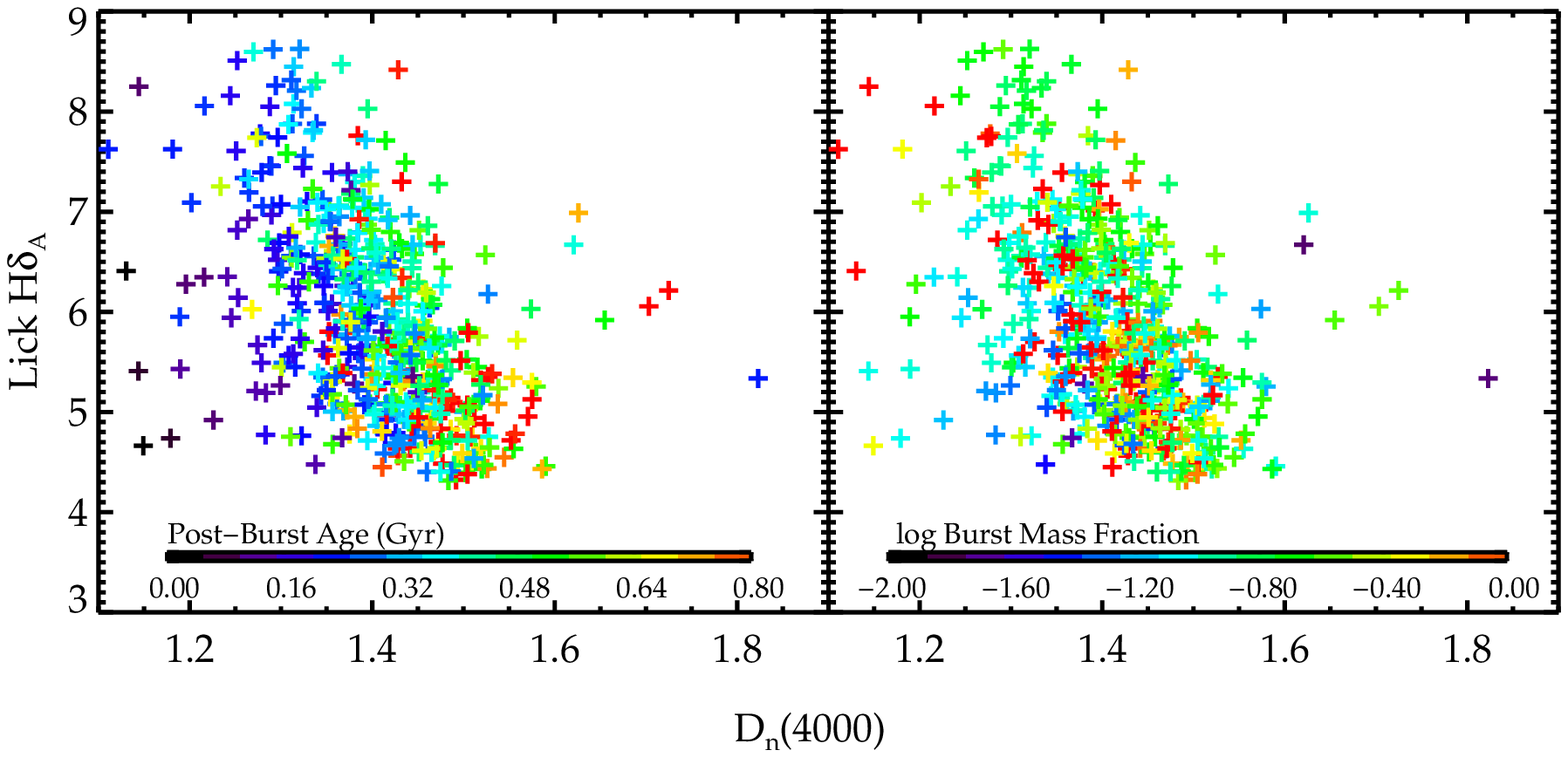}
\caption{$D_n(4000)$ vs. Lick H$\delta_{\rm A}$, colored by post-burst age (left) and burst mass fraction (right). $D_n(4000)$ and Lick H$\delta_{\rm A}$ are from the SDSS catalogs described in the text, not the best-fit values. This parameter space has been used a a proxy for the post-burst age and burst mass fraction \citep[e.g.,][]{Yagi2006}. Here, and in the next figure, we demonstrate that there is significant scatter in where post-starburst galaxies with a certain post-burst age and burst mass fraction lie. 
}
\label{fig:d4000_hd}
\end{figure*}

In the absence of detailed modeling, the indices $D_n(4000)$ and H$\delta$ are sometimes used as proxies for the post-burst age and burst mass fraction \citep[e.g.,][]{Yagi2006}. Using the results of our stellar population fits, we evaluate how accurate this method is in assessing the recent SFHs of post-starbursts. We show the standard plot of $D_n(4000)$ vs. Lick H$\delta_{\rm A}$, colored by post-burst age and burst mass fraction (Figure \ref{fig:d4000_hd}). 

\begin{figure*}
\includegraphics[width = 0.8\textwidth]{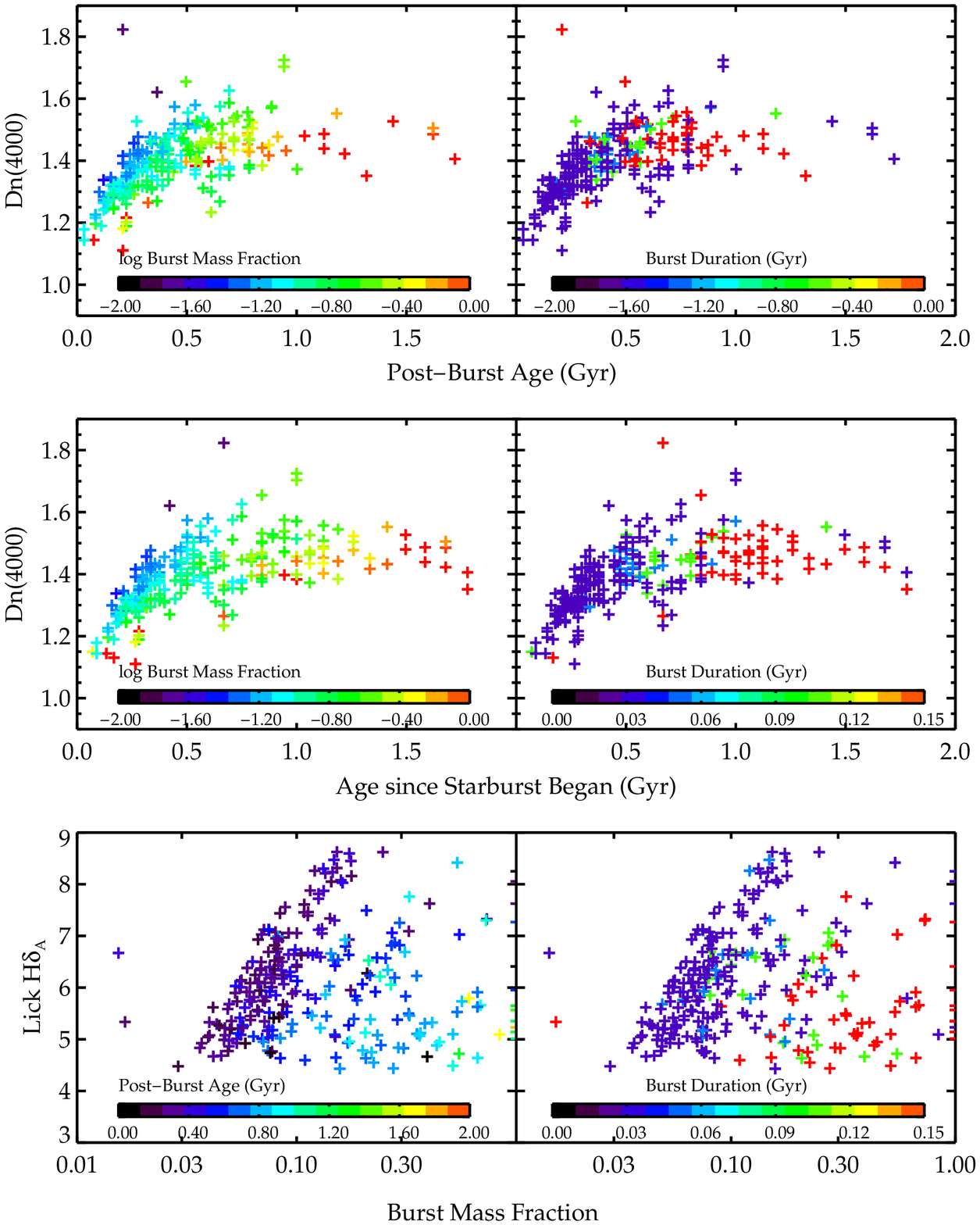}
\caption{Comparison of post-burst age, age since starburst began, and burst mass fraction to the indices $D_n(4000)$ and H$\delta$, which are sometimes used as proxies for the post-burst age and burst mass fraction. $D_n(4000)$ is from the SDSS catalogs described in the text, not the best-fit values. We plot only galaxies best fit by a single recent burst, and color code by the additional SFH parameters (post-burst age, burst mass fraction, or burst duration). The relation between $D_n(4000)$ and the post-burst age suffers from a degeneracy with the burst mass fraction and burst duration. For post-starbursts with $D_n(4000) < 1.3$, where post-starburst ages are typically younger than 300 Myr, the degeneracy is lessened, and $D_n(4000)$ is highly correlated with the post-burst age. However, if $D_n(4000) > 1.42$, where post-starburst ages typically range from 300-1500 Myr, $D_n(4000)$ is no longer significantly ($>3\sigma$) correlated with either the post-burst age or the age since the starburst began. We caution that these results depend on our post-starburst selection method. For the youngest post-burst ages, we only can select short-duration, low burst mass fraction starbursts. Selection methods that do not make the same H$\alpha$-H$\delta$ cuts may find increased scatter in the $D_n(4000)$-age at younger ages, where it currently appears more robust, if longer duration starbursts are allowed into the sample. For the Lick H$\delta_{\rm A}$ index, we see that the post-starbursts with the highest values of H$\delta_{\rm A}$ are not those with the strongest bursts, but those with short duration bursts of $m_{\rm burst} \sim 0.1$ observed within 500 Myr post-burst.}
\label{fig:d4000}
\end{figure*}

In Figure \ref{fig:d4000} we plot the post-burst age and burst mass fraction vs. $D_n(4000)$ and Lick H$\delta_{\rm A}$, for galaxies in our sample best fit by a single recent burst. The relation between $D_n(4000)$ and the post-burst age suffers from a degeneracy with the burst mass fraction and burst duration. For post-starbursts with $D_n(4000) < 1.3$, where post-starburst ages are typically younger than 300 Myr, the degeneracy is lessened, and $D_n(4000)$ is highly correlated with the post-burst age. However, if $D_n(4000) > 1.42$, where post-starburst ages typically range from 300-1500 Myr, $D_n(4000)$ is no longer significantly ($>3\sigma$) correlated with the post-burst age or the age since the starburst began. We caution that these results depend on our post-starburst selection method. For the youngest post-burst ages, we only can select short-duration, low burst mass fraction starbursts. Selection methods that do not make the same H$\alpha$-H$\delta$ cuts may find increased scatter in the $D_n(4000)$-age at younger ages, where it currently appears more robust, if longer duration starbursts are allowed into the sample. We further caution that the $D_n(4000)$ index often shows the most deviation between the data and best-fit model (see Appendix A). However, the large scatter observed between the measured $D_n(4000)$ and the post-burst age is much larger than the typical deviations seen in the best-fit models. For the Lick H$\delta_{\rm A}$ index, we see that the post-starbursts with the highest values of H$\delta_{\rm A}$ are not those with the strongest bursts, but those with short duration bursts of $m_{\rm burst} \sim 0.1$ observed within 500 Myr post-burst.

\section{Results: Constraints on Star Formation Histories}

\subsection{Derived Starburst Properties}
\label{sec:sbchars}

\begin{figure*}
\includegraphics[width = 1\textwidth]{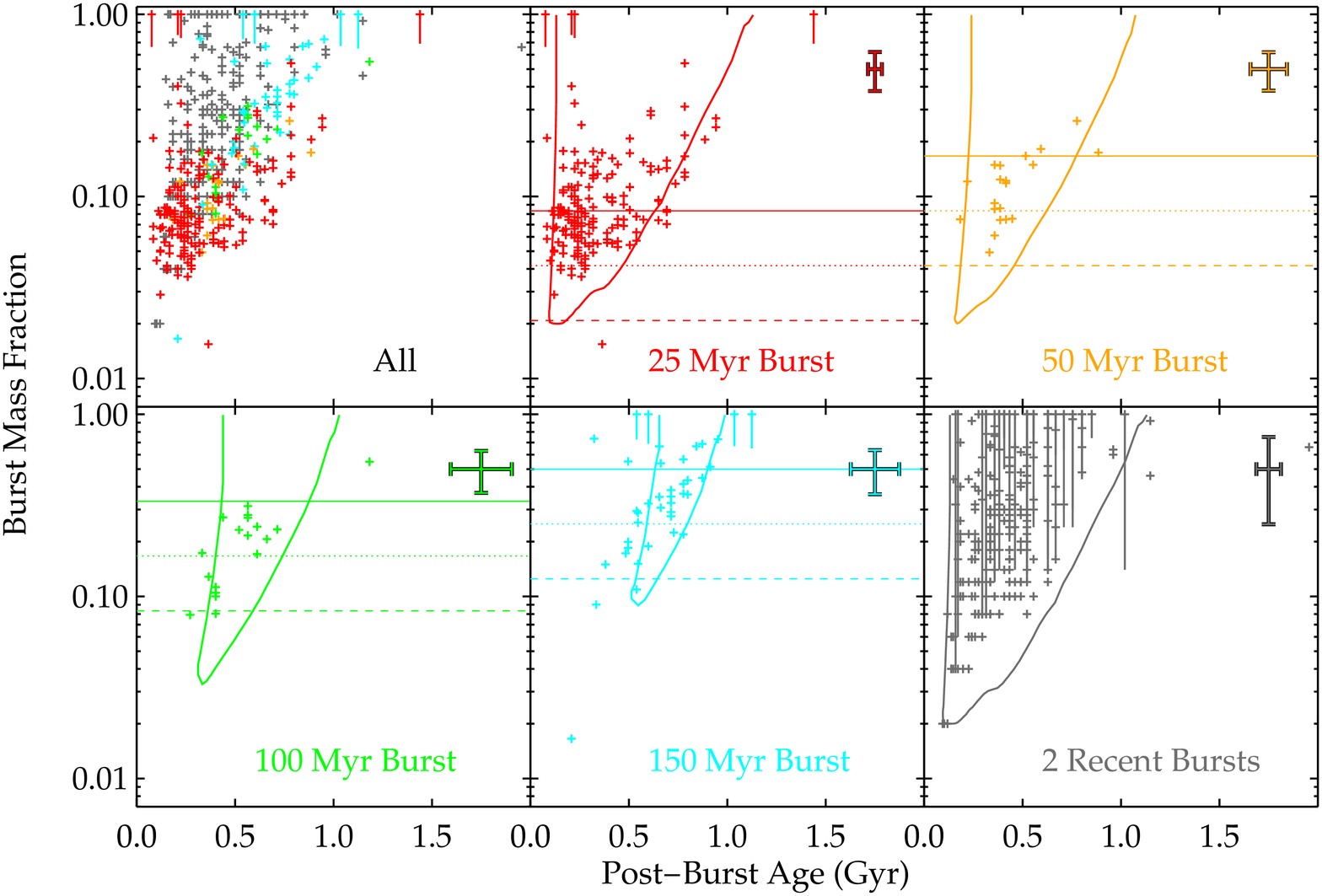}
\caption{Burst mass fraction and post-burst age for post-starburst galaxies, divided up by SFH. For galaxies which strongly prefer a single recent burst, we show only those with prefer a value for $\tau$. The final panel shows the galaxies best fit by two recent bursts. Colored contours mark where galaxies enter and leave the post-starburst H$\alpha$ EW and H$\delta$ absorption criteria. The observed lack of post-starburst galaxies at old age and low mass fraction is due to their lack of strong H$\delta$ absorption. The dearth of post-starburst galaxies at short $\tau$ and high mass fraction is due to the high burst (maximum) SFRs that would be required to produce so much mass in such a short time. Overplotted as solid, dotted, and dashed lines are the burst SFRs for a 1e10 M$_\odot$ galaxy at SFR=25, 50, and 100 M$_\odot$ yr$^{-1}$ respectively. The post-starburst galaxies uniformly fill the space within their selection criteria and constraints on the burst SFR, to within the formal fit errors. Characteristic error bars reflecting the fit uncertainties are shown in each panel, and we plot individual error bars for galaxies with unphysical burst mass fractions of 1, to show that the error bars extend down to much lower mass fractions. For galaxies best fit by the two recent burst model, we do not have the sensitivity to fit the individual durations or mass fractions of each burst. Because of this, we cannot accurately estimate the maximum SFR for these galaxies.}
\label{fig:selection_contour}
\end{figure*}

We plot the post-burst ages and burst mass fractions for the post-starburst galaxies best fit by a single recent burst in Figure \ref{fig:selection_contour}. 68\% of the burst mass fractions are within 7.0\%--68\%, and 68\% of the post-burst ages are within 220--650 Myr. The ages appear correlated with the burst mass fractions, although this is due to our selection of the post-starburst galaxies, rather than fitting degeneracies.

To illustrate how the selection of post-starburst galaxies influences the ages, burst mass fractions, and burst durations, we outline the parameter space in post-burst age, burst fraction, and burst duration where the a hypothetical galaxy would meet our post-starburst selection criteria after the starburst, as in \S\ref{sec:thesample}. For each burst duration, the remaining panels in Figure \ref{fig:selection_contour} plot these outlines. The absence of post-starburst galaxies at old age and low mass fraction is due to a lack of strong H$\delta$ absorption. The dearth of post-starburst galaxies at short $\tau$ and high mass fraction is real, arising from the too high burst sSFRs ($>1\times10^8$ yr$^{-1}$) required to produce such a high fraction of the stellar mass in such a short time. For the median stellar mass of our sample, 3$\times10^{10}$M$_\odot$, this corresponds to an absence of starbursts with SFRs ($>300$ M$_\odot$ yr$^{-1}$). This is the reason the burst fractions appear to be correlated with the burst durations. If such galaxies were common in the local universe, we would have selected their descendants as members of the post-starburst sample. The post-starburst galaxies uniformly fill the space within their selection criteria to within the formal fit errors, when considering the lack of starburst progenitors with exceptionally high sSFRs. 

\begin{figure}
\includegraphics[width = 0.5\textwidth]{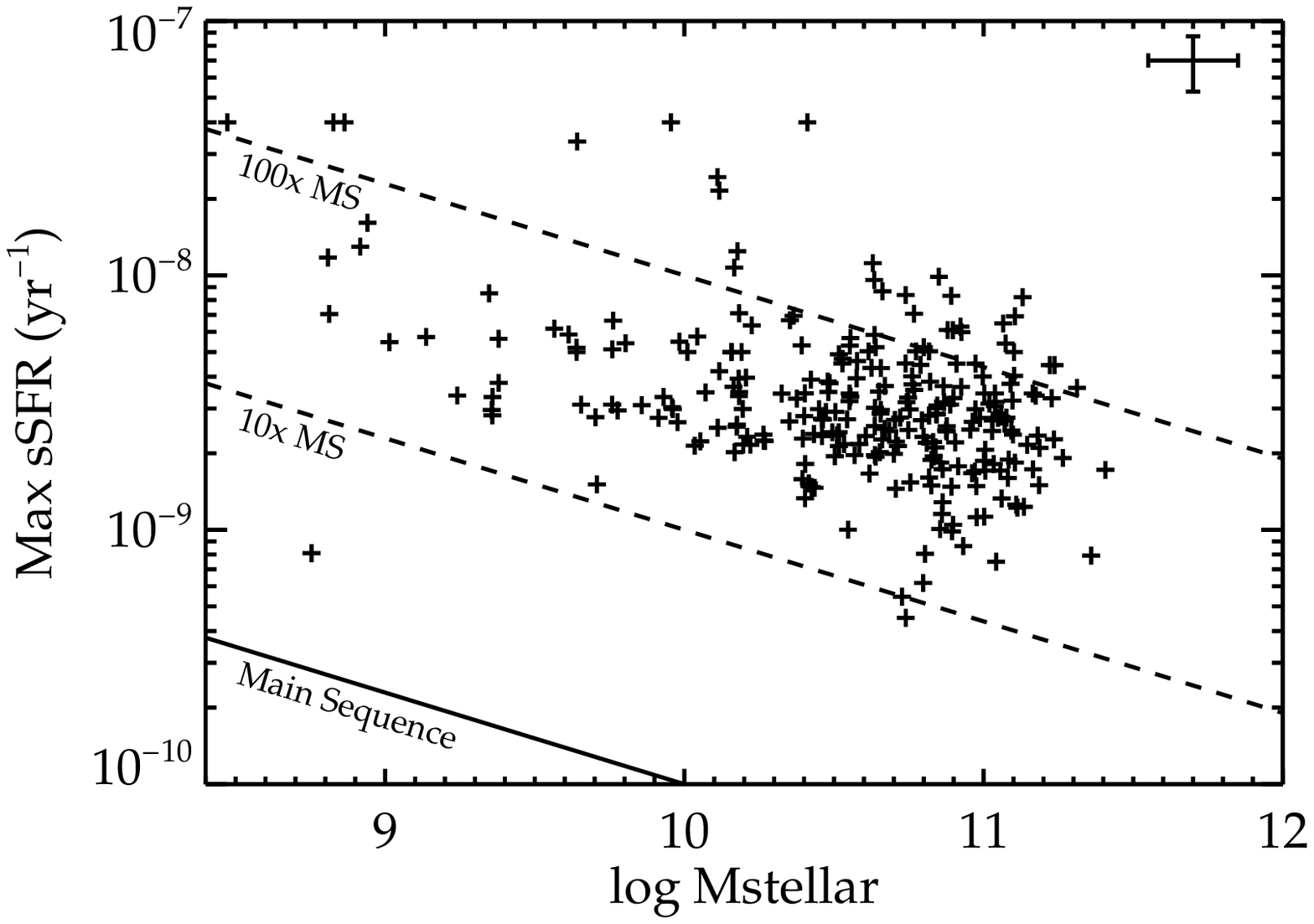}
\caption{Stellar mass vs. maximum sSFR ($m_{\rm burst}/\tau$) during the burst. The Main Sequence as fit by \citet{Schiminovich2007} is shown, along with multiples of 10-100$\times$, as is common for starbursting galaxies. A characteristic error bar is shown in the upper right. We plot sSFR (derived from the burst mass fraction $m_{\rm burst}$ and the burst duration $\tau$ from the stellar population fits) instead of SFR to avoid correlated errors. We only plot galaxies which strongly prefer a single recent burst, and prefer a value for $\tau$. The expected progenitors of the post-starburst sample have similar ranges of sSFRs.}
\label{fig:max_sfr}
\end{figure}

The fitted burst durations and mass produced in the burst provide an estimate of the maximum sSFR during the starburst. In Figure \ref{fig:max_sfr}, we plot the max sSFRs from the recent starburst and stellar masses in comparison to the star-forming Main Sequence fit from \citet{Schiminovich2007}. The starbursts experienced by our sample of post-starburst galaxies lie 10-100$\times$ above the Main Sequence and are typical of starbursting galaxies. 

In Figure \ref{fig:selection_contour}, we also plot the ages and burst mass fractions for post-starburst galaxies best fit by two recent bursts. Again, the galaxies fall within the bounds expected from the H$\alpha$ and H$\delta$ selection criteria. However, we do not have the sensitivity to fit the individual durations or mass fractions of each burst. Because of this, we cannot accurately estimate the maximum SFR for these galaxies.

\subsection{Single or Double Recent Burst?}
\label{singleordouble}

\begin{figure}
\includegraphics[width = 0.5\textwidth]{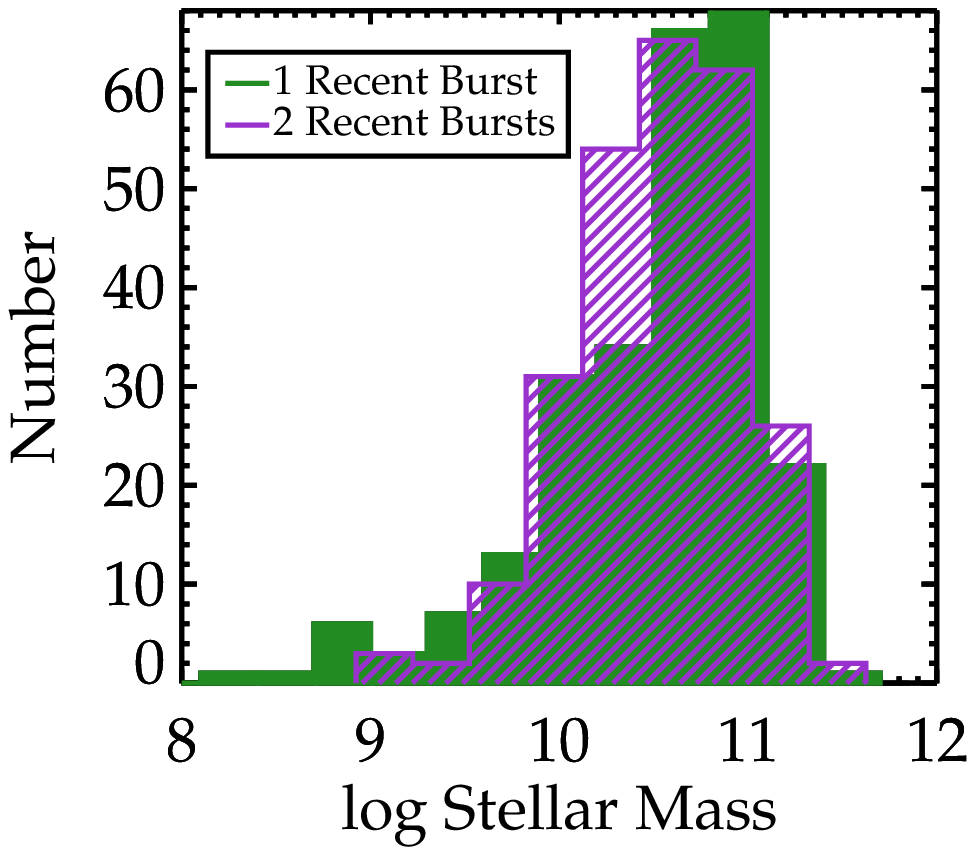}
\caption{Histograms of stellar masses for post-starburst galaxies preferring a single or double recent burst SFH. There is a significant shift toward lower stellar mass for galaxies preferring two recent bursts, compared to those best fit by one recent burst.}
\label{fig:best_fit2}
\end{figure}

In Section \ref{sec:sfh} we discussed two classes of recent SFHs for the post-starburst galaxies: one or two recent bursts. Now, we constrain the implications of which galaxies prefer each model. There are 266 post-starburst galaxies with one recent burst, and 255 with two\footnote{Eleven galaxies have statistically indistinguishable fits to the two SFH models. We exclude them for the analysis in this section.}. In Figure \ref{fig:best_fit2}, we show the distribution of stellar mass for galaxies preferring each model. Recent double burst galaxies are at systematically lower stellar mass than single recent burst galaxies. The separation in the median values is significant ($3.0\sigma$).

\begin{figure}
\includegraphics[width = 0.5\textwidth]{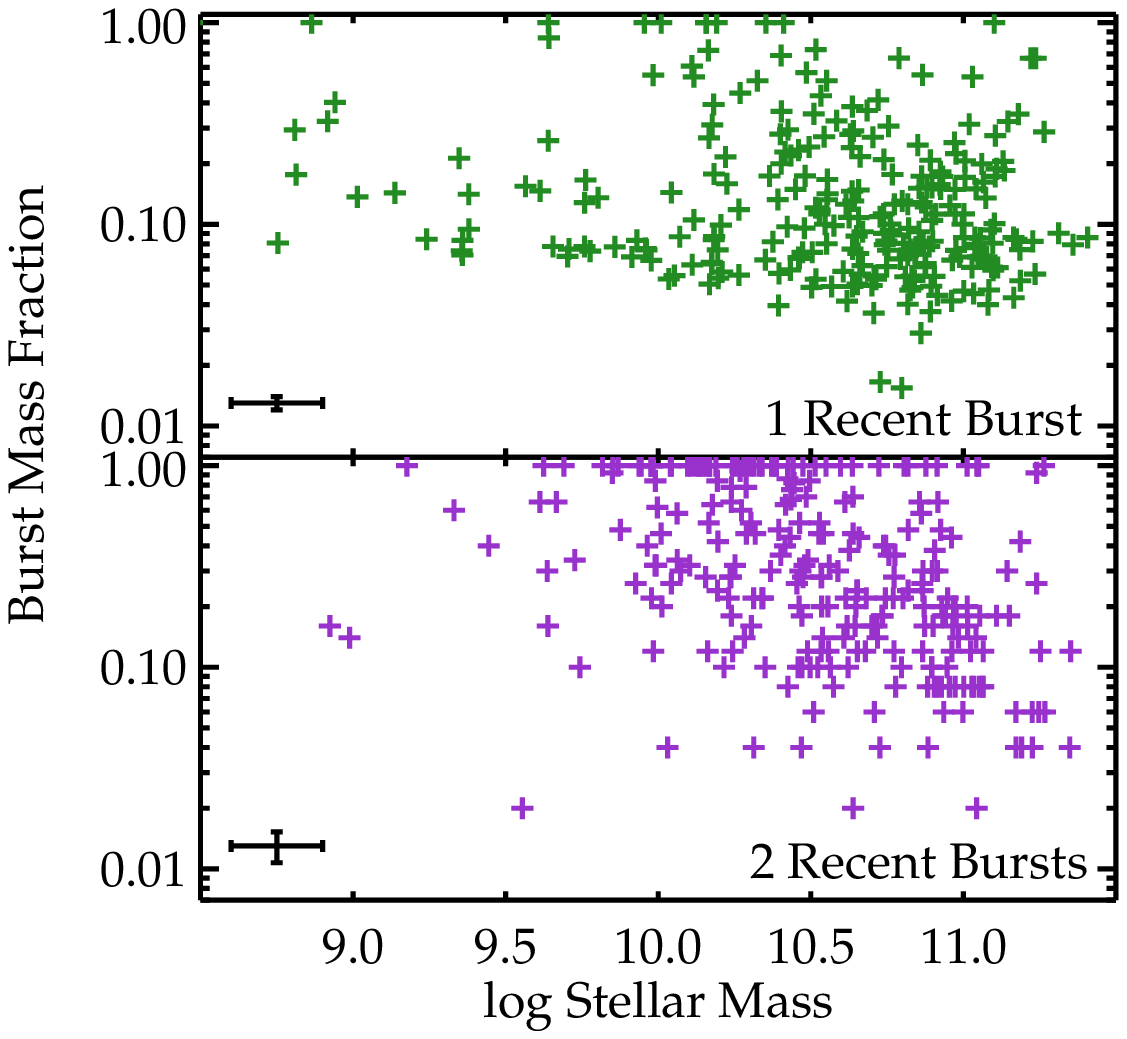}
\caption{Stellar mass vs. burst mass fraction, with galaxies color coded by best-fit SFH. A representative error bar reflecting the fit uncertainty and metallicity uncertainty is also shown. For the 266 galaxies with one recent burst, there is an (Spearman $\rho=-0.19$, 3.1$\sigma$) anti-correlation. For the 255 galaxies with two recent bursts, there is a stronger (Spearman $\rho=-0.45$, 7.2$\sigma$) anti-correlation.}
\label{fig:mfrac_mstellar}
\end{figure}

In Figure \ref{fig:mfrac_mstellar}, we compare the burst mass fractions to stellar mass for galaxies with either one or two recent bursts. Lower mass galaxies have on average higher burst mass fractions (see also \citealt{Bergvall2016}). This trend is strongly dependent on the number of recent bursts, considering both the fit error and systematic error due to the metallicity uncertainty. For the 266 galaxies with one recent burst, there is an (Spearman $\rho=-0.19$, 3.1$\sigma$) anti-correlation. For the 255 galaxies with two recent bursts, there is a stronger (Spearman $\rho=-0.45$, 7.2$\sigma$) anti-correlation. At a single stellar mass, the scatter in burst mass fraction is physical among the galaxies with well-determined mass fractions (i.e., fit errors less than 33\%).

Many of the galaxies at the highest burst mass fractions show a best-fit burst mass fraction of 1, which is unphysical. This is due to the effect discussed in Figure \ref{fig:yfracmfrac}, where small changes in light fraction can lead to large changes in burst mass fractions, especially at the highest burst mass fractions, leading to higher fit errors. The lower bounds of the fit errors on the mass fraction extend to lower burst mass fractions, which are more physical. Nonetheless, we also consider the strength of the anti-correlations with stellar mass excluding the galaxies with best-fit burst mass fractions of 1. The anti-correlation for the sample with two recent bursts remains significant (Spearman $\rho=-0.42$, 5.9$\sigma$, 203 galaxies), while the anti-correlation for the sample with single recent bursts drops below our significance threshold (Spearman $\rho=-0.13$, 2.0$\sigma$, 254 galaxies).

Why do the lower mass galaxies experience a greater number of recent bursts, with a greater fraction of their stellar mass created? There are two possibilities. First, if the lower mass galaxies are more susceptible to stellar feedback, and the first burst was interrupted by the expulsion and re-accretion of gas in the galaxy. \citet{Lee2009} study the duty cycle of starbursts in dwarf galaxies and find the typical burst duration is $\sim 100$ Myr, with such events happening every $1-2$ Gyr. This frequency is consistent with the burst durations and separations fit here for the low mass post-starburst galaxies. 

Another possibility is that the lower mass galaxies could have had progenitors with smaller bulge components, as discussed in our initial motivation of these models (see \S\ref{sec:sfh}). As galactic bulges can act to stabilize the gas, bulge-less progenitors can experience a burst of star formation during the first pericenter passage of the two galaxies, in addition to a burst upon final coalescence \citep{Mihos1994, Cox2008, Renaud2014}. We compare the stellar mass produced in the recent burst to estimates of the current bulge mass by \citet{Mendel2014}. The stellar mass produced in the recent burst is typically 10-100\% of the bulge mass. This implies that the bulges are not formed entirely by the new stars produced in the starburst and that the progenitors had pre-existing bulges. However, there are significant uncertainties in performing a bulge-disk decomposition on post-starburst galaxies using photometry alone. The high concentration of the newly formed stars can result in a smaller bulge radius \citep{Yang2006,Yang2008}, or the mass of the bulge can be underestimated if the SED fitting does not account for the extreme recent star formation history. Additionally, it is uncertain what fraction of the newly formed stars reside in the bulge. Spatially resolved spectroscopy is needed to determine the age of the bulge stellar populations.

We test whether these results could be driven by the magnitude-limited nature of this sample (see \S\ref{maglimit}). We define a volume-limited subset, which has $z<0.112$ and M$_{r} < -21.0$. In this volume-limited subset, there are 57 galaxies with one recent burst, and 76 galaxies with two recent bursts. We still find a significant difference between the stellar masses of galaxies preferring one vs. two recent bursts, and a significant trend in stellar mass vs. burst mass fraction for the galaxies with two recent bursts. Thus, our results are not likely to be driven by a scenario where only the lower stellar mass galaxies with the strongest bursts have been included in our sample.

\subsection{Comparison to Shocked Post-Starburst Galaxies}
\label{sec:spogs}

Our selection of post-starburst galaxies selects against all H$\alpha$ emission, regardless of the source. This biases us against post-starburst galaxies with Seyfert activity \citep{Yesuf2014} or strong shocks \citep{Alatalo2016}. Additionally, we only select galaxies which are truly \textit{post-}starburst. Galaxies with ongoing starbursts are excluded, as seen by the left hand limits in Figure \ref{fig:selection_buckets}. Identifying these galaxies before they enter the post-starburst criterion is of considerable interest, but requires more sophisticated selection methods. 

The age-dating method described here can be applied to both galaxies before they enter the post-starburst phase and to post-starburst galaxies with AGN activity. Our tests on galaxies with strong Balmer absorption and no cut on H$\alpha$ place the galaxies still within their burst (i.e., the post-burst age is negative), with similar burst mass fractions as the post-starburst sample. Our method is useful in matching progenitor starbursting galaxies to their likely post-starburst descendants. Galaxies with suspected AGN activity may benefit from an additional AGN template added to the SPS modeling to account for any contribution to the continuum light.

We use the age-dating method described here to test how a sample of post-starburst galaxies selected to allow emission lines from shocks \citep[SPOGs;][]{Alatalo2016} compares with the sample identified using our H$\alpha$-H$\delta$ cuts.  If the emission lines allowed by the SPOG selection are due to star formation, and the starburst has not truly ended, our best-fit model will show a starburst age $t_{SB}$ shorter than the duration $\tau$ of the most recent burst of star formation, resulting in a ``negative" post-burst age (i.e., the burst is still on-going). However, emission lines from star formation or other sources could impact the optical spectral indices using in these fits. We must correct the Lick H$\beta$, H$\gamma$ and H$\delta$ indices for emission filling. We estimate the emission in these lines, using the H$\alpha$ line flux from the MPA-JHU line catalogs \citep{Brinchmann2004}, the intrinsic line ratios H$\alpha$/H$\beta$ = 2.86, H$\alpha$/H$\gamma$ = 6.11, and H$\alpha$/H$\delta$ = 11.1, assuming case B recombination at T$=10^4$ K \citep{Osterbrock1974} . We use the average reddening from the SPOGs fit without this correction, A$_V = 0.8$ mag, and a Calzetti extinction law as before. The Lick index corrections are thus:

\begin{equation}
\Delta \textrm{H}\beta = \frac{f(\textrm{H}\alpha) 10^{-0.4 A_V/R_V (k_\lambda(H\beta)-k_\lambda(H\alpha))}}{2.86  f(\textrm{H}\beta)_{cont}}
\end{equation}

\begin{equation}
\Delta \textrm{H}\gamma = \frac{f(\textrm{H}\alpha) 10^{-0.4 A_V/R_V (k_\lambda(H\gamma)-k_\lambda(H\alpha))}}{6.11  f(\textrm{H}\gamma)_{cont}}
\end{equation}

\begin{equation}
\Delta \textrm{H}\delta = \frac{f(\textrm{H}\alpha) 10^{-0.4 A_V/R_V (k_\lambda(H\delta)-k_\lambda(H\alpha))}}{11.1  f(\textrm{H}\delta)_{cont}},
\end{equation}

where $f_{cont}$ is the continuum flux at each line. These corrections are large compared to the average error on the indices. For our post-starburst sample, $\langle \Delta$H$\beta \rangle = 0.25$ \AA, lower than the uncertainty $\sigma($Lick H$\beta) = 0.40$ \AA. However, for the SPOGs sample, $\langle \Delta$H$\beta \rangle = 2.71$ \AA, much higher than the uncertainty $\sigma($Lick H$\beta) = 0.64$ \AA. Additionally, we exclude the Fe5015 Lick index, as it is contaminated by the [OIII]$\lambda$5007\AA\ line in the SPOGs sample. 

If the emission lines in the SPOGs sample are due to type II AGN activity, this emission line treatment is sufficient, and the addition of an AGN component to fit the UV/optical photometry is not required. Except in the case of QSOs, the AGN contribution to the {\it NUV} photometry is small, at $\lesssim 15$\% \citep{Salim2007}, similar to the typical errors in our {\it NUV} photometry. The observed UV emission from type II AGN is observed to be extended \citep{Kauffmann2007} and thus not originating from the AGN. 

\begin{figure*}
\includegraphics[width =1\textwidth]{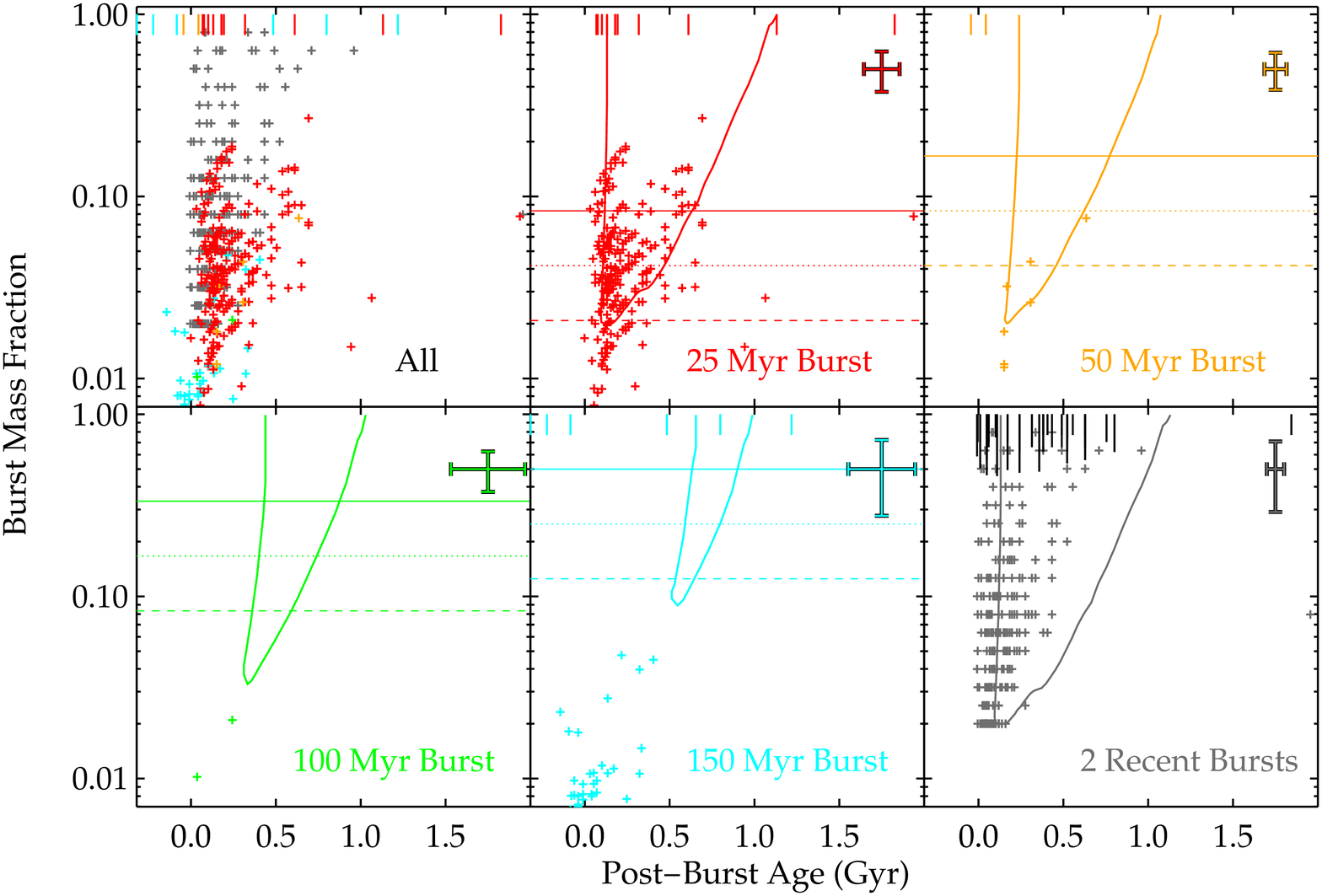}
\caption{Same as Figure \ref{fig:selection_contour}, but for shocked post-starburst galaxies \citep[SPOGs;][]{Alatalo2016}. We plot characteristic error bars representing the fit uncertainties, and we plot individual error bars for galaxies with unphysical burst mass fractions of 1. The contours represent the post-starburst selection contours for the traditionally-selected sample. The SPOGs are generally younger than the post-starburst galaxies, with 51\% too young to have been selected into our post-starburst criterion. SPOGs with similar mass fractions and durations as the post-starbursts may represent an evolutionary sequence. However, an additional population of SPOGs exists at long duration ($>100$ Myr) and small burst fraction ($<10$\%), which will not enter our post-starburst selection criteria.}
\label{fig:spogs}
\end{figure*}

Post-starburst ages and starburst properties for the SPOGs are shown in Table \ref{table:ages_spogs}. We plot the post-burst ages, mass fractions, and durations in Figure \ref{fig:spogs}. The SPOGs are generally younger than the post-starburst galaxies, with 51\% too young to have been selected by our post-starburst criteria, as they lay outside the selection contour lines from Figure \ref{fig:selection_contour}. SPOGs with similar mass fractions and durations as the post-starbursts may represent an evolutionary sequence. However, an additional population of SPOGs (10\% of the total SPOGs sample) exists at long duration ($>100$ Myr) and small burst fraction ($<10$\%), not consistent with our post-starburst selection criteria. We further explore the performance of our age-dating method on this sample in Appendix \ref{sec:appdxb}.

There are a number of galaxies in common between the three post-starburst samples considered here. The differences described above, where the Lick H$\beta$, H$\gamma$, and H$\delta$ indices are corrected for emission and the Fe5015 index is eliminated, result in small changes to the derived ages, that are nonetheless consistent within the derived uncertainties. However, we note that the derived number of bursts (whether the fit prefers the single or double recent burst model) is less stable to these changes. Thus, we adopt the fit results corrected for possible emission lines for the objects in multiple samples in the following analyses. We do not expect this correction to be important for the rest of our post-starburst sample, due to the lack of strong emission lines as detailed in \S\ref{sec:selectionsdss}.

\section{Results: Discovery of Gas and Dust Evolution}

\subsection{Molecular Gas Evolution}

Determining the post-burst ages of post-starburst galaxies, and identifying evolutionary sequences, is important in studying the mechanisms by which star formation shuts down and stays dormant in these galaxies. In this section, we combine molecular gas (CO $J$= 1--0 and CO $J$= 2--1) measurements of post-starburst galaxies to piece together the likely history of molecular gas depletion after the starburst. We combine three samples to extend our time baseline\footnote{We note that during the refereeing process, another paper with several CO observations of post-starburst galaxies was submitted to MNRAS by Yesuf et al.}. The first sample we consider, from \citet{French2015}, uses the selection method discussed in \S\ref{sec:selectionsdss}. These galaxies are contained within our parent sample of post-starburst galaxies, and have stellar masses $10^{9.96}-10^{11.31}$ and redshifts $z=0.0196 - 0.1129$. For those without {\it GALEX} photometry, we use upper limits when available. The second sample, by \citet{Rowlands2015}, uses the selection method by \citet{Wild2009}, which selects high burst mass fraction, long duration starbursts and post-starbursts. The third sample is the shocked post-starburst galaxy sample by \citet{Alatalo2016} described in \S\ref{sec:spogs}, with molecular gas measurements presented in \citet{Alatalo2016b}.

We exclude the \citet{French2015} galaxy labeled H01, as resolved CO (2--1) imaging with ALMA has shown that the gas is associated with a companion, non-post-starburst galaxy. The resolved CO is coincident with a 1.4 GHz FIRST \citep{Becker1995} detection, while the post-starburst component has no corresponding detection. Including this galaxy would not significantly change our results. Unlike H01, none of the other post-starbursts with multiple components inside the observed beam have possible star-forming companions, based on FIRST 1.4 GHz data.

How does the molecular gas and dust content of a galaxy evolve after the starburst ends? We would like to trace the evolution of a similar galaxy through time, although how finely we can divide the sample is limited by the number of post-starburst galaxies with molecular gas measurements in how finely we can divide the sample and by the fact that the measurements are from three differently-selected post-starburst samples. The galaxies considered here span a range of stellar masses from $10^{9.50} - 10^{11.31}$, with 68\% within $10^{10.04} - 10^{10.88}$. There is no significant difference amongst the median stellar masses of the three samples. Similarly, the range in redshifts is small, with all at $z>0.01$ to avoid issues of serious aperture bias, and no significant redshift evolution expected within the sample to $z\sim0.2$.

One of the main sources of scatter in tracing the molecular gas content with age is how much molecular gas was used up in the starburst. If the progenitors of the galaxies studied here had similar initial gas fractions, we should reduce the scatter caused by varying starburst efficiencies by dividing the sample by burst mass fraction. We split the samples into 2 classes of SFHs: those with burst mass fractions $\le 0.2$ and $>0.2$. The two classes have a mix of single and double recent bursts, although the first class has primarily short-duration bursts, and the second long-duration bursts (see \S\ref{sec:thesample} and \S\ref{sec:sbchars}).  We plot the molecular gas to stellar mass fractions vs. post-burst age for each class (Figure \ref{fig:gas}).  We observe a significant decline in the molecular gas fraction ($M(H_2)/M_\star$) with increasing post-burst age (Spearman $\rho=-0.53$ at 4.5$\sigma$ significance, 74 galaxies). 

The significance of this trend persists using either our derived post-starburst ages for the \citet{Rowlands2015} sample or our estimated post-starburst ages from the starburst ages in \citet{Rowlands2015}. Split up into low and high ($\le 0.2$ and $>0.2$) burst fraction bins, the significance is 4.4$\sigma$ (Spearman $\rho=-0.66$, 46 galaxies) for the low burst fraction sample and 1.9$\sigma$ (Spearman $\rho=-0.37$, 28 galaxies) for the high burst fraction sample. Because the errors are not uniform, we test the significance of these results by Monte-Carlo sampling from Gaussian estimates of the error ranges for both post-starburst age and molecular gas fraction. We similarly find a $>3\sigma$ significant trend for the full sample and the low burst mass fraction sample and no significant trend for the high burst mass fraction sample.

We consider several differences in how the various post-starburst samples were selected, which might influence the observed trend. We do not expect a selection bias in the \citet{Rowlands2015} sample against galaxies with low gas fractions. The \citet{Alatalo2016b} sample measured molecular gas content in only galaxies with SNR$>3$ in WISE 22$\mu$m, which would result in higher gas fractions if the dust traces the gas. However, in the \citet{French2015} sample, only 2/32 galaxies do not have WISE 22$\mu$m detections with SNR$>3$; neither has a molecular gas detection. Removing these two galaxies has no effect on the anti-correlations described, so the selection cut made by \citet{Alatalo2016b} should not affect the trend seen in these gas fractions with post-burst age. The \citet{French2015} sample was chosen with a cut on the median SNR of the SDSS spectra, while the \citet{Alatalo2016b} sample was not. However, only two of the \citet{Alatalo2016b} sample would not pass our cut of median SNR $>$10. 

All three post-starburst samples are affected by the magnitude limit of their SDSS parent sample. There is no additional limit imposed on their {\it GALEX} magnitudes, as we fit their stellar populations regardless of the presence of {\it GALEX} detections, unlike in the sample of post-starburst galaxies discussed earlier in this work. The lack of the {\it GALEX} information is reflected in the significantly larger errors in age in these cases. We established earlier (\S\ref{maglimit}) that the derived ages are not affected by biases in the SDSS parent sample. In combination with our normalization by stellar mass, our result is thus not due to the magnitude-limited nature of the SDSS parent sample. We caution that these results may differ for samples at dramatically different stellar masses or redshifts, or with differing star formation histories.

\subsection{Molecular Gas Depletion Mechanisms}

What processes could deplete the molecular gas reservoirs after the starburst has ended? We fit the observed trend, plotting the range of best fits in Figure \ref{fig:trends}, using a linear least-squares fit to ln $(M(H_2)/M_{\rm stellar})$ and the post-burst age, taking errors in both quantities into account. Using \texttt{fitexy} in IDL, we find an exponentially declining timescale of 117 Myr.

However, the intrinsic scatter around the best fit relation is large, so we also consider a fitting routine that assumes an intrinsic scatter term. Using \texttt{linmix\_err}, a significant intrinsic scatter term of $\sigma = 0.5$ dex is measured. This method produces a shallower slope, with an exponentially declining timescale of 230 Myr. If the scatter beyond the measurement error is driven by an intrinsic difference in the initial molecular gas fraction, the assumption of constant intrinsic scatter is valid. However, if the intrinsic scatter varies with time, as it might if the scatter beyond the measurement errors is driven by a time-dependent process, or if the scatter is driven by variation in measurement systematics like the choice of $\alpha_{\rm CO}$, the assumption of constant intrinsic scatter may be acting to flatten the observed slope. This has the effect of adding a systematic uncertainty to the molecular gas depletion timescale, and we consider the range of possibilities from 117-230 Myr. Similarly, we find the best fit molecular gas fraction at zero post-burst age (when 90\% of the stars in the recent starburst had formed) is $0.4 - 0.7$.

Mechanisms related to star formation or AGN activity could act to deplete the molecular gas. Star formation not only consumes gas, but could also expel or heat the gas through stellar feedback (from stellar winds and supernovae). Black holes could also consume some of the gas, and expel or heat the rest via AGN feedback. In some cases, strong outflows have been observed, but without sufficient energy to become unbound from the galaxy \citep{Alatalo2015}. Feedback processes could also result in a change of state in the gas, to atomic or hot ionized gas, removing the observed molecular gas signatures. 

We test these possibilities by comparing the timescales for gas depletion. Depletion this rapid cannot be due to star formation, given the upper limits on the current SFRs in these post-starbursts. Depleting a gas reservoir of gas fraction $0.1-1$, typical of starburst galaxies and consistent with the start of the observed post-starburst trend, in 100 Myr would require sSFR$\sim 10^{-9} - 10^{-8}$yr$^{-1}$. These sSFRs are as high as they were during the starburst (see Figure \ref{fig:max_sfr}). The sSFRs of the post-starburst galaxies are much lower, sSFR$ \le 5\times10^{-13} - 2\times10^{-10}$ yr$^{-1}$, even considering estimates from D$_n$(4000), which are sensitive to star formation over several hundred Myr. The current sSFRs of the \citet{Alatalo2016b} sample are also too low to account for the rapid gas depletion, with sSFR$= 1\times10^{-12} - 4\times10^{-10}$ yr$^{-1}$.  

We consider two pathological cases that would affect our gas depletion time estimates by affecting our inferred rate of gas consumption by stars: a bottom-heavy IMF or an unusual dust geometry. For an extremely bottom-heavy IMF (see \citealt{French2015}), the SFRs would only differ by $\sim20\times$, not the $\sim400\times$ required to account for the molecular gas depletion. In the case of an unusual dust geometry, we can estimate the SFRs using the 1.4 GHz emission. As discussed in \citet{French2015}, 1.4 GHz emission is sensitive to dust-obscured star formation, but may be enhanced by the LINER (LINER-like emission is commonly seen in post-starbursts, \citealt{Yan2006,Yang2006}) or by the recent starburst. For the \citet{French2015} galaxies with FIRST \citep{Becker1995} detections, the predicted median molecular gas depletion time is still only 1 Gyr, not enough to explain the observed rapid depletion.

The molecular gas depletion cannot be due to stellar feedback either, as the mass loading factors (mass outflow / SFR) from stellar feedback are expected to be 1-5 for the stellar mass range of this sample \citep{Muratov2015}, so the addition of stellar feedback is not sufficient to resolve this deficiency. Given the current low SFRs, mass loading factors of $\sim400$ would be required to explain the observed molecular gas depletion rates. While starburst-driven outflows can result in molecular gas depletion times of $\sim10^8$ yr, these outflow rates are observed scale with the SFRs \citep{Cicone2013}, so cannot explain the observed molecular gas depletion given the low limits on the current SFRs described above.

Direct accretion of gas by the black hole is unlikely to be a significant cause of the observed decline, as the black hole masses are expected to be of order $10^7-10^8$ M$_\sun$ for this sample, given bulge mass estimates from \citet{Mendel2014} and the best-fit black hole - bulge relation from \citet{McConnell2012}, and of order $10^9-10^{10}$ M$_\sun$ of molecular gas is lost.

The molecular gas reservoirs could be expelled or destroyed after the starburst has ended through AGN-driven outflows. The delay between the decline in star-formation and the onset of AGN feeding is expected to be $\sim$10-300 Myr \citep{Davies2007,Schawinski2009,Wild2010,Cales2015}, consistent with the durations of the starbursts observed here, such that the observed depletion of the gas takes place with similar delay times after the starburst began. Longer delays of several Gyr \citep{Curtis2016} may also occur, but none of the post-starburst sample selection methods would have selected such galaxies. We note that there may exist an additional time delay between AGN feeding and the loss of the molecular gas detection. The median rate of gas depletion required, given the $M(H_2)$ measurements and 117-230 Myr depletion time, is 15-31 M$_\sun$yr$^{-1}$ with a 68\% likelihood range of 2-150 M$_\sun$yr$^{-1}$. This rate is consistent with observed outflow rates driven by AGN, although the depletion timescales observed are much shorter, of only a few Myr \citep{Cicone2013}. If depletion times are this rapid, scatter in the time between the starburst and the onset of gas depletion could result in the observed depletion time. The LINERs observed in post-starburst galaxies \citep{Yan2006, Yang2006, French2015} may be related to the molecular gas depletion, as \citet{Cicone2013} find that outflows in LINERs have outflow rates of 10-100 M$_\sun$yr$^{-1}$, with molecular gas depletion times of 10-250 Myr, both consistent with what we observe during the post-starburst phase\footnote{see also \citet{Baron2017}}. Thus, the observed decline in molecular gas fraction during the post-starburst phase may be a smoking gun for AGN or LINER-related feedback in transitioning galaxies.

Post-starburst galaxies have stellar populations, color gradients, morphologies, and kinematics consistent with reaching the red sequence of early type galaxies in \citep{Norton2001, Yang2004, Yang2008, Pracy2013, Pawlik2015} in a few Gyr. Early type galaxies are typically gas-poor, with molecular gas fractions of $\lesssim 10^{-3}$ \citep{Young2011}. With the observed molecular gas depletion rate, the post-starburst galaxies should reach early-type levels of molecular gas in 700-1500 Myr, thus becoming gas-poor when their other properties start to resemble early-types. 

\subsection{Dust Evolution}

\begin{figure*}
\centering
\includegraphics[width =0.8\textwidth]{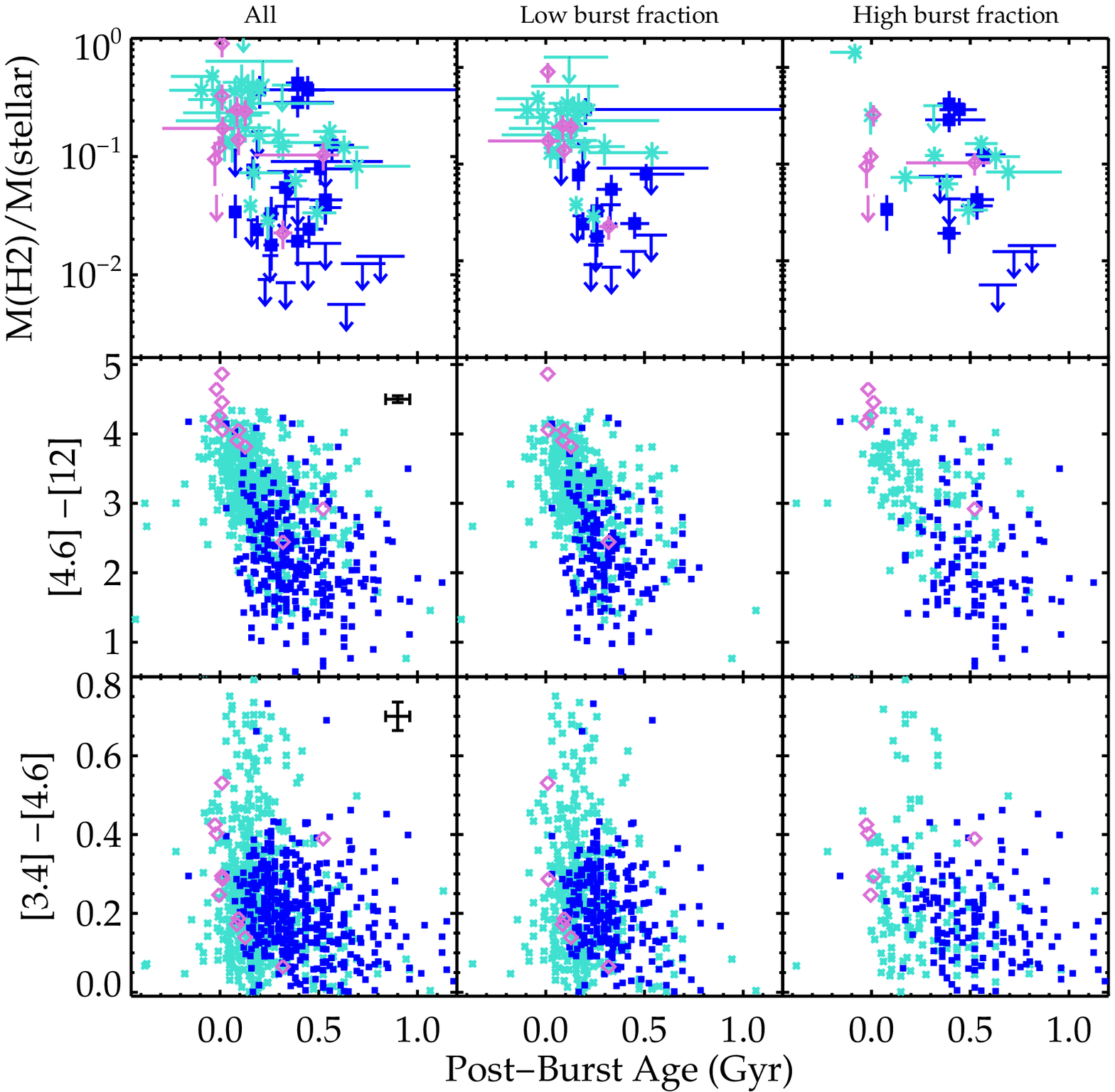}
\caption{ Post-burst ages vs. molecular gas fractions and WISE colors for three samples of starburst/post-starburst galaxies. The left panels show the full set of galaxies, broken down into those with burst mass fractions $\le 20$\% (middle panel) and $>20$\% (right panel). The top panels show molecular gas mass fractions vs. post-burst age for three samples: post-starburst galaxies (blue squares) from \citet{French2015}, shocked post-starbursts (teal stars) from \citet{Alatalo2016b}, and starburst/post-starbursts (pink diamonds) from \citet{Rowlands2015}. We observe a significant trend in the molecular gas fraction with the post-burst age, at 4.5$\sigma$ significance. Split up into low and high burst fraction bins, the significance drops for the high burst fraction bin.  Early type galaxies are typically gas-poor, with molecular gas fractions of $\lesssim 10^{-3}$ \citep{Young2011}. With the observed molecular gas depletion rate, the post-starburst galaxies should reach early-type levels of molecular gas in 700-1500 Myr. The middle panels show the WISE [4.6]-[12] $\mu$m colors vs. post-burst age for the post-starburst galaxies studied in this work, shocked post-starbursts from \citet{Alatalo2016}, and starburst/post-starbursts from \citet{Rowlands2015}. The bottom panels show the WISE [3.4]-[4.6] $\mu$m colors vs. post-burst age for the same samples. We see significant ($>3\sigma$) anti-correlations for each of the WISE colors with post-burst age, and for all of the burst mass fraction bins. These WISE colors represent a combination of the dust mass and dust heating. Various heating sources can act to change the WISE colors: star-formation, young (A) stars, evolved stars (post-AGB or TP-AGB), or AGN. Given the anti-correlations of bluer WISE colors with post-burst age, either the dust mass could be declining with the gas mass, and/or the sources of dust heating could be declining. Early type galaxies have WISE [3.4]-[4.6] colors of $-0.1 - 0$, and [4.6]-[12] colors of $0-1$. If the WISE [4.6]-[12] colors decline linearly with post-burst age, they should reach colors typical of early types at $\sim$1 Gyr post-burst. The WISE [3.4]-[4.6] colors have significantly more scatter, and it is not clear when post-burst the scatter decreases.
}
\label{fig:gas}
\end{figure*}

\begin{figure}
\includegraphics[width = 0.5\textwidth]{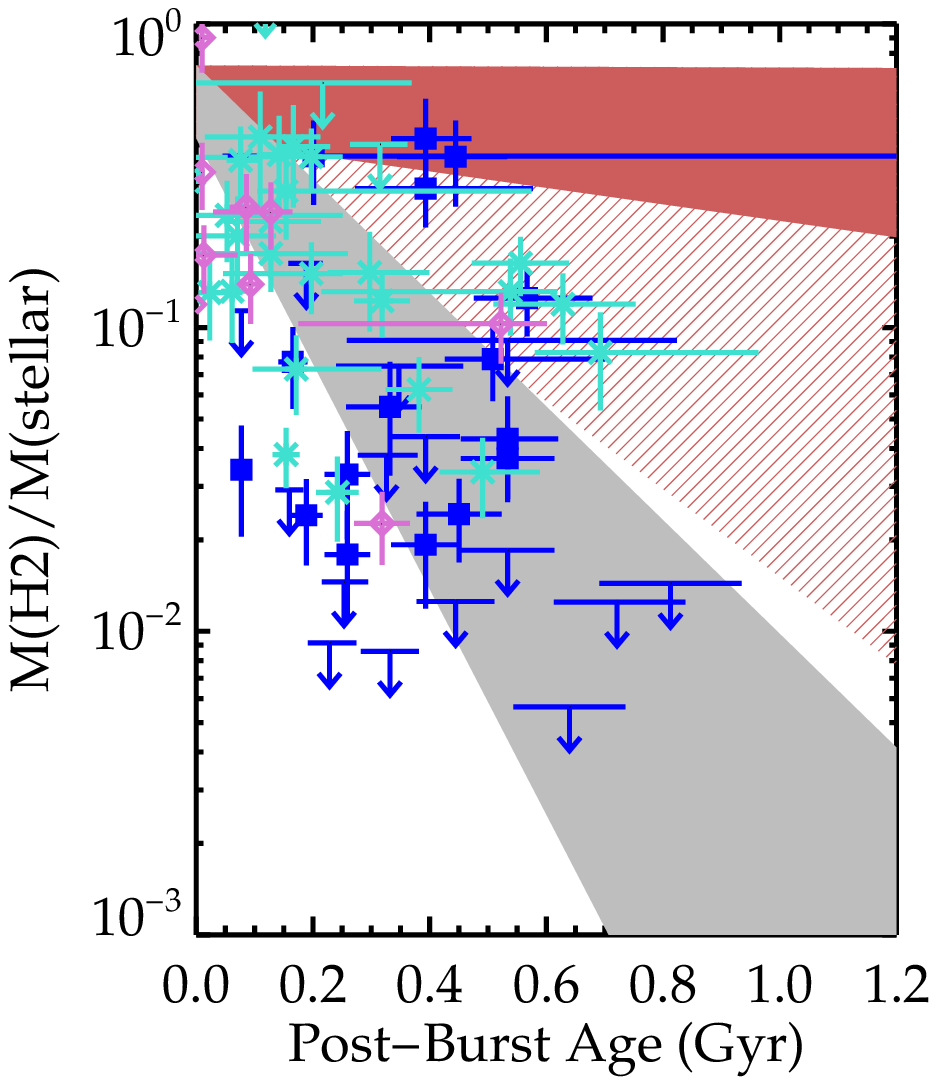}
\caption{Post-burst ages vs. molecular gas fractions for the same samples as in Figure \ref{fig:gas}. The best-fit line for an exponential depletion of the molecular gas reservoirs (grey region) has a timescale of 117-230 Myr. The post-burst age is the time since 90\% of the stars from the recent starburst were formed. The gas depletion cannot be due to on-going star formation or stellar feedback, which would have a much longer allowed depletion time (solid red region), even allowing for an unusual IMF or dust geometry (hashed red region). Early type galaxies are typically gas-poor, with molecular gas fractions of $\lesssim 10^{-3}$ \citep{Young2011}. With the observed molecular gas depletion rate, the post-starburst galaxies should reach early-type levels of molecular gas in 700-1500 Myr. 
}
\label{fig:trends}
\end{figure}

We test for dust evolution with post-burst age, using the WISE [3.4]-[4.6] and [4.6]-[12] colors. We take the SDSS DR12 - WISE all-sky matched catalog, and select the {\tt wxmpro} profile-fit magnitudes for the \citet{Rowlands2015} sample, the whole \citet{Alatalo2016} SPOGs sample, and for our entire post-starburst sample presented here\footnote{We note that the WISE magnitudes used are Vega magnitudes.}. We only use WISE data for which SNR$>3$ for both bands in each color used in the following analysis. We plot each color against the post-burst age in Figure \ref{fig:gas}, broken up into the same burst mass fraction categories as for the previous molecular gas analysis. For the WISE [4.6]-[12] color, there is a significant (Spearman $\rho=-0.55$, $16\sigma$, 828 galaxies) anti-correlation with the post-burst age that remains significant for the low and high burst mass fraction bins (Spearman $\rho=-0.55, -$0.57, 13$\sigma$, 9$\sigma$, 575, 253 galaxies, respectively). The typical WISE [4.6]-[12] colors decrease by $\sim3$ magnitudes over the post-starburst phase. For the WISE [3.4]-[4.6] color, there is a weaker, but still significant (Spearman $\rho=-0.26$, 8$\sigma$, 953 galaxies) anti-correlation with the post-burst age that remains significant for the low and high burst mass fraction bins (Spearman $\rho=$-0.21, -0.29, 6$\sigma$, $5\sigma$, 648, 305 galaxies respectively). However, the anti-correlation between post-burst age and WISE [3.4]-[4.6] color may be driven by increased scatter in WISE [3.4]-[4.6] color for the SPOGs sample compared to the older post-starburst sample. The typical WISE [3.4]-[4.6] colors decrease by $\sim0.5$ magnitudes over the post-starburst phase.

We note that the WISE [3.4]-[4.6] colors may be subject to redshift-dependent effects. We do not attempt to model the infrared spectra of the post-starburst galaxies (see \citep{Smercina2018}), so we do not k-correct the WISE colors. We do not expect this choice to drive the observed correlations with post-starburst age, because there is no significant correlation observed between post-starburst age and redshift. We also do not observe a significant correlation with WISE [4.6]-[12] color with redshift, though there is a significant redshift trend in WISE [3.4]-[4.6].

These WISE colors represent a combination of the dust mass and dust heating. Various heating sources can act to change them: star-formation, young (A) stars, evolved stars (post-AGB or TP-AGB), or AGN. PAH features can also potentially influence these colors. Starbursts, star-forming galaxies, and early type galaxies separate in the color-color space of WISE [4.6]-[12]  vs. [3.4]-[4.6] \citep{Wright2010, Yesuf2014, Alatalo2014}, with post-starburst galaxies lying near the infrared ``green valley" and star-forming regions in WISE [4.6]-[12] color and near star-forming galaxies in WISE [3.4]-[4.6] color \citep{Alatalo2016c}. Given the anti-correlations of bluer WISE colors with post-burst age, either the dust mass or sources of dust heating could be declining over time.

\citet{Rowlands2015} use additional observations from {\it Herschel} PACS and SPIRE of their sample of starbursts and post-starbursts to show that both the dust-to-stellar mass ratio and cold dust temperature declines with post-burst age. However, it is not clear if the observed trends in WISE [3.4]-[4.6] and [4.6]-[12] are driven by the same effects, as these colors are sensitive to hotter dust. Additionally, many of the shocked post-starbursts have higher WISE [3.4]-[4.6] colors than the \citet{Rowlands2015} sample and may be subject to additional sources of dust heating. The lack of a significant anti-correlation in the molecular gas mass with post-burst age in the \citet{Rowlands2015} sample alone may be due to small numbers, or a narrower post-burst age baseline. \citet{Rowlands2015} do see a trend in the dust-to-gas ratio with age, but only when excluding the galaxy with the longest post-burst age, one of only two from that sample with a sSFR consistent with our post-starburst selection. 
Further work studying the full MIR-FIR SED of post-starburst galaxies as they evolve through the post-starburst phase will be needed to disentangle the dust mass and various possible heating sources. 

Early type galaxies have narrow ranges of WISE [3.4]-[4.6] colors of $-0.1$ to 0, and [4.6]-[12] colors of 0 to 1. If the WISE [4.6]-[12] colors decline linearly with post-burst age, they should reach colors typical of early types at $\sim$1 Gyr post-burst. The WISE [3.4]-[4.6] colors have significantly more scatter, and it is not clear when the scatter decreases after the burst.

\section{Conclusions}

We fit stellar population models to 532 post-starburst galaxies, breaking the age - burst strength - burst duration degeneracy using a combination of UV photometry from {\it GALEX} \citep{Martin2004} and optical photometry and spectroscopy from SDSS \citep{Strauss2002, Ahn2013}. We present a catalog of post burst ages, burst mass fractions, and burst durations. We conclude the following:
\begin{enumerate}

\item We derive post-burst ages and burst mass fractions, with median errors of 22\% and 42\%, respectively. 68\% of the burst mass fractions are within 7.0\%--68\%, and 68\% of the post-burst ages are within 220--650 Myr. We define the ``post-burst" ages to be the time elapsed since 90\% of the stars from the recent starburst(s) formed.

\item This method is more accurate than using the $D_n(4000)$ - H$\delta$ parameter space to derive post-burst ages and burst mass fractions. The relation between $D_n(4000)$ and the post-burst age suffers from a degeneracy with the burst mass fraction and burst duration. For post-starbursts with $D_n(4000) < 1.3$, where post-starburst ages are typically younger than 300 Myr, the degeneracy is lessened, and $D_n(4000)$ is highly correlated with the post-burst age. However, if $D_n(4000) > 1.42$, where post-starburst ages typically range from 300-1500 Myr, $D_n(4000)$ is no longer significantly ($>3\sigma$) correlated with the post-burst age. 

\item The star formation rates experienced during the starburst were $\sim$10-100$\times$ above the stellar mass - specific star formation rate relation, consistent with those in starbursting galaxies \citep{Schiminovich2007}. 

\item Starbursts with specific star formation rates $>10^{-8}$ yr$^{-1}$ are rare ($<1$\% of the sample). As a consequence, those post-starburst galaxies with short duration (25 Myr) bursts typically generate 8\% of their stellar mass in the burst, whereas galaxies with longer duration ($\ge$150 Myr) bursts produce a median value of 36\% of their stellar mass in the burst. 

\item Many post-starbursts show signs of intermediate mass ($\sim$F) stellar populations; 50\% of the post-starbursts are best fit by a single recent burst, 48\% prefer a double recent burst, and 2\% do not have a statistical preference. Lower stellar mass galaxies are more likely to experience two recent bursts, and the fraction of mass produced in their recent burst(s) is more strongly anti-correlated with their stellar mass.

\item We compare the SFHs selected via our selection criteria with those from other, differently-selected 
post-starburst samples. \citet{Rowlands2015} use the approach of \citet{Wild2007, Wild2009} to select longer duration ($>$150 Myr) starbursts, at a wider range of post-burst ages than we do; they also are biased against shorter duration starbursts that have specific star formation rates similar to our sample.  Shocked POst-Starburst Galaxies \citep[SPOGs;][]{Alatalo2016} have generally younger post-burst ages than ours, with 51\% too young to have entered into our post-starburst selection. SPOGs with younger post-burst ages, but similar mass fractions and durations as our post-starbursts, may represent progenitors to our sample. However, an additional population (10\%) of SPOGs exists at long duration ($>100$ Myr) and small burst fraction ($<10$\%), which will not evolve into our post-starburst population.

\item Combining these three samples of post-starburst galaxies, we observe a significant decline in their molecular gas to stellar mass fraction with increasing post-burst age, at $4\sigma$ significance. This trend persists when we control for the fraction of stellar mass produced in the recent burst(s). The best fit exponentially declining timescale is 117-230 Myr, with the best fit initial molecular gas fractions 0.4-0.7 at a post-burst age of zero. With the observed molecular gas depletion rate, the post-starburst galaxies should reach early-type levels of molecular gas in 700-1500 Myr. The rapid depletion rate implied by this trend of 2-150 M$_\sun$yr$^{-1}$ cannot be due to current star formation, given the upper limits on the current SFRs in these post-starbursts, suggesting that the molecular gas is expelled or destroyed in AGN-driven outflows. 

\item We find significant ($>3\sigma$) anti-correlations of the WISE [4.6]-[12] and [3.4]-[4.6] $\mu$m colors with the post-burst age of the galaxy. Given the anti-correlations of bluer WISE colors with post-burst age, either the dust mass or sources of dust heating could be declining over time, as with the gas fraction. Various heating sources are possible during this phase: star-formation, young (A) stars, evolved stars (post-AGB or TP-AGB), or AGN.

\end{enumerate}

Post-starburst galaxies are a critical laboratory for studying the evolution of starbursts, of galaxies onto the red sequence, and of galaxies onto the black hole - bulge relation \citep{Ferrarese2000}. They are also the preferred hosts of tidal disruption events \citep{Arcavi2014,French2016}, and thus provide clues to what sets the rate of tidal disruption events \citep{French2017}.  The UV-optical stellar population fitting method presented here will be a useful tool in timing the detailed evolution of individual galaxies from star-forming, through the post-starburst phase, and eventually to quiescence.


\acknowledgements
We thank the referee for their detailed feedback and suggestions, which have improved this paper. We thank C. Conroy for his help regarding the FSPS models. We thank Stephen Shectman for providing the LCRS spectra. We thank Dennis Zaritsky, Zhihui Li and Adam Smercina for useful discussions. KDF is supported by Hubble Fellowship Grant HST-HF2-51391.001-A, provided by NASA through a grant from the Space Telescope Science Institute, which is operated by the Association of Universities for Research in Astronomy, Incorporated, under NASA contract NAS5-26555. KDF acknowledges support from NSF grant DGE-1143953, P.E.O., and the ARCS Phoenix Chapter and Burton Family. YY is supported by Basic Science Research Program through the National Research Foundation of Korea (NRF) funded by the Ministry of Science, ICT \& Future Planning (NRF-2016R1C1B2007782). AIZ acknowledges funding from NASA grant ADP-NNX10AE88G. Funding for SDSS-III has been provided by the Alfred P. Sloan Foundation, the Participating Institutions, the National Science Foundation, and the U.S. Department of Energy Office of Science. The SDSS-III web site is http://www.sdss3.org/.

\bibliographystyle{apj}
\bibliography{earefs.bib}

\clearpage


\begin{splitdeluxetable*}{cccccccccccBccccccccc}
\tabletypesize{\scriptsize}
\tablewidth{0pt}
\tablecaption{Post-burst Ages\label{table:ages}}
\tablehead{\colhead{RA (J2000)\tablenotemark{a}} & \colhead{Dec (J2000)} &  \colhead{z} &  \colhead{log M$_\star$/M$_\odot$} &  \colhead{SFH\tablenotemark{b}} & \multicolumn{3}{c}{Post-burst Age (Myr)} & \multicolumn{3}{c}{Age since Burst Start(Myr)} & \multicolumn{3}{c}{$\tau$ or $\Delta t$\tablenotemark{c}(Myr)} &  \multicolumn{2}{c}{Burst Light Fraction $y_f$ \tablenotemark{d}} &   \multicolumn{3}{c}{Burst Mass Fraction $m_{\rm burst}$} &  \colhead{$A_V$\tablenotemark{f}} \\
\colhead{} & \colhead{} &  \colhead{} &  \colhead{} &  \colhead{} & \colhead{(16\%)} & \colhead{(50\%)} &  \colhead{(84\%)} & \colhead{(16\%)} & \colhead{(50\%)} &  \colhead{(84\%)} & \colhead{(16\%)} & \colhead{(50\%)}  & \colhead{(84\%)} & \colhead{(burst 1)} & \colhead{(burst 2)\tablenotemark{e}} & \colhead{(16\%)} & \colhead{(50\%)} &  \colhead{(84\%)} & \colhead{(mag)}}
\startdata
       3.963556 &      -10.388311 &          0.1981 &            10.9 & 2 & 161 &  241 &  308 &  701 &  781 &  848 &  500 &  500 &  500 &            0.24 &            0.58 &            0.09 &            0.16 &            1.00 &             0.6 \\
       4.349453 &      -10.758440 &          0.1868 &            10.9 & 1 & 193 &  331 &  778 &  308 &  446 &  893 &   25 &   50 &  100 &            0.41 & &            0.04 &            0.05 &            0.06 &            0.5 \\
       5.275367 &       -1.226071 &          0.1071 &            10.0 & 1 &1460 & 1720 & 2045 & 1517 & 1778 & 2102 &   25 &   25 &  200 &            1.00 & &            0.75 &            1.00 &            1.00 &            0.3 \\
      10.414436 &        1.068420 &          0.0430 &            10.5 & 2 & 534 &  628 &  721 & 1574 & 1668 & 1761 & 1000 & 1000 & 1000 &            0.27 &            0.66 &            0.25 &            0.52 &            0.88 &             0.2 \\
      11.246839 &       -8.889684 &          0.0196 &            10.2 & 2 & 683 &  801 &  919 & 1723 & 1841 & 1959 & 1000 & 1000 & 1000 &            0.31 &            0.63 &            0.42 &            0.66 &            1.00 &             0.6 \\
      13.103527 &        0.734070 &          0.0699 &            10.5 & 2 & 235 &  294 &  373 & 1275 & 1334 & 1413 & 1000 & 1000 & 1000 &            0.25 &            0.72 &            0.08 &            0.58 &            1.00 &             1.1 \\
      17.773197 &       14.266233 &          0.0994 &            10.4 & 2 & 139 &  171 &  247 & 1179 & 1211 & 1287 & 1000 & 1000 & 1000 &            0.20 &            0.80 &            0.28 &            1.00 &            1.00 &             0.7 \\
      19.925944 &        1.131008 &          0.0899 &            10.4 & 1 & 189 &  224 &  260 &  246 &  281 &  317 &   25 &   25 &   25 &            0.46 & &            0.05 &            0.06 &            0.07 &            0.3 \\
      20.065347 &       -9.988923 &          0.1368 &            10.6 & 2 & 124 &  159 &  206 & 1164 & 1199 & 1246 & 1000 & 1000 & 1000 &            0.18 &            0.77 &            0.11 &            0.32 &            1.00 &             0.5 \\
      22.128778 &        0.965819 &          0.2541 &            11.4 & 1 & 325 &  386 &  447 &  440 &  501 &  562 &   25 &   50 &   50 &            0.51 & &            0.06 &            0.09 &            0.11 &            0.4 \\
\enddata
\tablenotetext{a} {Table truncated after 10 rows. Columns 1-4 are from SDSS DR8 \citep{Aihara2011} and MPA-JHU catalogs \citep{Brinchmann2004, Tremonti2004}.}
\tablenotetext{b} {Number of recent bursts.}
\tablenotetext{c} {If SFH = 1, burst duration ($\tau$). If SFH = 2, separation between bursts ($\Delta t$).}
\tablenotetext{d} {If SFH = 2 recent bursts, light fraction for each burst is shown (burst mass fractions are the same for each burst, so the light fractions will be different).}
\tablenotetext{e} {If SFH = 2 recent bursts, burst mass fraction shown is combined from both recent bursts.}
\tablenotetext{f} {Includes Galactic foreground extinction.}
\end{splitdeluxetable*}


\begin{splitdeluxetable*}{cccccccccccBccccccccc}
\tabletypesize{\scriptsize}
\tablewidth{0pt}
\tablecaption{Shocked Post-starburst Galaxy Ages (\citealt{Alatalo2016} sample) \label{table:ages_spogs}}
\tablehead{\colhead{RA (J2000)\tablenotemark{a}} & \colhead{Dec (J2000)} &  \colhead{z} &  \colhead{log M$_\star$/M$_\odot$} &  \colhead{SFH\tablenotemark{b}} & \multicolumn{3}{c}{Post-burst Age (Myr)} & \multicolumn{3}{c}{Age since Burst Start(Myr)} & \multicolumn{3}{c}{$\tau$ or $\Delta t$\tablenotemark{c}(Myr)} &  \multicolumn{2}{c}{Burst Light Fraction $y_f$ \tablenotemark{d}} &   \multicolumn{3}{c}{Burst Mass Fraction $m_{\rm burst}$} &  \colhead{$A_V$\tablenotemark{f}} \\
\colhead{} & \colhead{} &  \colhead{} &  \colhead{} &  \colhead{} & \colhead{(16\%)} & \colhead{(50\%)} &  \colhead{(84\%)} & \colhead{(16\%)} & \colhead{(50\%)} &  \colhead{(84\%)} & \colhead{(16\%)} & \colhead{(50\%)}  & \colhead{(84\%)} & \colhead{(burst 1)} & \colhead{(burst 2)\tablenotemark{e}} & \colhead{(16\%)} & \colhead{(50\%)} &  \colhead{(84\%)} & \colhead{(mag)}}
\startdata
       0.825899 &        0.812318 &          0.1390 &            10.8 & 1 & 105 &  153 &  574 &  162 &  211 &  631 &   25 &   25 &  150 &            0.63 & &            0.04 &            0.05 &            0.07 &            1.2 \\
       1.132990 &       -1.236591 &          0.0887 &            10.6 & 1 &  96 &  404 &  556 &  441 &  749 &  902 &   25 &  150 &  200 &            0.79 & &            0.01 &            0.04 &            0.06 &            1.4 \\
       2.614132 &      -10.728253 &          0.1328 &            10.7 & 2 & 200 &  335 &  529 & 1240 & 1375 & 1569 & 1000 & 1000 & 1000 &            0.21 &            0.58 &            0.07 &            0.16 &            0.32 &             0.4 \\
       2.938423 &       -0.908529 &          0.0476 &            10.2 & 1 & 586 &  692 &  960 &  644 &  749 & 1018 &   25 &   25 &  200 &            0.79 & &            0.21 &            0.27 &            0.33 &            1.2 \\
       7.370773 &       14.561913 &          0.1432 &            10.8 & 1 & 115 &  193 &  417 &  172 &  251 &  474 &   25 &   25 &   50 &            0.40 & &            0.02 &            0.02 &            0.03 &            1.2 \\
       7.512109 &       -0.885154 &          0.0598 &            10.1 & 1 & 765 & 1218 & 1485 & 1225 & 1678 & 1945 &   25 &  200 &  200 &            1.00 & &            0.77 &            1.00 &            1.00 &            0.6 \\
       8.039521 &       -9.225753 &          0.1674 &            10.8 & 1 &   6 &  323 &  657 &  351 &  668 & 1003 &   25 &  150 &  200 &            0.50 & &            0.01 &            0.01 &            0.01 &            0.8 \\
       8.511626 &       -9.705304 &          0.0125 &            11.1 & 1 &-488 & -428 & -368 &  -28 &   31 &   91 &  200 &  200 &  200 &            0.05 & &            0.00 &            0.00 &            0.00 &            0.2 \\
       8.541455 &       -0.288485 &          0.0580 &             9.2 & 1 &-371 &  -38 &  -12 &   89 &  421 &  447 &   25 &  200 &  200 &            0.50 & &            0.01 &            0.01 &            0.01 &            0.2 \\
       9.282580 &        0.410167 &          0.0806 &            10.2 & 2 & 261 &  314 &  364 & 1301 & 1354 & 1404 & 1000 & 1000 & 1000 &            0.26 &            0.71 &            0.45 &            1.00 &            1.00 &             0.8 \\
\enddata
\tablenotetext{a} {Table truncated after 10 rows. Columns 1-4 are from SDSS DR8 \citep{Aihara2011} and MPA-JHU catalogs \citep{Brinchmann2004, Tremonti2004}.}
\tablenotetext{b} {Number of recent bursts.}
\tablenotetext{c} {If SFH = 1, burst duration ($\tau$). If SFH = 2, separation between bursts ($\Delta t$).}
\tablenotetext{d} {If SFH = 2 recent bursts, light fraction for each burst is shown (burst mass fractions are the same for each burst, so the light fractions will be different).}
\tablenotetext{e} {If SFH = 2 recent bursts, burst mass fraction shown is combined from both recent bursts.}
\tablenotetext{f} {Includes Galactic foreground extinction.}
\end{splitdeluxetable*}


\begin{splitdeluxetable*}{cccccccccccBccccccccc}
\tabletypesize{\scriptsize}
\tablewidth{0pt}
\tablecaption{Post-burst Ages for \citet{Rowlands2015} sample \label{table:ages_rowlands}}
\tablehead{\colhead{RA (J2000)\tablenotemark{a}} & \colhead{Dec (J2000)} &  \colhead{z} &  \colhead{log M$_\star$/M$_\odot$} &  \colhead{SFH\tablenotemark{b}} & \multicolumn{3}{c}{Post-burst Age (Myr)} & \multicolumn{3}{c}{Age since Burst Start(Myr)} & \multicolumn{3}{c}{$\tau$ or $\Delta t$\tablenotemark{c}(Myr)} &  \multicolumn{2}{c}{Burst Light Fraction $y_f$ \tablenotemark{d}} &   \multicolumn{3}{c}{Burst Mass Fraction $m_{\rm burst}$} &  \colhead{$A_V$\tablenotemark{f}} \\
\colhead{} & \colhead{} &  \colhead{} &  \colhead{} &  \colhead{} & \colhead{(16\%)} & \colhead{(50\%)} &  \colhead{(84\%)} & \colhead{(16\%)} & \colhead{(50\%)} &  \colhead{(84\%)} & \colhead{(16\%)} & \colhead{(50\%)}  & \colhead{(84\%)} & \colhead{(burst 1)} & \colhead{(burst 2)\tablenotemark{e}} & \colhead{(16\%)} & \colhead{(50\%)} &  \colhead{(84\%)} & \colhead{(mag)}}
\startdata
     233.131990 &       57.882917 &          0.0394 &             9.8 & 1 & -20 &  -17 &  -15 &   37 &   39 &   42 &   25 &   25 &   25 &            1.00 & &            0.77 &            1.00 &            1.00 &            1.0 \\
     228.951270 &       20.022363 &          0.0363 &            10.6 & 2 &  -9 &   -4 &    5 &  130 &  135 &  145 &  100 &  100 &  100 &            0.27 &            0.73 &            0.57 &            1.00 &            1.00 &             1.6 \\
     225.401270 &       16.729684 &          0.0319 &             9.6 & 1 & -29 &  -25 &  -22 &   28 &   31 &   35 &   25 &   25 &   25 &            1.00 & &            0.77 &            1.00 &            1.00 &            1.4 \\
     246.455270 &       40.345214 &          0.0290 &            10.3 & 2 &   3 &   10 &   17 &  143 &  150 &  157 &  100 &  100 &  100 &            0.28 &            0.72 &            0.45 &            1.00 &            1.00 &             1.2 \\
     244.397560 &       14.052304 &          0.0338 &             9.8 & 2 &   1 &   10 &   24 &  141 &  150 &  164 &  100 &  100 &  100 &            0.27 &            0.64 &            0.09 &            0.13 &            0.63 &             1.6 \\
     252.923730 &       41.668378 &          0.0427 &            10.4 & 2 &  74 &   93 &  112 &  214 &  233 &  252 &  100 &  100 &  100 &            0.24 &            0.39 &            0.04 &            0.05 &            0.06 &             0.8 \\
     249.495290 &       13.859418 &          0.0469 &            10.2 & 2 &  54 &   85 &  103 &  194 &  225 &  243 &  100 &  100 &  100 &            0.25 &            0.41 &            0.05 &            0.06 &            0.08 &             1.2 \\
     232.702380 &       55.328833 &          0.0461 &            10.0 & 2 & 174 &  522 &  601 & 1214 & 1562 & 1641 & 1000 & 1000 & 1000 &            0.25 &            0.54 &            0.03 &            0.25 &            0.34 &             0.6 \\
     239.568480 &       52.489259 &          0.0486 &            10.3 & 2 &  27 &  127 &  167 &  167 &  267 &  307 &  100 &  100 &  100 &            0.15 &            0.21 &            0.02 &            0.02 &            0.03 &             0.4 \\
     247.178970 &       22.397116 &          0.0342 &            10.2 & 1 &-292 &   12 &   70 &  167 &  473 &  530 &  100 &  200 &  200 &            0.79 & &            0.01 &            0.02 &            0.03 &            1.6 \\
     237.803050 &       14.696400 &          0.0478 &            10.5 & 1 & 273 &  318 &  363 &  331 &  375 &  420 &   25 &   25 &   25 &            0.50 & &            0.04 &            0.05 &            0.06 &            0.6 \\

\enddata
\tablenotetext{a} {Columns 1-4 are from SDSS DR8 \citep{Aihara2011} and MPA-JHU catalogs \citep{Brinchmann2004, Tremonti2004}.}
\tablenotetext{b} {Number of recent bursts.}
\tablenotetext{c} {If SFH = 1, burst duration ($\tau$). If SFH = 2, separation between bursts ($\Delta t$).}
\tablenotetext{d} {If SFH = 2 recent bursts, light fraction for each burst is shown (burst mass fractions are the same for each burst, so the light fractions will be different).}
\tablenotetext{e} {If SFH = 2 recent bursts, burst mass fraction shown is combined from both recent bursts.}
\tablenotetext{f} {Includes Galactic foreground extinction.}
\end{splitdeluxetable*}



\begin{table}
\centering
\scriptsize
\caption{Fit residuals}
\label{table:resids}
\begin{tabular}{l r r}
\hline
\hline
Data & Mean & rms \\
\hline
FUV & 0.20 &  1.67 \\
NUV &-1.58 &  2.21 \\
u &-0.04 &  1.44 \\
g & 0.65 &  1.43 \\
r & 0.35 &  1.45 \\
i & 0.87 &  1.53 \\
z &-1.62 &  1.98 \\
D4000n &-2.76 &  2.91 \\
CN2 &-1.58 &  1.33 \\
Ca4227 & 0.48 &  1.04 \\
G4300 &-0.58 &  1.09 \\
Fe4383 & 1.22 &  1.43 \\
Ca4455 &-1.02 &  1.04 \\
Fe4531 &-0.62 &  1.21 \\
Ca4668 & 0.29 &  1.67 \\
Hb &-1.20 &  1.93 \\
Fe5015 &-0.90 &  1.92 \\
Mg1 & 0.14 &  1.38 \\
Mg2 & 0.54 &  1.48 \\
Mgb & 1.06 &  1.11 \\
Fe5270 &-0.12 &  1.22 \\
Fe5335 &-1.06 &  1.45 \\
Fe5406 &-0.44 &  1.16 \\
Fe5709 &-0.40 &  0.93 \\
Fe5782 &-0.25 &  0.93 \\
TiO1 &-0.25 &  1.01 \\
TiO2 & 1.87 &  1.49 \\
Hd A &-0.19 &  1.47 \\
Hg A &-1.59 &  1.45 \\

\hline
\end{tabular}
\end{table}

\begin{table*}
\centering
\scriptsize
\caption{Parameter residual covariance matrix}
\label{table:cov}
\begin{tabular}{l r r r r r r r r r r r r r r r }
 &FUV & NUV & u & g & r & i & z & D4000n & CN2 & Ca4227 & G4300 & Fe4383 & Ca4455 & Fe4531 & Ca4668\\
 \hline \hline
FUV &  1.00 &  0.28 & -0.02 & -0.18 & -0.23 & -0.04 &  0.14 &  0.12 & -0.08 & -0.08 & -0.17 & -0.13 & -0.08 & -0.10 & -0.05\\
NUV &  0.28 &  1.00 & -0.45 & -0.44 & -0.37 & -0.07 &  0.25 &  0.50 &  0.17 & -0.18 & -0.08 & -0.31 & -0.04 & -0.12 & -0.17\\
u & -0.02 & -0.45 &  1.00 &  0.37 &  0.09 & -0.10 & -0.18 & -0.35 & -0.07 &  0.00 & -0.02 &  0.14 &  0.02 &  0.08 &  0.08\\
g & -0.18 & -0.44 &  0.37 &  1.00 &  0.49 &  0.07 & -0.61 & -0.25 & -0.08 &  0.05 &  0.08 &  0.14 &  0.03 &  0.16 &  0.11\\
r & -0.23 & -0.37 &  0.09 &  0.49 &  1.00 &  0.63 & -0.80 & -0.10 &  0.02 &  0.20 &  0.22 &  0.26 &  0.09 &  0.15 &  0.14\\
i & -0.04 & -0.07 & -0.10 &  0.07 &  0.63 &  1.00 & -0.67 &  0.17 &  0.10 &  0.14 &  0.21 &  0.13 &  0.06 & -0.00 &  0.03\\
z &  0.14 &  0.25 & -0.18 & -0.61 & -0.80 & -0.67 &  1.00 & -0.01 & -0.06 & -0.07 & -0.20 & -0.16 & -0.01 & -0.12 & -0.08\\
D4000n &  0.12 &  0.50 & -0.35 & -0.25 & -0.10 &  0.17 & -0.01 &  1.00 &  0.13 & -0.20 & -0.05 & -0.34 & -0.05 & -0.18 & -0.15\\
CN2 & -0.08 &  0.17 & -0.07 & -0.08 &  0.02 &  0.10 & -0.06 &  0.13 &  1.00 & -0.11 &  0.18 & -0.12 &  0.17 &  0.11 & -0.12\\
Ca4227 & -0.08 & -0.18 &  0.00 &  0.05 &  0.20 &  0.14 & -0.07 & -0.20 & -0.11 &  1.00 & -0.02 &  0.23 &  0.07 &  0.03 &  0.16\\
G4300 & -0.17 & -0.08 & -0.02 &  0.08 &  0.22 &  0.21 & -0.20 & -0.05 &  0.18 & -0.02 &  1.00 & -0.13 &  0.17 &  0.17 & -0.08\\
Fe4383 & -0.13 & -0.31 &  0.14 &  0.14 &  0.26 &  0.13 & -0.16 & -0.34 & -0.12 &  0.23 & -0.13 &  1.00 &  0.20 &  0.13 &  0.17\\
Ca4455 & -0.08 & -0.04 &  0.02 &  0.03 &  0.09 &  0.06 & -0.01 & -0.05 &  0.17 &  0.07 &  0.17 &  0.20 &  1.00 &  0.23 &  0.01\\
Fe4531 & -0.10 & -0.12 &  0.08 &  0.16 &  0.15 & -0.00 & -0.12 & -0.18 &  0.11 &  0.03 &  0.17 &  0.13 &  0.23 &  1.00 &  0.07\\
Ca4668 & -0.05 & -0.17 &  0.08 &  0.11 &  0.14 &  0.03 & -0.08 & -0.15 & -0.12 &  0.16 & -0.08 &  0.17 &  0.01 &  0.07 &  1.00\\
Hb &  0.15 &  0.23 & -0.13 & -0.24 & -0.20 & -0.14 &  0.26 &  0.05 &  0.20 &  0.04 & -0.05 &  0.07 &  0.06 &  0.11 &  0.06\\
Fe5015 & -0.05 & -0.01 &  0.04 &  0.03 &  0.01 & -0.02 & -0.01 & -0.05 &  0.14 &  0.02 &  0.10 &  0.03 &  0.13 &  0.16 &  0.13\\
Mg1 &  0.03 & -0.15 &  0.12 &  0.11 &  0.09 &  0.08 & -0.11 & -0.23 & -0.07 &  0.06 &  0.05 & -0.03 & -0.04 &  0.02 &  0.11\\
Mg2 &  0.07 & -0.24 &  0.10 &  0.10 &  0.11 &  0.07 & -0.05 & -0.33 & -0.22 &  0.21 & -0.04 &  0.11 & -0.07 & -0.03 &  0.24\\
Mgb &  0.01 & -0.23 &  0.03 &  0.07 &  0.09 &  0.07 & -0.01 & -0.25 & -0.29 &  0.23 & -0.11 &  0.17 & -0.07 & -0.03 &  0.17\\
Fe5270 & -0.03 & -0.15 & -0.00 &  0.05 &  0.15 &  0.07 & -0.03 & -0.20 & -0.01 &  0.26 &  0.02 &  0.15 &  0.22 &  0.21 &  0.12\\
Fe5335 & -0.00 &  0.01 & -0.07 & -0.02 &  0.07 &  0.02 &  0.03 & -0.04 &  0.05 &  0.12 &  0.07 &  0.08 &  0.29 &  0.28 &  0.04\\
Fe5406 &  0.00 & -0.02 & -0.06 & -0.01 &  0.03 &  0.05 &  0.03 & -0.13 &  0.04 &  0.14 &  0.00 &  0.04 &  0.12 &  0.04 &  0.02\\
Fe5709 &  0.00 & -0.01 &  0.01 & -0.02 &  0.04 &  0.04 &  0.00 & -0.02 &  0.07 &  0.10 &  0.11 & -0.01 &  0.17 &  0.09 & -0.04\\
Fe5782 &  0.02 & -0.02 &  0.06 & -0.02 &  0.05 &  0.06 & -0.03 & -0.04 & -0.01 &  0.03 & -0.06 &  0.04 &  0.06 &  0.06 &  0.01\\
TiO1 & -0.01 & -0.08 &  0.12 &  0.07 &  0.02 & -0.01 &  0.01 & -0.09 & -0.05 &  0.07 & -0.02 &  0.08 &  0.10 &  0.02 &  0.06\\
TiO2 & -0.08 & -0.27 &  0.22 &  0.07 &  0.10 &  0.03 & -0.02 & -0.35 & -0.27 &  0.25 & -0.11 &  0.23 & -0.09 & -0.01 &  0.12\\
Hd A &  0.33 &  0.35 & -0.12 & -0.17 & -0.30 & -0.12 &  0.25 &  0.25 & -0.31 & -0.00 & -0.20 & -0.12 & -0.06 & -0.06 &  0.06\\
Hg A &  0.28 &  0.41 & -0.15 & -0.24 & -0.41 & -0.32 &  0.38 &  0.30 &  0.07 & -0.06 & -0.44 & -0.24 & -0.02 & -0.04 & -0.00\\

\hline
\end{tabular}
\end{table*}

\begin{table*}
\centering
\scriptsize
\caption{Parameter residual covariance matrix (continued)}
\begin{tabular}{l r r r r r r r r r r r r r r}
 &Hb & Fe5015 & Mg1 & Mg2 & Mgb & Fe5270 & Fe5335 & Fe5406 & Fe5709 & Fe5782 & TiO1 & TiO2 & Hd A & Hg A\\
 \hline \hline
FUV &  0.15 & -0.05 &  0.03 &  0.07 &  0.01 & -0.03 & -0.00 &  0.00 &  0.00 &  0.02 & -0.01 & -0.08 &  0.33 &  0.28\\
NUV &  0.23 & -0.01 & -0.15 & -0.24 & -0.23 & -0.15 &  0.01 & -0.02 & -0.01 & -0.02 & -0.08 & -0.27 &  0.35 &  0.41\\
u & -0.13 &  0.04 &  0.12 &  0.10 &  0.03 & -0.00 & -0.07 & -0.06 &  0.01 &  0.06 &  0.12 &  0.22 & -0.12 & -0.15\\
g & -0.24 &  0.03 &  0.11 &  0.10 &  0.07 &  0.05 & -0.02 & -0.01 & -0.02 & -0.02 &  0.07 &  0.07 & -0.17 & -0.24\\
r & -0.20 &  0.01 &  0.09 &  0.11 &  0.09 &  0.15 &  0.07 &  0.03 &  0.04 &  0.05 &  0.02 &  0.10 & -0.30 & -0.41\\
i & -0.14 & -0.02 &  0.08 &  0.07 &  0.07 &  0.07 &  0.02 &  0.05 &  0.04 &  0.06 & -0.01 &  0.03 & -0.12 & -0.32\\
z &  0.26 & -0.01 & -0.11 & -0.05 & -0.01 & -0.03 &  0.03 &  0.03 &  0.00 & -0.03 &  0.01 & -0.02 &  0.25 &  0.38\\
D4000n &  0.05 & -0.05 & -0.23 & -0.33 & -0.25 & -0.20 & -0.04 & -0.13 & -0.02 & -0.04 & -0.09 & -0.35 &  0.25 &  0.30\\
CN2 &  0.20 &  0.14 & -0.07 & -0.22 & -0.29 & -0.01 &  0.05 &  0.04 &  0.07 & -0.01 & -0.05 & -0.27 & -0.31 &  0.07\\
Ca4227 &  0.04 &  0.02 &  0.06 &  0.21 &  0.23 &  0.26 &  0.12 &  0.14 &  0.10 &  0.03 &  0.07 &  0.25 & -0.00 & -0.06\\
G4300 & -0.05 &  0.10 &  0.05 & -0.04 & -0.11 &  0.02 &  0.07 &  0.00 &  0.11 & -0.06 & -0.02 & -0.11 & -0.20 & -0.44\\
Fe4383 &  0.07 &  0.03 & -0.03 &  0.11 &  0.17 &  0.15 &  0.08 &  0.04 & -0.01 &  0.04 &  0.08 &  0.23 & -0.12 & -0.24\\
Ca4455 &  0.06 &  0.13 & -0.04 & -0.07 & -0.07 &  0.22 &  0.29 &  0.12 &  0.17 &  0.06 &  0.10 & -0.09 & -0.06 & -0.02\\
Fe4531 &  0.11 &  0.16 &  0.02 & -0.03 & -0.03 &  0.21 &  0.28 &  0.04 &  0.09 &  0.06 &  0.02 & -0.01 & -0.06 & -0.04\\
Ca4668 &  0.06 &  0.13 &  0.11 &  0.24 &  0.17 &  0.12 &  0.04 &  0.02 & -0.04 &  0.01 &  0.06 &  0.12 &  0.06 & -0.00\\
Hb &  1.00 &  0.31 & -0.04 &  0.04 & -0.04 &  0.04 &  0.09 &  0.09 & -0.04 &  0.01 & -0.07 & -0.01 &  0.10 &  0.40\\
Fe5015 &  0.31 &  1.00 &  0.05 &  0.04 & -0.01 &  0.10 &  0.11 &  0.09 & -0.06 &  0.04 & -0.02 & -0.09 & -0.11 &  0.03\\
Mg1 & -0.04 &  0.05 &  1.00 &  0.51 &  0.12 &  0.09 & -0.08 &  0.07 & -0.04 &  0.04 &  0.02 &  0.07 &  0.07 & -0.04\\
Mg2 &  0.04 &  0.04 &  0.51 &  1.00 &  0.53 &  0.16 &  0.02 &  0.17 & -0.07 &  0.08 &  0.08 &  0.32 &  0.20 & -0.00\\
Mgb & -0.04 & -0.01 &  0.12 &  0.53 &  1.00 &  0.12 &  0.04 &  0.11 & -0.04 &  0.01 &  0.06 &  0.29 &  0.13 & -0.05\\
Fe5270 &  0.04 &  0.10 &  0.09 &  0.16 &  0.12 &  1.00 &  0.36 &  0.20 &  0.17 &  0.18 &  0.10 &  0.06 &  0.04 &  0.04\\
Fe5335 &  0.09 &  0.11 & -0.08 &  0.02 &  0.04 &  0.36 &  1.00 &  0.25 &  0.15 &  0.13 &  0.01 & -0.03 &  0.07 &  0.12\\
Fe5406 &  0.09 &  0.09 &  0.07 &  0.17 &  0.11 &  0.20 &  0.25 &  1.00 &  0.03 &  0.11 & -0.02 &  0.08 &  0.08 &  0.07\\
Fe5709 & -0.04 & -0.06 & -0.04 & -0.07 & -0.04 &  0.17 &  0.15 &  0.03 &  1.00 &  0.06 &  0.08 &  0.05 &  0.00 & -0.02\\
Fe5782 &  0.01 &  0.04 &  0.04 &  0.08 &  0.01 &  0.18 &  0.13 &  0.11 &  0.06 &  1.00 &  0.04 &  0.05 &  0.08 &  0.02\\
TiO1 & -0.07 & -0.02 &  0.02 &  0.08 &  0.06 &  0.10 &  0.01 & -0.02 &  0.08 &  0.04 &  1.00 &  0.12 &  0.05 &  0.03\\
TiO2 & -0.01 & -0.09 &  0.07 &  0.32 &  0.29 &  0.06 & -0.03 &  0.08 &  0.05 &  0.05 &  0.12 &  1.00 &  0.01 & -0.10\\
Hd A &  0.10 & -0.11 &  0.07 &  0.20 &  0.13 &  0.04 &  0.07 &  0.08 &  0.00 &  0.08 &  0.05 &  0.01 &  1.00 &  0.51\\
Hg A &  0.40 &  0.03 & -0.04 & -0.00 & -0.05 &  0.04 &  0.12 &  0.07 & -0.02 &  0.02 &  0.03 & -0.10 &  0.51 &  1.00\\

\hline
\end{tabular}
\end{table*}

\clearpage

\appendix

\section{Tests of Parameter Errors}
\label{sec:appdxa}
Here, we consider several tests of the robustness of the presented uncertainties on the derived ages, burst mass fractions, and burst durations. We test whether the derived parameter errors are reasonable, despite the high $\chi^2$ values (median reduced $\chi^2/\nu \sim2.5$) by performing a jackknife test as an additional way to calculate the parameter errors. We fit each galaxy many times, removing one datapoint each time, then compare the mean and variance of the derived parameters to the original. The mean of the ratio of the initial errors to the jackknife-derived errors is consistent with 1. This indicates that the derived parameter errors are meaningful, and not severely over- or under-estimated.

We also test the validity of the errors using a bootstrap approach. In Table \ref{table:resids} we show the mean and rms of the fit residuals, and in Table \ref{table:cov} we show the full covariance matrix of the fit residuals. Because many of the parameters and parameter residuals are correlated, the jackknife test may not be valid. Using a bootstrap approach, we fit each galaxy many times, sampling randomly with replacement from the set of datapoints, then comparing the mean and variance of the derived parameters to the original. The mean of the ratio of the initial errors to the bootstrap-derived errors is 0.86.

We consider an additional test on the parameter errors by fitting synthetic data using our method. We use the same set of synthetic data described in \S\ref{sec:synth}, produced by adding noise to simulated data, and apply the age-dating method to recover the input parameters. We draw randomly from the fit grid in age, burst light fraction, and burst duration, keeping a constant A$_V$ and metallicity. Random errors are drawn from the distribution of uncertainties of the real data and applied to these simulated data. We apply the age-dating method to these synthetic data, testing whether the method recovers the original parameters or adds any systematic biases. The fit ages are within the 68\%ile uncertainty range of the input ages 94\% of the time, the fit burst light fractions are within the 68\%ile uncertainty range of the input burst light fractions 91\% of the time, and the fit burst durations are within the 68\%ile uncertainty range of the input burst durations 92\% of the time. The SFH (single or double burst) is the same as the input 77\% of the time. The median $\chi^2/\nu  = 1.09$ for the best-fit models of the simulated data. This test demonstrates that our method reliably finds the best-fitting model.

Because the residual of D$_n$(4000) is the largest, we explore whether aperture bias between the fiber-based D$_n$(4000) measure and the continuum shape traced by the total magnitudes is responsible. However, we find no change in the mean D$_n$(4000) residual with redshift, and removing this datapoint does not significantly alter the results. Neither the $\chi^2$ of the fits nor the parameter errors show any dependence on redshift, excluding the possibility of size discrepancies driving the goodness of fit. We also test the use of various measures of the UV photometry, and find no evidence that aperture bias is responsible for the large D$_n$(4000) mean residual or other non-zero mean residuals between the data and best-fit models. 

Because mismatches in velocity dispersion can also affect our line measurements of the model relative to the data, we test convolving our model spectra with 100, 150, and 200 km/s velocity dispersions. We find no cases where the derived post-burst ages, burst light fractions, or burst durations fall outside of our inferred 68\%ile error ranges. The only parameter with a non-zero change in the median difference from changing the velocity dispersion is the burst light fraction, with a median difference of 0.01, well below our typical uncertainty.

We test the effects of any additional parameters outside of the SFH that could influence the quality of our fits (IMF, dust law, metallicity, SPS model, recent SF, the shape of the old stellar population), as described in \S3. Our primary goal in this modeling and fitting procedure is to determine parameterized quantities relating to the time elapsed since a starburst and the nature of that starburst. While better $\chi^2$ fits would be obtained by adding more parameters to the SFH descriptions, the results would be less useful. That approach would be valuable if the goal were accurate stellar masses (for example). Nonetheless, our derived stellar masses are within the fit errors of those by the MPA-JHU group.

We derive parameter errors on the post-starburst ages using those determined for the ages since the starburst began; here we assess the robustness of that choice. We have been using the fit error on the age since the starburst began for the post-starburst age error, which is conservative. While the post-starburst age is the derived quantity after the fit, it is the easier physical quantity to measure, as the starburst age is correlated with the burst duration, but the post-starburst age is more independent. We test whether including both the uncertainties in the starburst age and in the burst duration results in significantly higher errors for the post-starburst ages. We use a Monte-Carlo method to propagate the errors in the starburst age and burst duration by sampling from the error ranges for each quantity, and compare these errors to the fit errors on the starburst age. Assuming the starburst age and burst duration errors are uncorrelated is a conservative approach, likely to overestimate the post-starburst age errors. However, we find that the starburst age errors and estimated post-starburst age errors have a median difference of 15 Myr, much less than the typical fit uncertainty of around 50 Myr. Thus, the errors we present on the post-starburst ages are robust to our method of deriving post-starburst ages.

\section{Tests on other post-starburst samples}
\label{sec:appdxb}

We test the robustness of our age-dating method for our sample of post-starburst galaxies in a variety of ways in \S3, but are several differences between our post-starburst sample and those of \citet{Alatalo2016} and \citet{Rowlands2015} that may violate our assumption that the 
method can be used for all three samples without introducing systematic differences. The primary difference between the samples is the presence of emission lines in the \citet{Alatalo2016} and \citet{Rowlands2015} post-starbursts. We alter our method to account for this, as described in \S\ref{sec:spogs}. We have established that the star formation histories of the \citet{Alatalo2016} and \citet{Rowlands2015} galaxies are similar to our post-starbursts, and thus the choices of stellar population model, dust law, and IMF should be the same for all samples. The remaining uncertainties then are whether the younger ages of the other samples affect the systematic uncertainties with metallicity and whether the different redshift ranges violate any assumptions we have made about the lack of significant aperture bias.

We test the impact of the metallicity uncertainty, as we did for our main post-starburst sample (see \S\ref{sec:metallicity} for more detail). For our post-starburst sample, the metallicity uncertainty adds a systematic uncertainty of 14\% to the post-starburst age, or a median uncertainty of 56 Myr, and a systematic uncertainty of 23\% to the burst mass fraction, or a median uncertainty of 0.021. The SPOGs \citep{Alatalo2016} and \citet{Rowlands2015} samples have lower median systematic uncertainty, but higher fractional uncertainties on the post-starburst age when we consider only those with post-burst ages greater than zero. For the SPOGs and \citet{Rowlands2015} samples, the metallicity uncertainty adds a systematic uncertainty of 18\% to the post-starburst age (when post-burst age is greater than zero), or a median uncertainty of 33 Myr (for the full samples), and a systematic uncertainty of 28\% to the burst mass fraction, or a median uncertainty of 0.012.

We test to make sure the aperture bias considerations for our post-starburst galaxies in \S\ref{sec:apbias} are also true for the \citet{Alatalo2016} and \citet{Rowlands2015} samples. The redshift ranges for the three samples are not identical; the 68 percentile ranges are $0.054<z<0.189$ for our post-starburst galaxies, $0.034<z<0.147$ for the SPOGs, and $0.032<z<0.049$ for the \citet{Rowlands2015} sample.  As for our post-starburst galaxies, most of the \citet{Alatalo2016} and \citet{Rowlands2015} samples have galaxies with $z>0.017$, where a 500 pc radius would be included in the 3\arcsec\ SDSS fiber. 97\% of the SPOGs and all of the \citet{Rowlands2015} sample meet this criterion. The difference in the total and fiber magnitude $u-r$ colors is similarly small. For the SPOGs sample, the difference in the $u-r$ {\tt modelmag} and {\tt fibermag} colors is less than (12\% of) the typical uncertainties in the colors. For the \citet{Rowlands2015} sample, the difference in the $u-r$ {\tt modelmag} and {\tt fibermag} colors is also less than (29\% of) the typical uncertainties in the colors.

\clearpage

\end{document}